\documentclass[11pt,a4paper]{article}

\usepackage[colorlinks=true, linkcolor=black!50!blue, urlcolor=blue, citecolor=blue, anchorcolor=blue]{hyperref}
\usepackage[font=small,labelfont=bf,margin=0mm,labelsep=period,tableposition=top]{caption}
\usepackage[a4paper,top=3cm,bottom=2.5cm,left=2.5cm,right=2.5cm,bindingoffset=0mm]{geometry}

\usepackage{graphicx}
\usepackage{float}
\usepackage{afterpage}
\usepackage{epsfig,cite}
\usepackage{amssymb}
\usepackage{amsmath}
\usepackage{bm}
\usepackage{dsfont}
\usepackage{multirow}
\usepackage{url}
\usepackage{xcolor}
\usepackage{float}
\usepackage{afterpage}
\usepackage{ulem}

\usepackage{url}
\usepackage{hyperref}

\usepackage{multirow,booktabs,multirow}

\newcommand{\be}{\begin{equation}}
\newcommand{\ee}{\end{equation}}
\newcommand{\bea}{\begin{eqnarray}}
\newcommand{\eea}{\end{eqnarray}}
\newcommand{\bi}{\begin{itemize}}
\newcommand{\ei}{\end{itemize}}
\newcommand{\ben}{\begin{enumerate}}
\newcommand{\een}{\end{enumerate}}

\newcommand{\lc}{\left[}
\newcommand{\rc}{\right]}
\newcommand{\lp}{\left(}
\newcommand{\rp}{\right)}

\def\frac#1#2{{{#1}\over {#2}}}
\def\gsim{\mathrel{\rlap{\lower4pt\hbox{\hskip1pt$\sim$}}
    \raise1pt\hbox{$>$}}}       
\def\lsim{\mathrel{\rlap{\lower4pt\hbox{\hskip1pt$\sim$}}
    \raise1pt\hbox{$<$}}}

\newcommand{\rep}{\mathrm{rep}}

\newcommand{\draft}[1]{}

\def\beq{\begin{equation}}
\def\eeq{\end{equation}}

\def\lapprox{\lower .7ex\hbox{$\;\stackrel{\textstyle <}{\sim}\;$}}
\def\gapprox{\lower .7ex\hbox{$\;\stackrel{\textstyle >}{\sim}\;$}}

\numberwithin{equation}{section}
\numberwithin{figure}{section}
\numberwithin{table}{section}

\usepackage{tabularx}
\newcolumntype{C}[1]{>{\centering\arraybackslash}p{#1}}

\begin{document}
\newgeometry{top=1.5cm,bottom=1.5cm,left=2.5cm,right=2.5cm,bindingoffset=0mm}
\begin{figure}[h]
  \includegraphics[width=0.32\textwidth]{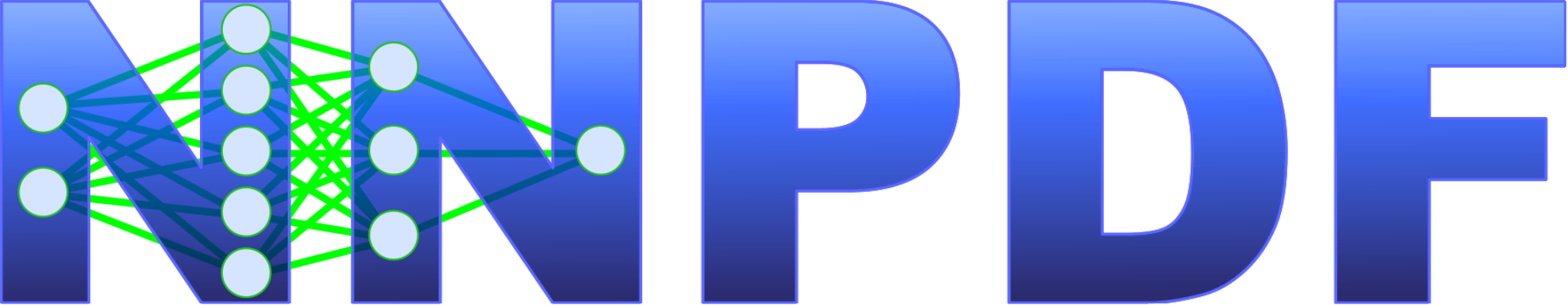}
\end{figure}
\vspace{-2.0cm}
\begin{flushright}
Nikhef/2019-005\\
\end{flushright}
\vspace{2cm}

\begin{center}
  {\Large \bf
    Nuclear Parton Distributions from Lepton-Nucleus Scattering \\[0.3cm]
   and the Impact of an Electron-Ion Collider
  }
\vspace{1.4cm}

{\bf  The NNPDF Collaboration:} \\[0.1cm]

 Rabah Abdul Khalek, Jacob J. Ethier, and Juan Rojo\\

\vspace{0.4cm}
       {\it
             Department of Physics and Astronomy, Vrije Universiteit
  Amsterdam, 1081 HV
  Amsterdam, \\
  Nikhef Theory Group, Science Park 105, 1098 XG Amsterdam, The
  Netherlands.
       }

       \vspace{1.0cm}

       {\bf \large Abstract}
       
\end{center}

We present a first determination of the nuclear parton distribution functions (nPDF) 
based on the NNPDF methodology: nNNPDF1.0.
This analysis is based on neutral-current deep-inelastic
structure function data and is performed up to NNLO
in QCD calculations with heavy quark mass effects.
For the first time in the NNPDF fits,
the $\chi^2$ minimization is achieved using
stochastic gradient descent
with reverse-mode automatic 
differentiation (backpropagation).
We validate the robustness
of the fitting methodology through closure tests,
assess the perturbative stability of the resulting nPDFs, and
compare them with other recent analyses.
The nNNPDF1.0 distributions satisfy the boundary condition whereby the
NNPDF3.1 proton PDF central values and uncertainties are reproduced at $A=1$,
which introduces important constraints particularly for low-$A$ nuclei.
We also investigate the information that would be provided
by an Electron-Ion Collider (EIC), finding that EIC measurements would
significantly constrain the nPDFs down to $x\simeq 5\times 10^{-4}$.
Our results represent the first-ever nPDF determination obtained
using a Monte Carlo methodology consistent with that of state-of-the-art
proton PDF fits, and provide the foundation for a subsequent
global nPDF analyses including also proton-nucleus
data.

\clearpage

\tableofcontents

\section{Introduction}
\label{sec:introduction}

It has been known for more than three decades~\cite{Aubert:1983xm} that the
parton distribution functions (PDFs) of nucleons bound within 
nuclei, more simply referred to as nuclear PDFs (nPDFs)~\cite{Zurita:2018vrs,Paukkunen:2018kmm}, 
can be modified with respect to their free-nucleon counterparts~\cite{Gao:2017yyd}.
Since MeV-scale nuclear binding effects were expected to be negligible 
compared to the typical momentum transfers ($Q\gsim 1$ GeV) in hard-scattering reactions
such as deep-inelastic lepton-nucleus scattering, such a phenomena came as a surprise 
to many in the physics community. 
Despite active experimental
and theoretical investigations, the underlying mechanisms that drive
in-medium modifications of nucleon substructure are yet to be
fully understood. 
The determination of nPDFs is therefore relevant
to improve our fundamental understanding of the strong 
interactions in the nuclear environment.

In addition to pinning down
the dynamics of QCD in heavy nuclei,
nPDFs are an essential ingredient for the interpretation of
heavy ion collisions at RHIC and the LHC, in particular
for the characterization of the Quark-Gluon
Plasma (QGP)~\cite{Abreu:2007kv,Adams:2005dq} via hard probes.
Moreover, a reliable determination of the nuclear PDFs 
is required to separate the hot nuclear matter (QGP) from the cold nuclear
matter effects that will in general already be present in the initial stages
of the heavy ion collision.

The importance of nPDF fits is further highlighted by their
contribution to the quark flavor separation in global PDF analyses
of the proton~\cite{Ball:2014uwa,Dulat:2015mca,Harland-Lang:2014zoa,Alekhin:2017kpj}.
Even with constraints from related
processes such as gauge boson production
at the Tevatron and the LHC, information provided by 
neutrino-induced charged current
deep-inelastic scattering on heavy nuclear
targets play a critical role in disentangling the 
proton's quark and antiquark distributions.
However, given the current precision of proton PDF fits, neglecting the nuclear
uncertainties associated with neutrino-nucleus scattering may not be well justified
anymore~\cite{Ball:2018twp}, as opposed to the situation some years ago~\cite{Ball:2009mk}.

Lastly, nPDF extractions can sharpen
the physics case of future high-energy lepton-nucleus colliders
such as the Electron-Ion Collider (EIC)~\cite{Aschenauer:2017oxs} and the
Large Hadron electron Collider (LHeC)~\cite{AbelleiraFernandez:2012cc,Helenius:2016hcu}, 
which will probe nuclear
structure deep in the region of small parton momentum fractions, $x$,
and aim to unravel novel
QCD dynamics such as non-linear (saturation) effects.
The latter will only be possible provided that a faithful
estimate of the nPDF uncertainties at small $x$ can be attained, similar
to what was required for the recent discovery of BFKL dynamics from
the HERA structure function data~\cite{Ball:2017otu}.

Unfortunately, the determination
of the nuclear PDFs is hampered by the rather limited experimental dataset
available.
In fact, until a few years ago, most nPDF analyses~\cite{Eskola:2009uj,deFlorian:2003qf,
Hirai:2007sx,Kovarik:2015cma,deFlorian:2011fp,Khanpour:2016pph}
were largely based on
fixed-target DIS structure functions in
lepton-nucleus scattering (with kinematic coverage
restricted to $x\gsim 0.01$) supplemented by some Drell-Yan cross-sections.
A major improvement in this respect
has been 
the recent availability of data on hard-scattering cross-sections from proton-lead 
collisions at the LHC, with processes ranging from
jet~\cite{Adam:2015hoa,Adam:2016dau,Adam:2015xea,Aad:2016zif,Chatrchyan:2014hqa} and electroweak boson
production~\cite{1742-6596-612-1-012009,ATLAS-CONF-2015-056,Aad:2015gta,Khachatryan:2015pzs,Khachatryan:2015hha}, to heavy quark production~\cite{CMS-PAS-HIN-15-012,Adam:2016mkz,Adam:2015qda,Abelev:2014hha,Khachatryan:2015sva,Khachatryan:2015uja,Aaij:2017gcy,Aaij:2019lkm} among several
others.
Indeed, 
measurements of hard probes in p+Pb collisions provide useful
information to constrain the nPDFs, as was
demonstrated by a few recent studies~\cite{Eskola:2016oht,Kusina:2016fxy}.

On the other hand, a survey of 
various nPDF determinations reveals 
limitations that are of methodological origin as well.
First of all, current nuclear PDF fits
rely on model-dependent assumptions
for the parameterization of the non-perturbative $x$ and 
atomic mass number $A$ dependence,
resulting in a theoretical bias whose magnitude is difficult to assess.  
Moreover, several nPDF sets are extracted in terms of a proton baseline (to which the former must reduce in the $A\to 1$ limit) that have been determined by other groups based
on fitting methodologies and theoretical settings which might not fully equivalent,
for instance, in the prescriptions used to estimate the nPDF uncertainties.
Finally, PDF uncertainties are often estimated using the Hessian method,
which is restricted to a Gaussian approximation
with {\it ad hoc} tolerances, introducing 
a level of arbitrariness in their statistical
interpretation.

Motivated by this need for a reliable and consistent
determination of nuclear PDFs and their uncertainties,
we present in this work a first nPDF analysis
based on the NNPDF methodology~\cite{Forte:2002fg,DelDebbio:2004qj,DelDebbio:2007ee,Ball:2008by,Rojo:2008ke,Ball:2009qv,Ball:2010de,Ball:2011mu,Ball:2011uy,Ball:2012cx}: nNNPDF1.0.
In this initial study, we restrict our analysis to 
neutral-current nuclear deep-inelastic
structure function measurements, and compute the corresponding predictions
in QCD up to NNLO in the $\alpha_s$ expansion.
Moreover, heavy quark mass effects are included using
the FONLL general-mass variable-flavor number scheme~\cite{Forte:2010ta}.
Since the nPDFs are determined using the same theoretical and methodological 
framework as the NNPDF3.1 proton PDFs, we are able to impose the 
boundary condition in a consistent manner so that the nNNPDF1.0 results 
reproduce both the NNPDF3.1 central values and uncertainties when 
evaluated at $A=1$.

The nNNPDF1.0 sets are constructed following
the general fitting methodology outlined in previous NNPDF studies,
which utilizes robust Monte Carlo techniques to obtain 
a faithful estimate of nPDF uncertainties.
In addition, in this study we implement for
the first time stochastic gradient descent 
to optimize the model parameters.
This is performed using {\tt TensorFlow}~\cite{tensorflow2015-whitepaper}, an open 
source machine learning library in which the gradients of the $\chi^2$ 
function can be computed via automatic differentiation.
Together with several other 
improvements, we present a validation of the 
nNNPDF1.0 methodology through closure tests.

As a first phenomenological application of the nNNPDF1.0 sets, 
we quantify the impact of future lepton-nucleus scattering
measurements provided by an Electron-Ion Collider.
Using pseudo-data generated with different electron and nucleus 
beam energy configurations, we perform fits to
quantify the effect on the nNNPDF1.0
uncertainties and discuss the extent to which
novel QCD dynamics can be revealed.
More specifically, we demonstrate how the EIC would lead to a 
significant reduction of
the nPDF uncertainties at small $x$, paving the way for
a detailed study of nuclear matter in a presently unexplored 
kinematic region.

The outline of this paper is the following.
In Sect.~\ref{sec:expdata} we present
the input experimental data used in this analysis, namely
ratios of neutral-current deep-inelastic structure functions,
followed by a discussion of the corresponding
theoretical calculations.
The description of the fitting strategy, including
the updated minimization procedure and choice
of parameterization, is presented in Sect.~\ref{sec:fitting}.
We discuss the validation of our fitting methodology 
via closure tests in Sect.~\ref{sec:closuretests}.
The main results of this work, the nNNPDF1.0 nuclear
parton distributions, are then presented in Sect.~\ref{sec:results}.
In Sect.~\ref{sec:eic} we quantify the impact on the nPDFs
from future EIC measurements of nuclear structure functions.
Lastly, in Sect.~\ref{sec:summary} we summarize and
discuss the outlook for future studies.

\section{Experimental data and theory calculations}
\label{sec:expdata}
\label{sec:theory}

In this section we review the formalism that describes
deep-inelastic scattering (DIS) of charged leptons off of nuclear targets.
We then present
the data sets that have been used
in the present determination of the nuclear PDFs, discussing also
the kinematical cuts and the treatment of experimental
uncertainties.
Lastly, we discuss the theoretical framework for the evaluation of
the DIS structure functions, including the quark
and anti-quark flavor decomposition,
the heavy quark mass effects, and the software tools used for
the numerical calculations.

\subsection{Deep-inelastic lepton-nucleus scattering}

The description of hard-scattering collisions involving
nuclear targets begins with collinear factorization theorems
in QCD that are identical to those in free-nucleon scattering.

For instance, in deep inelastic lepton-nucleus scattering,
the leading power
contribution to the cross section can be 
expressed in terms of a hard partonic cross section 
that is unchanged with respect to the corresponding lepton-nucleon 
reaction, and the 
nonperturbative PDFs of the nucleus. 
Since these nPDFs are defined by the same leading twist 
operators as the free nucleon PDFs
but acting instead on nuclear states, the modifications from 
internal nuclear effects are naturally contained within 
the nPDF definition and the factorization 
theorems remain valid assuming power suppressed corrections are 
negligible in the perturbative regime, $Q^2 \gsim$ 1 GeV$^2$.
We note, however, that this assumption may not hold for 
some nuclear processes, and therefore must be 
studied and verified through the analysis of relevant physical 
observables.

We start now by briefly reviewing the
definition of the DIS structure functions and of the associated 
kinematic variables which are relevant for the description of 
lepton-nucleus scattering.
The double differential cross-section for scattering
of a charged lepton off a nucleus with 
atomic mass number $A$ is given by
\be
\label{eq:sec2fullcrosssection}
\frac{d^2 \sigma^{{\rm NC},l^\pm}}{dx dQ^2}(x,Q^2,A) =
\frac{2\pi \alpha^2}{xQ^4} \lc Y_+\,F_2^{\rm NC}(x,Q^2,A)
\mp Y_-\, xF_3^{\rm NC}(x,Q^2,A) - y^2\,F_L^{\rm NC}(x,Q^2,A)\ \rc 
\ee
where $Y_\pm = 1 \pm (1-y)^2$ and the usual DIS
kinematic variables can be expressed in Lorentz-invariant form as
\be
\label{eq:kinvariables}
x = \frac{Q^2}{2P\cdot q} \, ,\qquad Q^2 = -q^2 \, , \qquad y = \frac{q\cdot P}{k\cdot P} \,.
\ee
Here the four-momenta of the target nucleon, the incoming charged lepton, and the
exchanged virtual boson ($\gamma^*$ or $Z$) are denoted by $P$, $k$, and $q$, 
respectively. 
The variable $x$ is defined here to be the standard Bjorken scaling variable, 
which at leading order can be interpreted as the fraction of the nucleon's momentum 
carried by the struck parton, and $y$ is known as the inelasticity.
Lastly, the virtuality of the exchanged boson is $Q^2$,
which represents the hardness of the scattering reaction. 

As will be discussed below,
the maximum value of the momentum transfer $Q^2$
in the nNNPDF1.0 input dataset is $Q^2_{\rm max} \simeq 200$ GeV$^2$ 
(see Fig.~\ref{figkinplot}).
Given that $Q^2_{\rm max}\ll M_Z^2$,
the contribution from the parity-violating $xF_3$ structure functions
and the contributions to $F_2$ and $F_L$ arising
from $Z$ boson exchange can be safely neglected.
Therefore, for the kinematic range relevant to the description
of available nuclear DIS data, Eq.~(\ref{eq:sec2fullcrosssection})
simplifies to
\be
\label{eq:sec2fullcrosssection2}
\frac{d^2 \sigma^{{\rm NC},l^\pm}}{dx dQ^2}(x,Q^2,A) =
\frac{2\pi \alpha^2}{xQ^4} Y_+F_2^{\rm NC}(x,Q^2,A) \lc 1
 - \frac{y^2}{1+(1-y)^2}\,\frac{F_L^{\rm NC}(x,Q^2,A)}{F_2^{\rm NC}(x,Q^2,A)}\ \rc \, ,
 \ee
 where only the photon-exchange contributions
 are retained for the $F_2$ and $F_L$ structure functions.
 In Eq.~(\ref{eq:sec2fullcrosssection2})
 we have isolated the dominant $F_2$ dependence, since the second
 term is typically rather small.
 Note that since the center of mass energy of the lepton-nucleon
 collision $\sqrt{s}$ is determined by
 \be
 s = \lp k\,+\,P\rp^2 \simeq 2 k\cdot P = \frac{Q^2}{xy} \, ,
 \ee
 where hadron and lepton masses have been neglected, measurements
 with the same values for $x$ and $Q^2$ but different center of mass energies 
 $\sqrt{s}$ will lead to a different value of the prefactor
 in front of the $F_L/F_2$ ratio in Eq.~(\ref{eq:sec2fullcrosssection2}),
 allowing in principle the separation of the two structure functions
 as in the free proton case.
 
\subsection{Experimental data}

In this analysis,
we include all available inclusive DIS measurements
of neutral-current structure functions on nuclear targets.
In particular, we use data from the
EMC~\cite{Aubert:1987da,Ashman:1988bf,Arneodo:1989sy,Ashman:1992kv}, 
NMC~\cite{Amaudruz:1995tq,Arneodo:1995cs,Arneodo:1996rv,Arneodo:1996ru},
and BCDMS experiments at CERN, E139 measurements from 
SLAC~\cite{PhysRevD.49.4348}, and E665 data
from Fermilab.
The measurements of nuclear structure functions
are typically presented as ratios of the form
\be
\label{eq:Sect2Rf2}
R_{F_2}\lp x, Q^2, A_1, A_2\rp \equiv
\frac{F_2(x,Q^2,A_2)}{F_2(x,Q^2,A_1)} \, ,
\ee
where $A_1$ and $A_2$
are the atomic mass numbers of the two different nuclei.
Some of the experimental measurements included in this analysis
are presented instead as ratios of DIS cross-sections.
As discussed earlier, the double-differential
DIS cross-sections are related
to the $F_2$ and $F_L$ structure functions by
\be
\label{eq:ratioReducedXsec}
\frac{d^2 \sigma^{\rm NC}}{dx dQ^2}(x,Q^2,A) \propto F_2\lc
1-\frac{y^2}{1+(1-y)^2}\frac{F_L}{F_2}
\rc  \,.
\ee
Therefore, one should in principle account for the contributions from the
longitudinal structure function $F_L$ to cross-section ratios measured
by experiment.
However, it is well known that the ratio $F_L/F_2$ exhibits a very weak dependence
with $A$~\cite{Amaudruz:1992wn,Dasu:1988ru}, and
therefore the second term in Eq.~(\ref{eq:ratioReducedXsec})
cancels out to a good approximation when taking
ratios between different nuclei.
In other words, we can exploit the fact that
\be
\frac{d^2 \sigma^{\rm NC}(x,Q^2,A_2)/dxdQ^2}{
d^2 \sigma^{\rm NC}(x,Q^2,A_1)/dxdQ^2} \simeq
\frac{F_2(x,Q^2,A_2)}{F_2(x,Q^2,A_1)} = R_{F_2}\lp x, Q^2, A_1, A_2\rp \, ,
\ee
in which then the ratios
of DIS cross-sections for $Q \ll M_Z$ in the form
of Eq.~(\ref{eq:ratioReducedXsec}) are equivalent
to ratios of the $F_2$ structure functions.
Lastly, it is important to note that whenever 
the nuclei involved in the 
measurements are not isoscalar,
the data is corrected to give isoscalar ratios and an 
additional source of systematic error is added as a
result of this conversion.

Summarized in Table~\ref{dataset} are the different types of nuclei 
measured by the experiments included in the nNNPDF1.0 analysis.
  For each dataset, we indicate the nuclei $A_1$ and $A_2$
  that are used to construct the
  structure function ratios in Eq.~\ref{eq:Sect2Rf2}, 
  quoting
  explicitly the corresponding atomic mass numbers.
  We also display the number of data points that
  survive the baseline kinematical cuts, and give
  the corresponding publication references.

\begin{table}[p]
  \centering
  \small
   \renewcommand{\arraystretch}{1.45}
\begin{tabular}{c c c c}
Experiment & ${\rm A}_1/{\rm A}_2$ & ${\rm N}_{\rm dat}$ & Reference\\
\toprule
  SLAC E-139 & $^4$He/$^2$D & 3 & \cite{Gomez:1993ri} \\
  NMC 95, re. & $^4$He/$^2$D & 13 & \cite{Amaudruz:1995tq}\\
\midrule
  NMC 95 & $^6$Li/$^2$D & 12 & \cite{Arneodo:1995cs}\\
\midrule
  SLAC E-139 & $^9$Be/$^2$D & 3 & \cite{Gomez:1993ri}\\
  NMC 96 & $^9$Be/$^{12}$C & 14 & \cite{Arneodo:1996rv}\\
\midrule
  EMC 88, EMC 90 & $^{12}$C/$^2$D & 12 & \cite{Ashman:1988bf,Arneodo:1989sy}\\
  SLAC E-139 & $^{12}$C/$^2$D & 2 & \cite{Gomez:1993ri}\\
  NMC 95, NMC 95, re.  & $^{12}$C/$^2$D & 26 & \cite{Arneodo:1995cs,Amaudruz:1995tq}\\
  FNAL E665 & $^{12}$C/$^2$D & 3 & \cite{Adams:1995is}\\
  NMC 95, re. & $^{12}$C/$^6$Li & 9 & \cite{Amaudruz:1995tq}\\
\midrule
  BCDMS 85 & $^{14}$N/$^2$D & 9 & \cite{Alde:1990im}\\
\midrule
  SLAC E-139 & $^{27}$Al/$^2$D & 3 & \cite{Gomez:1993ri}\\
  NMC 96 & $^{27}$Al/$^{12}$C & 14 & \cite{Arneodo:1996rv}\\
\midrule
  SLAC E-139 & $^{40}$Ca/$^2$D & 2 & \cite{Gomez:1993ri}\\
  NMC 95, re. & $^{40}$Ca/$^2$D & 12 & \cite{Amaudruz:1995tq}\\
  EMC 90 & $^{40}$Ca/$^2$D & 3 & \cite{Arneodo:1989sy}\\
  FNAL E665 & $^{40}$Ca/$^2$D & 3 & \cite{Adams:1995is}\\
  NMC 95, re. & $^{40}$Ca/$^6$Li & 9 & \cite{Amaudruz:1995tq}\\
  NMC 96 & $^{40}$Ca/$^{12}$C & 23 & \cite{Arneodo:1996rv}\\
\midrule
  EMC 87 & $^{56}$Fe/$^2$D & 58 & \cite{Aubert:1987da}\\
  SLAC E-139 & $^{56}$Fe/$^2$D & 8 & \cite{Gomez:1993ri}\\
  NMC 96 & $^{56}$Fe/$^{12}$C & 14 & \cite{Arneodo:1996rv}\\
  BCDMS 85, BCDMS 87 & $^{56}$Fe/$^2$D & 16 & \cite{Alde:1990im,Benvenuti:1987az}\\
\midrule
  EMC 88, EMC 93 & $^{64}$Cu/$^2$D & 27 & \cite{Ashman:1988bf,Ashman:1992kv}\\
\midrule
  SLAC E-139 & $^{108}$Ag/$^2$D & 2 & \cite{Gomez:1993ri}\\
\midrule
 EMC 88 & $^{119}$Sn/$^2$D & 8 & \cite{Ashman:1988bf}\\
 NMC 96, $Q^2$ dependence  & $^{119}$Sn/$^{12}$C & 119 & \cite{Arneodo:1996ru}\\
\midrule
 FNAL E665 & $^{131}$Xe/$^2$D & 4 & \cite{Adams:1992vm}\\
\midrule
  SLAC E-139 & $^{197}$Au/$^2$D & 3 & \cite{Gomez:1993ri}\\
\midrule
 FNAL E665 & $^{208}$Pb/$^2$D & 3 & \cite{Adams:1995is}\\
 NMC 96 & $^{208}$Pb/$^{12}$C & 14 & \cite{Arneodo:1996rv}\\
 \midrule
 \midrule
 {\bf Total} & & {\bf 451} & \\
\bottomrule
\end{tabular}
\vspace{4mm}
\caption{\small The input datasets included in the present analysis.
  For each dataset, we give the nuclei $A_1$ and $A_2$ which
  have been used in the measurement with
  their atomic mass number.
  We also list the number of data points that
  survive the baseline kinematical cuts, and 
  provide the corresponding publication reference.
}
\label{dataset}
\end{table}

\begin{figure}[t]
\begin{center}
  \includegraphics[width=0.90\textwidth]{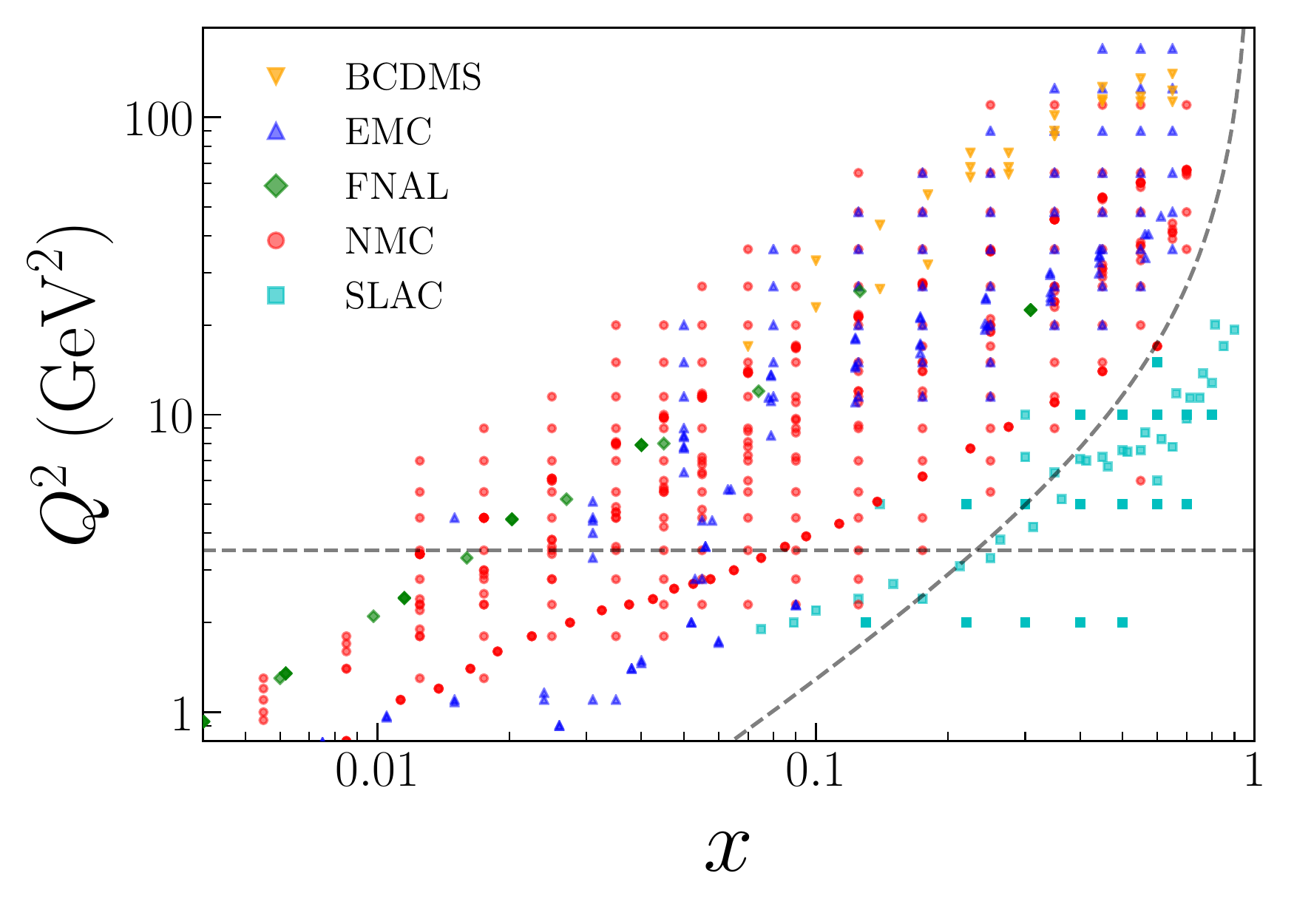}
 \end{center}
\vspace{-0.3cm}
\caption{\small Kinematical coverage in the $(x,Q^2)$ plane
  of the DIS neutral-current nuclear structure function data included in nNNPDF1.0,
  as summarized in Table~\ref{dataset}.
  The horizontal dashed and curved dashed lines correspond
  to $Q^2 = 3.5$ GeV$^2$ and $W^2 = 12.5$ GeV$^2$, respectively,
  which are the kinematic cuts imposed in this analysis.
  \label{figkinplot}
}
\end{figure}

In Fig.~\ref{figkinplot} we show the
kinematical coverage in the $(x,Q^2)$ plane
of the DIS nuclear data included in nNNPDF1.0.
To minimize the contamination from low-scale non-perturbative corrections and 
higher-twist effects, and also to remain
consistent with the baseline proton PDF analysis (to be discussed
in Sect.~\ref{sec:fitting}), we impose the same kinematical cuts on $Q^2$ and
the invariant final state mass squared $W^2=(P+q)^2$ as in the 
NNPDF3.1 global fit~\cite{Ball:2017nwa},
namely
 \be
 \label{eq:kincuts}
 Q^2 \ge Q^2_{\rm min}=3.5~{\rm GeV}^2 \, , \qquad 
 W^2 \ge W^2_{\rm min}=12.5~{\rm GeV}^2 \, ,
\ee
which are represented by the 
dashed lines in Fig.~\ref{figkinplot}.
In Table~\ref{tab:kincuts}, we compare our kinematics cuts in $W^2$ and
 $Q^2$ to those implemented in the nCTEQ15 and EPPS16 fits.
We find that our cuts are very similar to those of
the nCTEQ15 analysis~\cite{Kovarik:2015cma}, and as a result
our neutral-current DIS nuclear structure function dataset is similar
to that used in their analysis.
On the other hand, our choice of both the $Q^2_{\rm min}$ and $W^2_{\rm min}$ 
cut is more stringent than that made in the EPPS16 analysis \cite{Eskola:2016oht},
where they set $Q^2_{\rm min} = 1.69$~GeV$^2$ and do not impose any cut
in $W^2$.

\begin{table}[H]
  \centering
     \renewcommand{\arraystretch}{1.90}
  \begin{tabular}{cccc}
    &  nNNPDF1.0  &   nCTEQ15  &  EPPS16 \\
    \toprule
    $W^2_{\rm min}$  &  12.5 GeV$^2$  &  12.25 GeV$^2$  &  n/a \\
    \midrule
    $Q^2_{\rm min}$  & 3.5 GeV$^2$  &  4 GeV$^2$  &  1.69 GeV$^2$ \\
\bottomrule
  \end{tabular}
  \vspace{0.3cm}
  \caption{\label{tab:kincuts} The kinematics cuts in $W^2$ and
    $Q^2$ imposed in the nNNPDF1.0 analysis compared to those
    used in the nCTEQ15 and EPPS16 fits.
  }
  \end{table}

After imposing the kinematical cuts in Eq.~(\ref{eq:kincuts}),
we end up with $N_{\rm dat}=451$ data points.
As indicated in Table~\ref{dataset}, around half of these points
correspond to ratios of heavy nuclei with respect to
to deuterium, namely $R_{F_2}(A_1,A_2=2)$ in the notation
of Eq.~(\ref{eq:Sect2Rf2}).
For the rest of the data points, the values of $A_1$ and $A_2$
both correspond to heavier nuclei, with $A_2 \ge 6$.
It is worth noting that the measurements
from the NMC collaboration contain a significant amount of points
for which the carbon structure function is in the denominator,
$R_{F_2}(A_1,A_2=12)$.
In particular, we have $N_{\rm dat}=119$ data points
for the $Q^2$ dependence of the tin to carbon
ratio, $R_{F_2}(119,12)$.
These measurements provide valuable constraints on the $A$ dependence of the 
nuclear PDFs, since nuclear effects enter both the numerator
and denominator of Eq.~(\ref{eq:Sect2Rf2}).

Concerning the treatment of the experimental uncertainties,
we account for all correlations among data points whenever 
this information is provided
by the corresponding experiments.
This information is then encoded into the experimental
covariance matrix, constructed using the $t_0$ 
prescription~\cite{Ball:2009qv}:
\begin{eqnarray}
  ({\rm cov_{t_0}})^{(\rm exp)}_{ij}\equiv
  \lp \sigma^{\rm (stat)}_i\,R_{i}^{(\rm exp)}\rp^2\delta_{ij}
  &+&
  \Bigg(\sum_{\alpha=1}^{N_{\rm add}}\sigma_{i,\alpha}^{\rm (sys,a)}
  \sigma_{j,\alpha}^{\rm (sys,a)}R_{i}^{(\rm exp)}R_{j}^{(\rm exp)} \nonumber\\
  &+&\sum_{\beta=1}^{N_{\rm mult}}\sigma_{i,\beta}^{\rm (sys,m)}
  \sigma_{j,\beta}^{\rm (sys,m)}R_{i}^{(\rm th,0)}R_{j}^{(\rm th,0)}\Bigg)\; ,
  \label{eq:t0covmat}
\end{eqnarray}
where one treats the $N_{\rm add}$ additive (`sys,a') relative experimental 
systematic errors separately from the $N_{\rm mult}$ multiplicative (`sys,m') ones.
In the additive case, the central value of the experimental
measurement is used for the structure function ratio, $R_i^{({\rm exp})}$.
In the multiplicative case, e.g.
for overall normalization uncertainties, a fixed set of theoretical 
predictions for the ratios, $\{R_{i}^{(\rm th,0)}\}$, is constructed.
These predictions are
typically obtained from a previous fit which is
then iterated until convergence is reached.
The use of the $t_0$ covariance matrix defined in Eq.~(\ref{eq:t0covmat})
for the $\chi^2$ minimization (to be discussed in Sect.~\ref{sec:fitting})
avoids the bias associated with multiplicative uncertainties, which lead 
to a systematic underestimation of the best-fit values compared to their true 
values~\cite{dagos}.

For the case in which correlated systematic
uncertainties are not available, we simply add statistical and 
systematic errors in quadrature and Eq.~(\ref{eq:t0covmat}) reduces to
\be
  ({\rm cov_{t_0}})^{(\rm exp)}_{ij} = 
  \lp \sigma^{\rm (stat)2}_i + \sum_{\alpha=1}^{N_{\rm sys}}\sigma_{i,\alpha}^{\rm (sys)2}
  \rp \,R_{i}^{(\rm exp)2} \, \delta_{ij} \, ,
\ee
where $N_{\rm sys}=N_{\rm add}+N_{\rm mult}$.
It turns out that for all of the measurements listed in 
Table~\ref{dataset}, the detailed
break-up of the experimental systematic errors is not available
(in most cases these partially or totally cancel out when taking
ratios of observables), and the only systematic error that
enters the $t_0$ covariance matrix Eq.~(\ref{eq:t0covmat}) is
the multiplicative normalization error.

\subsection{Numerical implementation}

We turn now to discuss the numerical implementation of the calculations
of the DIS structure functions and their ratios $R_{F_2}$ relevant
for the nPDF interpretation of the nuclear DIS data.
In the framework of collinear QCD factorization, the $F_2$ structure function
can be decomposed in terms of hard-scattering coefficient functions and
nuclear PDFs as,
\begin{align} 
\label{eq:ev} 
F_2(x,Q^2,A) &= \sum_i^{n_f} C_i(x,Q^2) \otimes f_i(x,Q^2,A) \nonumber \\
&= \sum_{i,j}^{n_f} C_i(x,Q^2) \otimes \Gamma_{ij}(Q^2,Q_0^2) \otimes f_j(x,Q_0^2,A),
\end{align}
where $C_i(x,Q^2)$ are the process-dependent coefficient functions which
can be computed perturbatively as an expansion in the QCD and QED
couplings;  $\Gamma_{ij}(Q^2,Q_0^2)$ is an evolution operator, determined by the
solutions of the DGLAP equations, which evolves the nPDF from the initial
parameterization scale $Q_0^2$ into the hard-scattering scale $Q^2$,
$f_i(x,Q^2_0,A)$ are the nPDFs at the parameterization scale, and
$\otimes$ denotes the Mellin convolution.
The sum over flavors $i,j$ runs over the $n_f$ active quarks and antiquarks flavors at a given
scale $Q$, as well as over the gluon.

The direct calculation of Eq.~(\ref{eq:ev}) during the nPDF fit is not practical
since it requires first solving the DGLAP evolution equation for each new boundary
condition at $Q_0$ and then convoluting with the coefficient
functions.
To evaluate Eq.~(\ref{eq:ev}) in a more computationally efficient way, it is better 
to precompute all the perturbative information, i.e. the coefficient functions $C_i$
and the evolution operators $\Gamma_{ij}$, with a suitable
interpolation basis.
Several of these approaches have been made available in the context of
PDF fits~\cite{DelDebbio:2013kxa,amcfast,Carli:2010rw,Wobisch:2011ij}.
Here we use the {\tt APFELgrid} tool~\cite{Bertone:2016lga} to precompute the perturbative
information of the nDIS structure functions provided by the {\tt APFEL} program~\cite{Bertone:2013vaa}.

Within this approach,
we can factorize the dependence on the nPDFs at the input scale $Q_0$ from
the rest of Eq.~(\ref{eq:ev}) as follows.
First, we introduce
an expansion over a set of interpolating functions $\{ I_{\beta}\}$ spanning both $Q^2$ and $x$ such that
\begin{equation}
  f_i(x,Q^2,A) = \sum_{\beta} \sum_{\tau} f_{i,\beta \tau} I_{\beta}(x) I_{\tau}(Q^2) \, ,
\end{equation}
where the nPDFs are now tabulated
in a grid in the $(x,Q^2)$ plane, $f_{i,\beta \tau}\equiv f_i(x_\beta,Q^2_{\tau},A)$.
We can express this result in terms of the PDFs at the input evolution scale
using the (interpolated) DGLAP evolution operators,
\begin{equation}
  f_{i,\beta \tau} = \sum_j \sum_{\alpha} \Gamma^{\tau}_{ij,\alpha \beta}\,f_j(x_{\alpha},Q_0^2,A) \, ,
\end{equation}
so that the nuclear DIS structure function can be
evaluated as
\begin{equation}
  F_2(x,Q^2,A) = \sum_i^{n_f} C_i(x,Q^2) \otimes \lc
  \sum_{\alpha,\beta,\tau} \sum_j \Gamma^{\tau}_{ij,\alpha \beta}\,f_j(x_{\alpha},Q_0^2) I_{\beta}(x) I_{\tau}(Q^2)\rc\, .
\end{equation}
This can be rearranged to give
\begin{align}
  \label{eq:ev_interp}
  F_2(x,Q^2) &= \sum_i^{n_f} \sum_{\alpha}^{n_x} {\tt FK}_{i,\alpha}(x,x_{\alpha},Q^2,Q^2_0) \, f_i(x_{\alpha},Q_0^2) \end{align}
where all of the information about the partonic cross-sections and the DGLAP
evolution operators is now encoded into the so-called FK table, ${\tt FK}_{i,\alpha}$.
Therefore, with the {\tt APFELgrid} method we are able to
express the series of convolutions in Eq.(\ref{eq:ev}) by a matrix
multiplication in Eq.~(\ref{eq:ev_interp}), increasing the numerical 
calculation speed of the DIS structure functions by up to several orders
of magnitude.

In this work, the FK tables (and thus the nDIS structure functions) are computed
up to NNLO in the QCD coupling expansion, with heavy quark effects evaluated
by the FONLL general-mass variable flavor number scheme~\cite{Forte:2010ta}.
Specifically, we use the FONLL-B scheme for the NLO fits and the FONLL-C for
the NNLO fits.
The value of the strong coupling constant is set to be $\alpha_s(m_Z)=0.118$, consistent
with the PDG average~\cite{Tanabashi:2018oca} and with recent high-precision
determinations~\cite{Ball:2018iqk,Verbytskyi:2019zhh,Bruno:2017gxd,Zafeiropoulos:2019flq}
(see~\cite{Pich:2018lmu} for an overview).
Our variable flavor number scheme has a maximum of $n_f=5$ active quarks,
where the heavy quark pole masses are taken to be $m_c=1.51$ GeV and $m_b=4.92$ GeV
following the Higgs Cross-Sections Working Group recommendations~\cite{deFlorian:2016spz}.
The charm and bottom PDFs are generated dynamically from the gluon and the light
quark PDFs starting from the thresholds $\mu_c=m_c$ and $\mu_b=m_b$.
Finally, since all of these theoretical settings are the same as in the NNPDF3.1 global proton PDF
analysis, we choose this set to represent our nPDFs at $A=1$, which we explain in 
more detail in Sect.~\ref{sec:fitting}.

In Table~\ref{table:APFELvsFK} we show a
comparison between the deep-inelastic structure function $F_2(x,Q^2,A)$
computed with the {\tt APFEL} program and with the 
{\tt FK} interpolation method, Eq.~(\ref{eq:ev_interp}), using the theoretical settings
given above.
    The predictions have been evaluated using the EPPS16 sets for two different
    perturbative orders, FONLL-B and FONLL-C, at sample values of $x$ and $Q^2$ given
    by carbon ($A=12$) and lead ($A=208$) data.
    We also indicate the relative difference between the two calculations, $\Delta_{\rm rel}\equiv
    |{\tt APFEL}-{\tt FK}|/{\tt APFEL}$.
    Here we see that the agreement is excellent with residual differences much smaller
    than the typical uncertainties of the experimental data, and thus suitable
    for our purposes.
 
\begin{table}[H]
  \centering
  \small
     \renewcommand{\arraystretch}{1.65}
  \begin{tabular}{ccc|ccc|ccc} 
  \multirow{2.5}{*}{$A$} & \multirow{2.5}{*}{$x$} & \multirow{2.5}{*}{$Q^2$ (GeV$^2$)} &  \multicolumn{3}{c|}{FONLL-B (NLO)} &  \multicolumn{3}{c}{FONLL-C (NNLO)}\\ 
  & & & {\tt APFEL} & {\tt FK} & $\Delta_{\rm rel}~(\%)$  & {\tt APFEL} & {\tt FK} & $\Delta_{\rm rel}~(\%)$\\ [0.5ex] \toprule
  \multirow{5}{*}{12} 
  & 0.009 & 1.7  & 0.2895 & 0.2893 &  0.0694   &  0.2534 & 0.2534    &0.027\\ 
  & 0.013 & 2.3   & 0.3057 & 0.3052 & 0.1521  &     0.2837 & 0.283   & 0.228   \\ 
  & 0.13 & 14 &  0.27 & 0.2715 & 0.5642    &     0.2655 & 0.2647  &0.277       \\ 
  & 0.35 & 26 &  0.1292 & 0.1274 & 1.3683     &     0.1217 & 0.1213  &  0.308       \\ 
  & 0.65 & 42  & 0.0165 & 0.0168 & 1.8437  &      0.016 & 0.0163  & 2.347       \\
  \midrule
  \multirow{4}{*}{208} & 0.012 & 2.42 &  0.2795 & 0.279 & 0.1553       & 0.2581 & 0.2573  & 0.293      \\ 
  & 0.02 & 4.45 &   0.309 & 0.3103 & 0.3885     & 0.3041 & 0.3043      &0.084\\ 
  & 0.04 & 7.91 &   0.32 & 0.3214 & 0.3253     &  0.3181 & 0.3177     &0.104\\ 
  & 0.31 & 22.5 &   0.1467 & 0.1445 & 1.4583      &  0.1388 & 0.1382     &0.401 \\
 \bottomrule
  \end{tabular}
  \vspace{0.3cm}
  \caption{\small Comparison between the deep-inelastic structure function $F_2(x,Q^2,A)$
    computed with the {\tt APFEL} program and with the corresponding
    {\tt FK} interpolation tables.
    The predictions are given for two different
    perturbative orders, FONLL-B and FONLL-C, and are computed 
    using the EPPS16 nPDF set with theoretical settings described in the text.
    The values of $x$ and $Q^2$ correspond to representative measurements
     for carbon ($A=12$) and lead ($A=208$) nuclei.
    Also given are the relative differences between the two calculations, $\Delta_{\rm rel}\equiv
    |{\tt APFEL}-{\tt FK}|/{\tt APFEL}$.
  }
 \label{table:APFELvsFK} 
\end{table}


\subsection{Quark flavor decomposition}
\label{sec:quarkdecomposition}

With the {\tt APFELgrid} formalism, we can express any DIS structure function
in terms of the nPDFs at the initial evolution scale $Q_0^2$ using
Eq.~(\ref{eq:ev_interp}).
In principle, one would need to parameterize 7 independent PDFs: the
up, down, and strange quark and antiquark PDFs and the gluon.
Another two input PDFs would be required if in addition the 
charm and anti-charm PDFs
are also parameterized, as discussed in~\cite{Ball:2016neh}.
However, given that our input dataset  in this analysis is restricted
to DIS neutral current structure functions, a full quark flavor separation
of the fitted nPDFs is not possible.
In this section we discuss the specific quark flavor decomposition that is adopted
in the nNNPDF1.0 fit.

We start by expressing the neutral-current DIS structure function $F_2(x,Q^2,A)$
at leading order in terms of the nPDFs.
This decomposition is carried out for $Q^2 < m_c^2$ and therefore the charm PDF is absent.
In this case, one finds for the $F_2$ structure function, 
\begin{equation}
  F^{(\rm LO)}_2 (x,Q^2,A) =  x \sum_{i=1}^{n_f}e_i^2f_i^+(x,Q^2,A) = x
  \lc \frac{4}{9} u^+(x,Q^2,A) + \frac{1}{9}\lp d^+ + s^+\rp (x,Q^2,A) \rc \, ,
\end{equation}
where for consistency the DGLAP evolution has been performed at LO, and the 
quark and antiquark PDF combinations are given by
\be
f_i^\pm (x,Q^2,A) \equiv \lc f_i (x,Q^2,A) \, \pm \, \bar{f}_i (x,Q^2,A)\rc \,.
\qquad  i = u,d,s \, .
\ee
In this analysis, we will work in the PDF evolution basis, which is defined
as the basis composed by the eigenstates of the DGLAP evolution equations.
If we restrict ourselves to the  $Q < m_c$ ($n_f=3$) region,
the quark combinations are defined in this basis as
\bea
         \Sigma(x,Q^2,A) & \equiv& \sum_{i=1}^{n_f=3}f^+_i(x,Q^2,A)\qquad \text{(quark singlet)} \, ,\\
   \label{eq:def}      T_{3}(x,Q^2,A) & \equiv& \lp u^+ - d^+\rp (x,Q^2,A)\qquad \text{(quark triplet) }  \, , \\
        T_{8}(x,Q^2,A) & \equiv& \lp u^+ + d^+ - 2s^+\rp (x,Q^2,A) \qquad \text{(quark octet)}  \, .
 \eea
It can be shown that the neutral current DIS structure functions depend only on these
three quark combinations: $\Sigma$, $T_3$, and $T_8$.
Other quark combinations in the evolution basis, such as the valence distributions
$V=u^-+d^-+s^-$ and $V_3=u^--d^-$, appear only at the level
of charged-current structure functions, as well as in hadronic observables
such as $W$ and $Z$ boson production.

In the evolution basis, the $F_2$ structure function for a proton and a neutron target
at LO in the QCD expansion can be written as
\bea
  \label{eq:F2_p_lo}
    F^{({\rm LO}),p}_2 (x,Q^2) &=& x \lc \frac{2}{9} \Sigma + \frac{1}{6} T_3 + \frac{1}{18}T_8 \rc \, , \\
    F^{({\rm LO}),n}_2 (x,Q^2) &=& x \lc  \frac{2}{9} \Sigma - \frac{1}{6} T_3 + \frac{1}{18}T_8 \rc \, . \nonumber
\eea
Therefore, since the nuclear effects are encoded in the nPDFs,
the structure function for a nucleus with atomic number $Z$ and mass number $A$
will be given by a simple sum of the proton and neutron structure functions,
\be
  \label{eq:F2_A}
  F^{({\rm LO})}_2 (x,Q^2,A) = \frac{1}{A}\left( Z F_2^{({\rm LO}),p}(x,Q^2) + (A-Z) F_2^{({\rm LO}),n}(x,Q^2) \right)\\ .
\ee
Inserting the decomposition of Eq.~(\ref{eq:F2_p_lo}) into Eq.~(\ref{eq:F2_A}), we find
\be
\label{eq:F2_p_lo_v2}
F^{({\rm LO})}_2 (x,Q^2,A)= x \left[ \frac{2}{9} \Sigma - \left(\frac{Z}{3A}-\frac{1}{6}\right) T_3 + \frac{1}{18}T_8\right](x,Q^2,A) \, .
\ee
Note that nuclear effects, driven by QCD, are electric-charge blind and therefore
depend only on the total number of nucleons $A$ within a given nuclei, in addition to
$x$ and $Q^2$.
The explicit dependence on $Z$ in Eq.~(\ref{eq:F2_p_lo_v2}) arises from QED effects, since
the virtual photon $\gamma^*$ in the deep-inelastic scattering couples more strongly
to up-type quarks ($|e_q|=2/3$) than to down-type quarks ($|e_q|=1/3$).

From Eq.~(\ref{eq:F2_p_lo_v2}) we see that at LO the  $F_2^p$ structure function
in the nuclear case depends on three independent quark combinations: the
total quark singlet $\Sigma$, the quark triplet $T_3$, and the quark octet $T_8$.
However, the dependence on the non-singlet triplet combination is very weak, since 
its coefficient is given by
\be
\left(\frac{Z}{3A}-\frac{1}{6}\right) = \left(\frac{Z}{3(2Z+\Delta A)}-\frac{1}{6}\right)
\simeq -\frac{\Delta A}{12Z} \, ,
\ee
where $\Delta A \equiv A - 2Z$ quantifies the deviations from nuclear isoscalarity ($A=2Z$).
This coefficient is quite small for nuclei in which data is available,
and in most cases nuclear structure functions are corrected
for non-isoscalarity effects.
In this work, we will assume $\Delta A=0$ such that we have only
isoscalar nuclear targets.
The dependence on $T_3$ then drops out and the nuclear structure function
$F_2$ at LO is given by
\be
\label{eq:f2lo_v3}
F^{({\rm LO})}_2 (x,Q^2,A)= x \lc  \frac{2}{9} \Sigma + \frac{1}{18}T_8\rc(x,Q^2,A) \,,
\ee
where now the only relevant quark combinations are the quark singlet $\Sigma$
and the quark octet $T_8$.
Therefore, at LO, neutral-current structure function measurements
on isoscalar targets below
the $Z$ pole can only constrain a single quark combination, namely
\be
\label{eq:singlequarkcombination}
F^{({\rm LO})}_2 (x,Q^2,A) \propto \lp \Sigma +\frac{1}{4} T_8\rp(x,Q^2,A) \,.
\ee

At NLO and beyond, the dependence on the gluon PDF enters
and the structure function Eq.~(\ref{eq:f2lo_v3}) becomes
\be
\label{eq:f2lo_v4}
F^{({\rm NLO})}_2 (x,Q^2,A)= C_{\Sigma}\otimes \Sigma(x,Q^2,A) + C_{T_8}\otimes T_8(x,Q^2,A)
+ C_{g}\otimes g(x,Q^2,A) \, ,
\ee
where $C_{\Sigma}$, $C_{T_8}$, and $C_{g}$ are the coefficient functions associated
with the singlet, octet, and gluon respectively.
In principle one could aim to disentangle $\Sigma$ from $T_8$ due to 
their different $Q^2$ behavior, but in practice this is not possible
given the limited kinematical coverage of the available experimental data.
Therefore, only the $\Sigma+ T_8/4$ quark combination is effectively constrained
by the experimental data used in this analysis, as indicated by
Eq.~(\ref{eq:singlequarkcombination}).

\begin{figure}[t]
\begin{center}
  \includegraphics[width=0.9\textwidth]{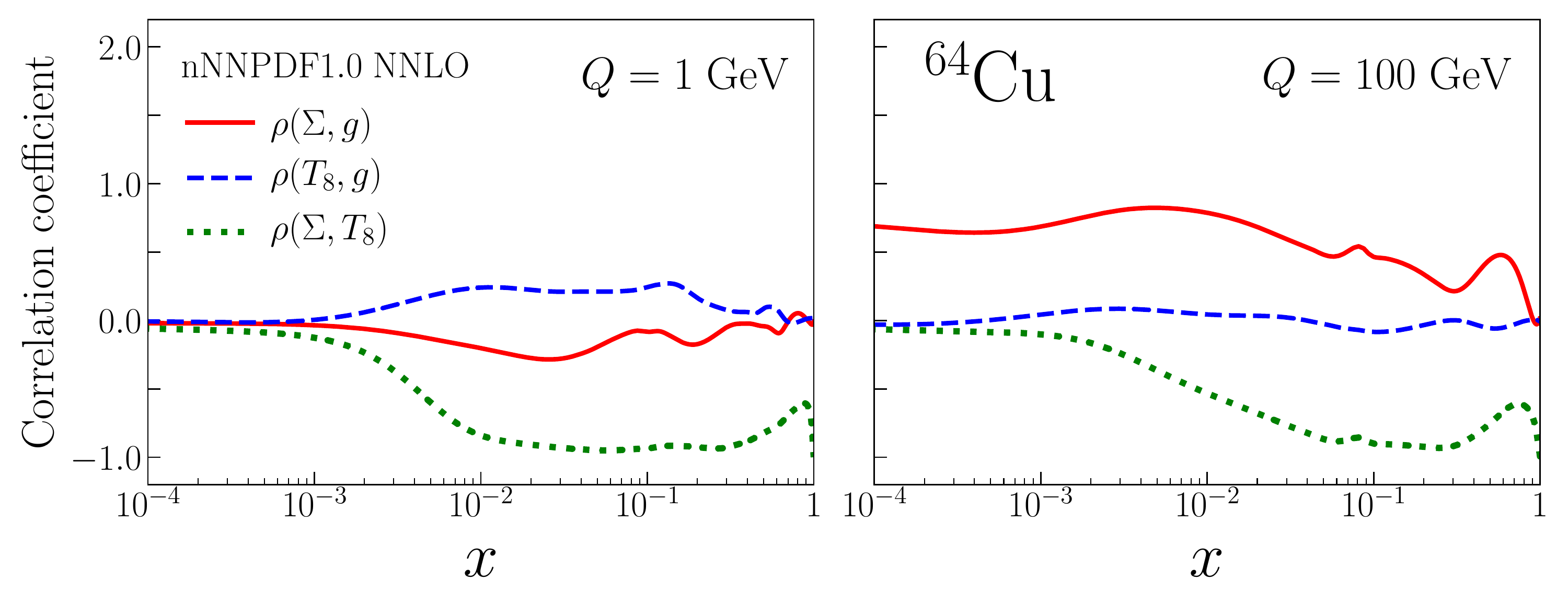}
 \end{center}
\vspace{-0.3cm}
\caption{\small The correlation coefficient $\rho=\langle \left(f_i - \langle f_i \rangle\right)\left(f_j - \langle f_j \rangle\right)\rangle /  \left(\sigma_i \sigma_j\right)$ between the
  the quark singlet $\Sigma$ and gluon $g$ (solid red line), 
  the quark octet $T_8$ and $g$ (dashed blue line),
  and between $\Sigma$ and $T_8$ (dotted green line).
  The coefficients are computed with $N_{\rm rep}=200$ replicas
  of the copper ($A=64$)
  nNNPDF1.0 NNLO set
  at $Q=1$ GeV (left) and $Q=100$ GeV (right).
}
\label{fig:correlations}
\end{figure}

Putting together all of this information, we will consider the following
three independent PDFs at the initial parameterization scale $Q_0$:
\begin{itemize}

\item the total quark singlet   $\Sigma(x,Q^2_0,A) = \sum_{i=1}^{3}f^+_i(x,Q^2_0,A)$,

\item the quark octet   $T_{8}(x,Q^2,A) = \lp u^+ + d^+ - 2s^+\rp (x,Q^2,A)$,

\item and the gluon nPDF $g(x,Q_0,A)$.

\end{itemize}
In Sect.~\ref{sec:fitting} we discuss the parameterization of these three
nPDFs using neural networks.
In Fig.~\ref{fig:correlations} we show the results for
the correlation coefficient between the
nPDFs that are parameterized in the nNNPDF1.0 fit (presented
in Sect.~\ref{sec:results}),
specifically the NNLO set for copper ($A=64$) nuclei.
  The nPDF correlations are computed at both $Q=1$ GeV and $Q=100$ GeV,
  the former of which contains experimental data in the region $ 0.01 \lsim x \lsim 0.4$ (illustrated in Fig.~\ref{figkinplot}).
 In the data region, there is a strong anticorrelation
  between $\Sigma$ and $T_8$, consistent with
  Eq.~(\ref{eq:singlequarkcombination}) which implies that only their weighted
  sum can be constrained.
  As a result, we will show in the following sections only results of the 
combination $\Sigma+T_8/4$ which can be meaningfully
determined given our input experimental data.
 From Fig.~\ref{fig:correlations}, one can also observe the strong correlation between $\Sigma$ and $g$
  for $x\lsim 0.01$ and $Q=100$ GeV, arising from the fact that
  these two PDFs are coupled via the DGLAP evolution equations 
  as opposed to $T_8$ and $g$ where the correlation is very weak.
    
\section{Fitting methodology}
\label{sec:fitting}

In this section we describe the fitting methodology that has been adopted
in the nNNPDF1.0 determination of nuclear parton distributions.
While much of this methodology follows
from previous NNPDF 
analyses, a number of significant improvements have been
implemented in this work.
Here we discuss these developments, together with
relevant aspects of the NNPDF framework 
that need to be modified or improved
in order to deal with the determination of the nuclear PDFs,
such as the parameterization of the $A$ dependence or imposing
the $A=1$ proton boundary condition.

Following the NNPDF methodology,
the uncertainties associated with the nPDFs are
estimated using the Monte Carlo replica method, where
a large number of $N_{\rm rep}$ replicas of the
experimental measurements are generated in a way that 
they represent a sampling of the probability
distribution in the space of data.
An independent fit is then performed for each of these replicas,
and the resulting ensemble of nPDF samples correspond to a 
representation of the probability
distribution in the space of nPDFs for which any statistical
estimator such as central values, variances, correlations,
and higher moments can be computed~\cite{Gao:2017yyd}.

\begin{figure}[t]
\begin{center}
  \includegraphics[width=0.9\textwidth]{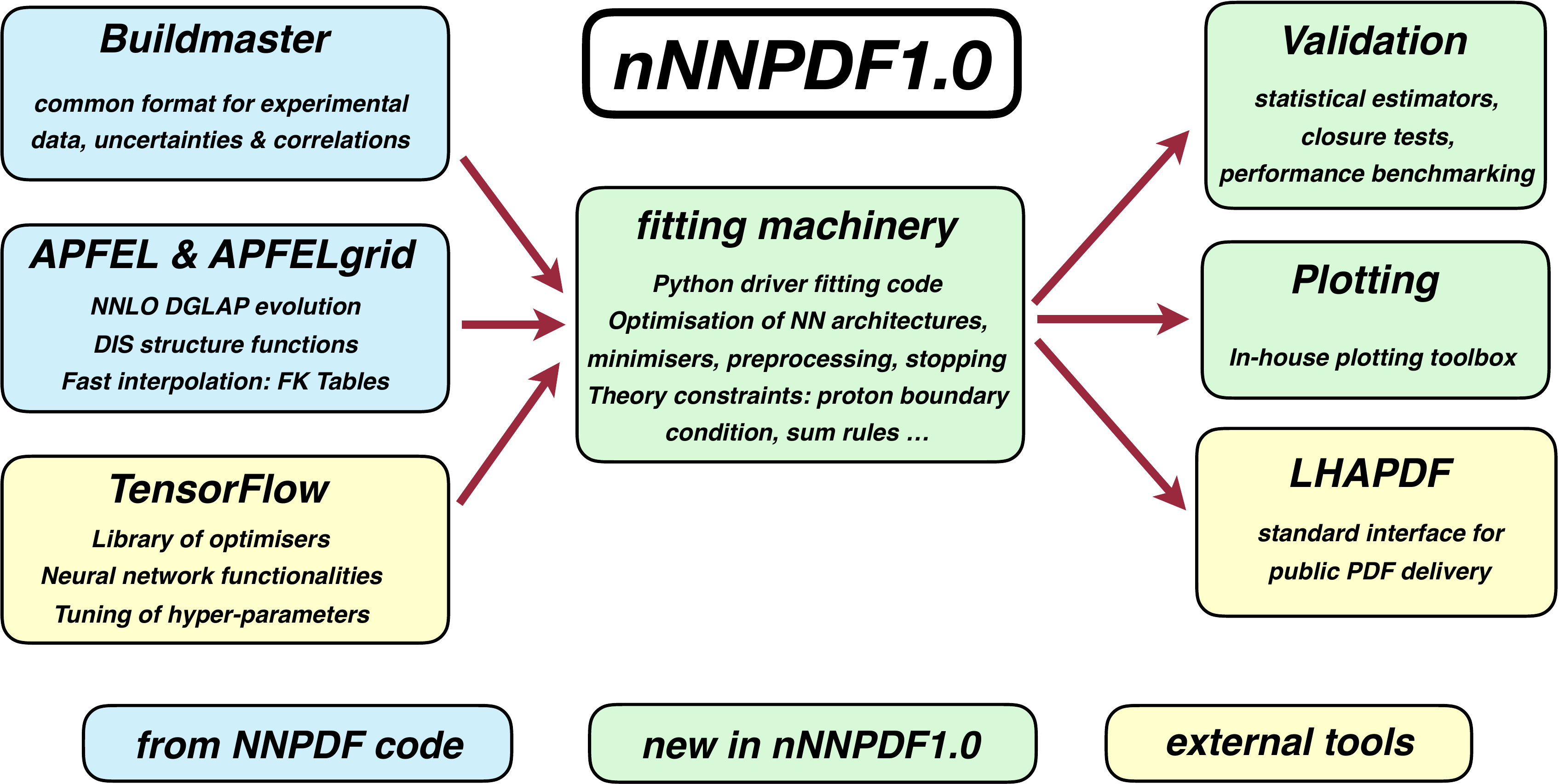}
 \end{center}
\vspace{-0.3cm}
\caption{\small Schematic representation of the tool-chain
  implemented in the nNNPDF1.0 analysis.
  The items in blue correspond to aspects that were inherited from
  the NNPDF code, those in green represent new programs, 
  and those in yellow
  represent external tools.
}
\label{fig:nNNPDF10}
\end{figure}

In order to illustrate the novel ingredients of the present
study as compared to the standard NNPDF framework,
we display in Fig.~\ref{fig:nNNPDF10} a schematic representation of 
the tool-chain adopted to construct the nNNPDF1.0 sets.
The items in blue correspond to components
of the fitting methodology inherited from the
NNPDF code, those in green represent new code modules
developed specifically
for this project, and those in yellow
indicate external tools.
As highlighted in Fig.~\ref{fig:nNNPDF10}, the main development is
the application of {\tt TensorFlow}~\cite{tensorflow2015-whitepaper},
an external machine learning library that allows us access 
to an extensive number of
tools for the efficient determination of the best-fit 
weights and thresholds of the neural network. 
The ingredients of Fig.~\ref{fig:nNNPDF10} will be discussed
in more detail in the following and subsequent sections. 

The outline of this section is the following.
We start first with a discussion of our strategy for the parameterization
of the nPDFs in terms of artificial neural networks.
Then we present the algorithm used for the minimization
of the cost function, defined to be the $\chi^2$, which is based 
on stochastic gradient
descent.
We also briefly comment on the performance improvement obtained
in this work as compared to previous NNPDF fits.

\subsection{Nuclear PDF parameterization}
\label{sec:nPDFparam}

As mentioned in Sect.~\ref{sec:theory}, the non-perturbative 
distributions that enter the collinear factorization framework
in lepton-nucleus scattering
are the PDFs of a nucleon within an isoscalar nucleus with atomic mass 
number $A$, $f_i(x,Q^2,A)$.
While the dependence of the nPDFs on the scale $Q^2$ is determined
by the perturbative DGLAP evolution equations,
the dependence on both Bjorken-$x$ and the atomic
mass number $A$ is non-perturbative and needs
to be extracted from experimental data through
a global analysis.\footnote{See~\cite{Lin:2017snn}
  for an overview of recent efforts in the first-principle
  calculations of PDFs by means of lattice QCD.}
Taking into account the flavor decomposition
presented in Sect.~\ref{sec:quarkdecomposition},
we are required to parameterize the $x$ and $A$ dependence
of the quark singlet $\Sigma$, the quark octet $T_8$, and the gluon $g$,
as indicated
by Eq.~(\ref{eq:f2lo_v3}) at LO and by Eq.~(\ref{eq:f2lo_v4})
for NLO and beyond.

\begin{figure}[t]
\begin{center}
  \includegraphics[width=0.8\textwidth]{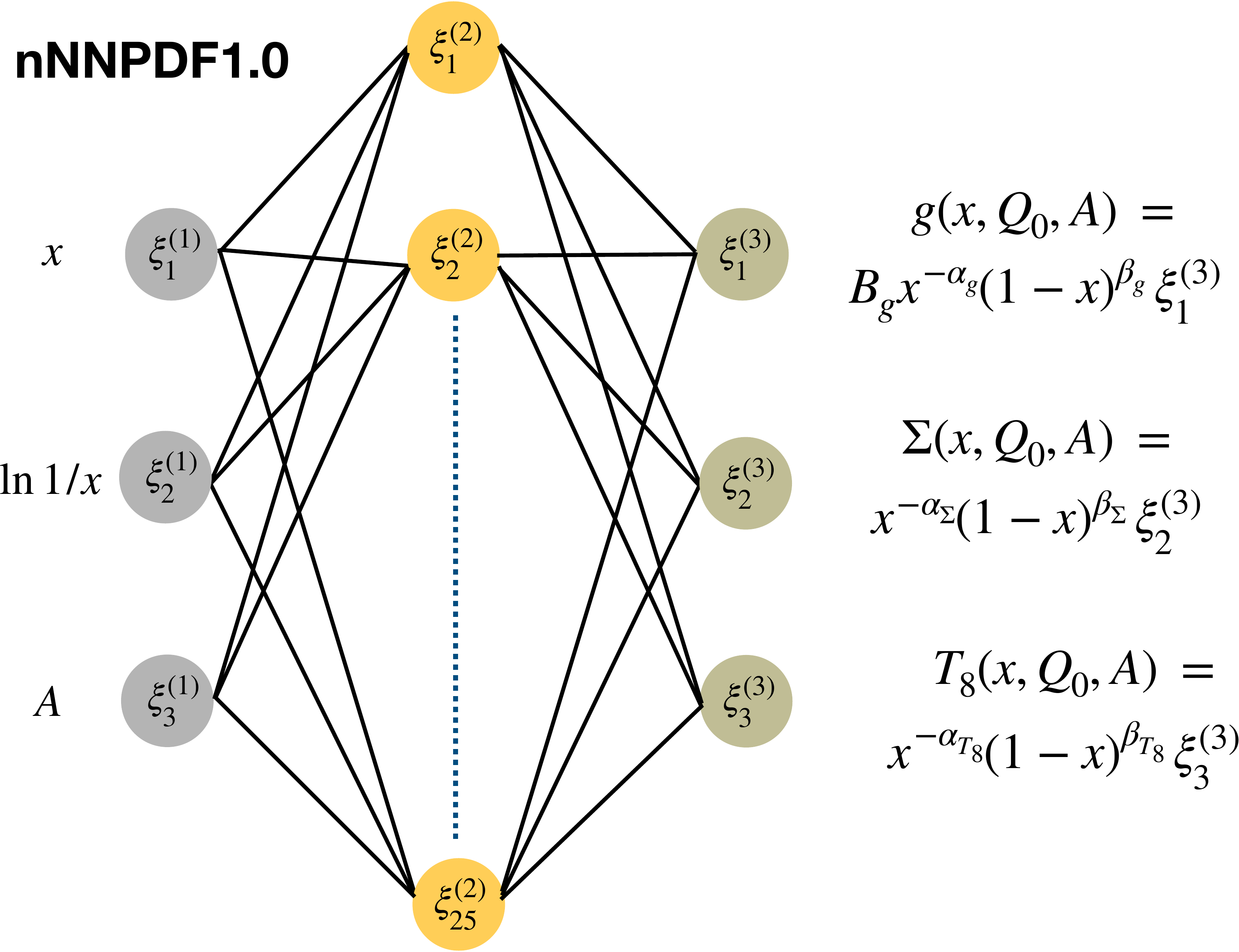}
 \end{center}
\vspace{-0.3cm}
\caption{\small Schematic representation of the architecture
  of the feed-forward neural network used in the nNNPDF1.0 analysis to
  parameterize the $x$ and $A$ dependence of $\Sigma$, $T_8$, and $g$
  at the initial scale $Q_0$.
  The architecture is 3-25-3, where the values of the three input
  neurons are $x$, $\ln 1/x$, and $A$, and the
  values of the output layer neurons
  correspond to the input nPDFs: $g(x,Q_a,A)$,
  $\Sigma(x,Q_0,A)$, and $T_8(x,Q_0,A)$.
  For the input and hidden layer, a sigmoid function is used for neuron activation, 
  and a linear activation is used for the final output layer.
   \label{fig:architecture}
}
\end{figure}

The three distributions listed above are parameterized
at the input scale $Q_0$ by the output of a
neural network NN$_f$ multiplied by an $x$-dependent polynomial functional form.
In previous NNPDF analyses, a different multi-layer feed-forward
neural network was used for each of
the parameterized PDFs so that in this case, three independent
neural networks would be required:
\bea
x\Sigma(x,Q_0,A) &=&x^{-\alpha_\Sigma} (1-x)^{\beta_\Sigma} {\rm NN}_\Sigma(x,A) \, , \nonumber \\
xT_8(x,Q_0,A) &=&x^{-\alpha_{T_8}} (1-x)^{\beta_{T_8}} {\rm NN}_{T_8}(x,A) \, , \label{eq:param2} \\
xg(x,Q_0,A) &=&B_gx^{-\alpha_g} (1-x)^{\beta_g} {\rm NN}_g(x,A) \, . \nonumber
\eea
However, in this work we use instead a single artificial neural network
consisting of an input layer, one hidden layer, and an output layer. 
In Fig.~\ref{fig:architecture}
we display a schematic representation of the architecture
of the feed-forward neural network used in the present analysis.
The input layer 
contains three neurons which take as input the values
of the momentum fraction $x$, $\ln(1/x)$, and 
atomic mass number $A$, respectively.
The subsequent hidden layer contains 
25 neurons, which feed into the final output layer of three neurons, 
corresponding to the three fitted distributions $\Sigma, T_8$ and $g$. 
A sigmoid activation function is used for the neuron activation in the first
two layers, while a linear activation is used for the output layer. This latter 
choice ensures that the network output will not be bounded and
can take any value required to reproduce experimental data. 
The output from the final layer of neurons is then used to construct the full
parameterization:
\bea
x\Sigma(x,Q_0,A) &=&x^{-\alpha_\Sigma} (1-x)^{\beta_\sigma} \,\xi_1^{(3)}(x,A) \, , \nonumber \\
xT_8(x,Q_0,A) &=&x^{-\alpha_{T_8}} (1-x)^{\beta_{T_8}} \,\xi_2^{(3)}(x,A) \, , \label{eq:param} \\
xg(x,Q_0,A) &=&B_gx^{-\alpha_g} (1-x)^{\beta_g} \,\xi_3^{(3)}(x,A) \, , \nonumber
\eea
where $\xi^{(3)}_i$ represent the values of the $i$-th neuron's activation state
in the third and final layer of the neural network.

Overall, there are a total of $N_{\rm par}=178$ free parameters
(weights and thresholds) in the neural network represented in Fig.~\ref{fig:architecture}.
These are supplemented
by the normalization coefficient $B_g$
for the gluon nPDF and by the six preprocessing exponents $\alpha_f$ and $\beta_f$.
The latter are fitted simultaneously with the network parameters, while
the former is fixed by the momentum sum rule, described in more detail below.
Lastly, the input scale $Q_0$ is set to 1 GeV to
maintain consistency 
with the settings of the
baseline proton fit, chosen to be the NNPDF3.1 set with perturbative charm.

\paragraph{Sum rules.}
Since the nucleon energy must be distributed among 
its constituents in a way that ensures energy conservation,
the PDFs are required to obey the momentum sum rule
given by
\be
\label{eq:MSR}
\int_0^1 dx x \left(\Sigma(x,Q_0,A) + g(x,Q_0,A)\right) = 1 \, , \quad \forall \, A \, .
\ee
Note that this expression needs only to be implemented at the input scale 
$Q_0$, since the properties of DGLAP evolution guarantees 
that it will also be satisfied for any $Q > Q_0$.
In this analysis, Eq.~(\ref{eq:MSR}) is applied by setting the overall 
normalization of the gluon nPDF to
\be
\label{eq:NormG}
B_g(A) = \frac{1 - \int_0^1 dx\, x\Sigma(x,Q_0,A)}{\int_0^1 dx\, xg(x,Q_0,A)}. 
\ee
where the denominator of Eq.~(\ref{eq:NormG}) is computed using 
Eq.~(\ref{eq:param}) and setting $B_g=1$.
Since the momentum sum rule requirement must
be satisfied for any value of $A$,
the normalization factor for the gluon distribution $B_g$ needs to be
computed separately for each
value of $A$ given by the experimental data (see Table~\ref{dataset}).

In addition to the momentum sum rule, 
nPDFs must satisfy other sum rules such as those for the valence distributions,
\be
\int_0^1 dx~ \left(u(x,Q_0,A) - \bar{u}(x,Q_0,A)\right) =
\int_0^1 dx~ \left(d(x,Q_0,A) - \bar{d}(x,Q_0,A)\right)=\frac{3}{2} \, , \quad \forall \, A \, ,
\ee
as well as
\be
\int_0^1 dx~ \left(s(x,Q_0,A) - \bar{s}(x,Q_0,A)\right) =0\, , \quad \forall \, A \, ,
\ee
given the quark flavor quantum numbers of isoscalar nucleons.
These valence sum rules involve quark combinations
which are not relevant for the description of neutral-current DIS structure
functions, and therefore do not need to be used in the present analysis.
However, they will become necessary in future updates of the nNNPDF fits 
in which, for instance, charged-current DIS measurements are also included.

\paragraph{Preprocessing.}
The polynomial preprocessing functions $x^{-\alpha_f}(1-x)^{\beta_f}$ in Eq.~(\ref{eq:param}) have
long been 
known to approximate well the general asymptotic behavior of the PDFs
at small and large $x$~\cite{Ball:2016spl}.
Therefore, they help to increase 
the efficiency of parameter optimization since the neural
networks have to learn smoother functions.
Note that the preprocessing exponents $\alpha_f$ and $\beta_f$
are independent of $A$, implying that the entire $A$ dependence of the input
nPDFs will arise from the output of the neural networks.

In previous NNPDF analyses, the preprocessing exponents $\alpha_f$ and $\beta_f$ were 
fixed to a randomly chosen value from a range that was determined iteratively.
Here instead we will fit their values for each Monte Carlo replica, so that they
are treated simultaneously with the weights and thresholds of the
neural network.
The main advantage of this approach is that one does not need to iterate the fit to find
the optimal range for the exponents, since now their best-fit values are automatically
determined for each replica.

Based on basic physical requirements, as well as
on empirical studies, we impose some additional constraints
on the range of allowed values that the exponents $\alpha_f$ and $\beta_f$
can take.
More specifically, we restrict the parameter values to
\be
\label{eq:preprocessing}
\alpha_f \in [-5,1] \, , \qquad \beta_f \in [0,10] \, ,\qquad f=\Sigma,T_8,g \, .
\ee
Concerning the large-$x$ exponent $\beta_f$, the lower limit
in Eq.~(\ref{eq:preprocessing}) guarantees that the nPDFs vanish 
in the elastic limit $x\to 1$; the upper limit follows from the observation that it
is unlikely for the nPDFs to be more strongly suppressed at large $x$.~\cite{Ball:2016spl}.
With respect to the small-$x$ exponent $\alpha_f$, the upper limit follows
from the nPDF integrability condition, given that for $\alpha_f > 1$
the momentum integral Eq.~(\ref{eq:MSR}) becomes divergent.

In addition to the conditions encoded in Eq.~(\ref{eq:preprocessing}), we 
also set $\beta_\Sigma = \beta_{T_8}$,
namely we assume that the two quark distributions $\Sigma$ and $T_8$ share
a similar large-$x$ asymptotic behavior.
The reason for this choice is two-fold.
First, we know that these two distributions are
highly (anti-) correlated for
neutral-current nuclear DIS observables (see Eq.~(\ref{eq:singlequarkcombination})).
Secondly, the large-$x$ behavior of these distributions is expected to be approximately
the same, given that the strange distribution $s^+$ is known to be suppressed
at large $x$ compared to the corresponding $u^+$ and $d^+$ distributions.
In any case, it is important to emphasize that
the neural network has the ability to compensate 
for any deviations in the shape of the preprocessing function, 
and therefore can in principle
distinguish any differences
between $\Sigma$ and $T_8$ in the large-$x$ region.

\begin{figure}[t]
\begin{center}
  \includegraphics[width=0.9\textwidth]{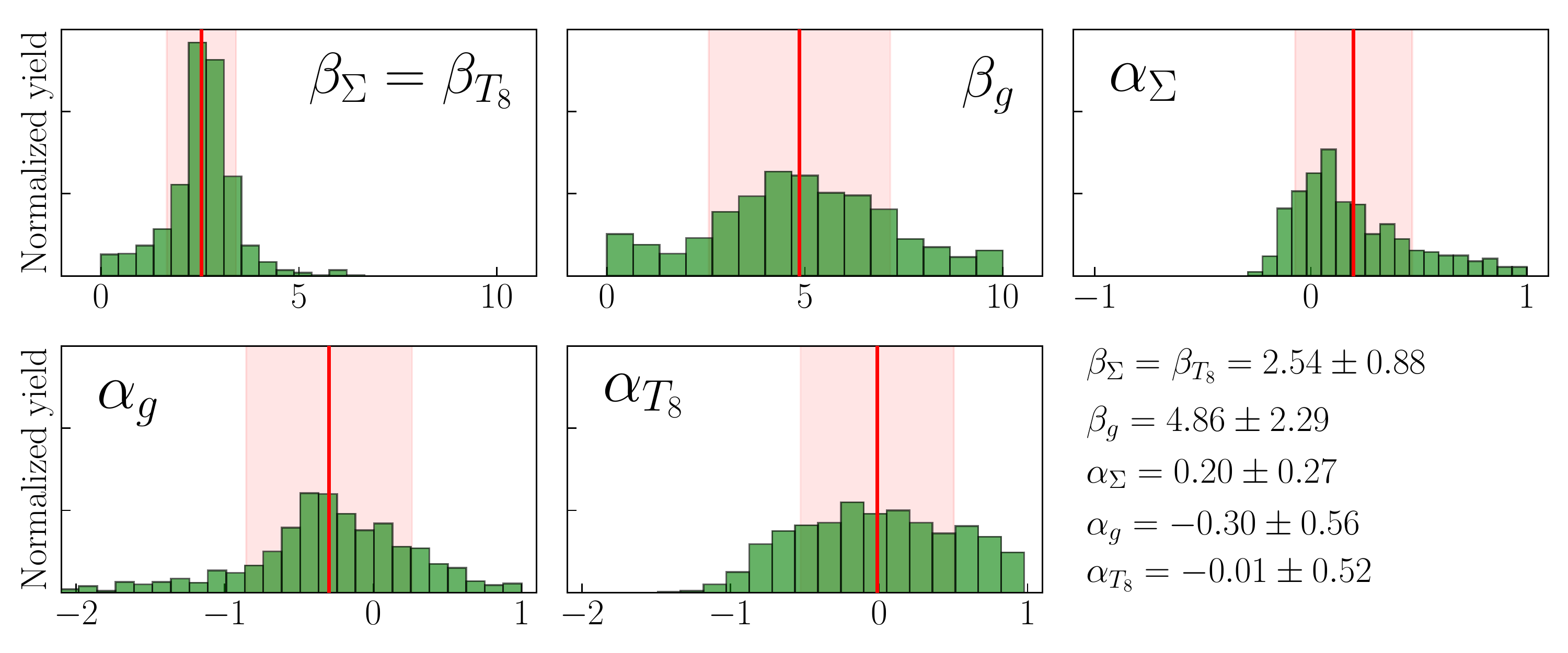}
 \end{center}
\vspace{-0.5cm}
\caption{\small The probability distribution of the fitted preprocessing
  exponents computed with the $N_{\rm rep}=1000$ replicas of the nNNPDF1.0
  NLO set.
  The vertical red line indicates the mean value and the transparent red band
  the 1-$\sigma$ range corresponding to each exponent.
}
\label{fig:preprocessing}
\end{figure}

To illustrate the results of fitting the small and
large-$x$ preprocessing exponents, we display in Fig.~\ref{fig:preprocessing}
the probability distributions associated with the $\alpha_f$ and $\beta_f$
exponents computed using the $N_{\rm rep}=1000$ replicas of the nNNPDF1.0
NLO set, to be discussed in Sect.~\ref{sec:results}.
Here the mean value of each exponent is marked by the solid red line, 
and the transparent 
red band describes the 1-$\sigma$ deviation.
Note that these exponents are restricted to vary only in the interval given by
Eq.~(\ref{eq:preprocessing}).
Interestingly, the resulting distributions for each of the $\alpha_f$ and $\beta_f$
exponents turn out to be quite different, for instance $\beta_{\Sigma}$ is Gaussian-like
while $a_{\Sigma}$ is asymmetric.

\paragraph{The $A=1$ limit of the nPDFs.}
An important physical requirement that must be satisfied by the nPDFs
is that they should reproduce the $x$ dependence
of the PDFs corresponding to isoscalar free nucleons when evaluated at $A=1$.
Therefore, the following boundary conditions needs to be satisfied
for all values of $x$ and $Q^2$:
\be
\label{eq:constraintprotonPDFs}
f(x,Q,A=1) = \frac{1}{2}\lc f_{p}(x,Q^2) + f_{n}(x,Q^2) \rc \, , \quad f=\Sigma,T_8,g \, ,
\ee
where $f_p$ and $f_n$ indicate the parton distributions of the free proton
and neutron, respectively, and are related by isospin symmetry (which is assumed
to be exact).
As opposed to other approaches adopted
in the literature, we do not implement Eq.~(\ref{eq:constraintprotonPDFs})
at the nPDF parameterization level,
but rather we impose it as a restriction in the allowed parameter space at the
level of $\chi^2$ minimization, as will be discussed below.
Our strategy has the crucial advantage that it assures that both central values
and uncertainties of the free-nucleon PDFs will be reproduced by the nNNPDF1.0
nuclear set in the $A\to 1$ limit.

\subsection{Minimization strategy}
\label{sec:minimstrategy}

Having described the strategy
for the nPDF parameterization in terms
of neural networks, we 
turn now to discuss how the best-fit values
of these parameters, namely the weights and thresholds
of the neural network and the preprocessing exponents $\alpha_f$
and $\beta_f$, are determined.
We also explain how we impose the $A=1$ boundary condition, Eq.~(\ref{eq:constraintprotonPDFs}).

In this analysis, the best-fit parameters are determined from the minimization
of a $\chi^2$ function defined as
\bea
\label{eq:chi2}
\chi^2 &\equiv & \sum_{i,j=1}^{N_{\rm dat}} \lp R_i^{\rm (exp)}-R_i^{\rm (th)}(\{ f_m\})\rp
 \lp {\rm cov_{t_0}}\rp_{ij}^{-1} \lp R_j^{\rm (exp)}-R_j^{\rm (th)}(\{ f_m\})\rp \\
&+&
 \lambda \sum_{m=g,\Sigma,T_8}\sum_{l=1}^{N_x} \lp f_m(x_l,Q_0,A=1) - f_m^{(p+n)/2}(x_l,Q_0) \rp^2 \, .
 \nonumber
 \eea
Here,
  $R_i^{\rm (exp)}$ and $R_i^{\rm (th)}(\{ f_m\})$ stand for the experimental
 data and the corresponding theoretical predictions for the
 nuclear ratios, respectively, the latter of which depend on the
 nPDF fit parameters. 
 The $t_0$ covariance matrix ${\rm cov_{t_0}}$ has been defined in
 Eq.~(\ref{eq:t0covmat}), and $N_{\rm dat}$ stands for the total
 number of data points included in the fit.
 Therefore, the first term above is the same
 as in previous NNPDF fits.
 Note that the first row in Eq.~(\ref{eq:chi2}) could also be expressed in terms of
 shifts to the data or theory allowed by the 
 correlated systematic errors~\cite{Gao:2017yyd}.

\paragraph{Reproducing the proton PDF baseline.}
The second term in Eq.~(\ref{eq:chi2}) is a new feature in nNNPDF1.0.
It corresponds to a quadratic penalty that forces
the fit to satisfy the boundary condition in Eq.~(\ref{eq:constraintprotonPDFs}),
namely that the fitted nPDFs for $A=1$ reproduce the PDFs of an isoscalar
free nucleon constructed as the average of the proton and neutron PDFs.
 In order to impose this constraint in a fully consistent way, it is necessary
 for the proton PDF baseline to have been determined using theoretical settings
 and a fitting methodology that best match those of the current
 nPDF analysis.
 This requirement is satisfied by the NNPDF3.1 global analysis~\cite{Ball:2017nwa},
 a state-of-the-art determination of the proton PDFs based on a wide range
 of hard-scattering processes together with higher-order QCD calculations.
 Crucially, NNPDF3.1 shares most of the methodological choices of nNNPDF1.0
 such as the use of neural networks for the PDF
 parameterization and of the Monte Carlo replica method
 for error propagation and estimation.

 As can be seen from Eq.~(\ref{eq:chi2}), this constraint is only 
 imposed at the initial scale $Q_0$. 
 This is all that is required, since the properties of DGLAP 
 evolution will result in distributions at $Q > Q_0$ that 
 automatically satisfy the constraint.
 The $A=1$ boundary condition is then constructed with a grid of $N_x=60$ values of $x$, where
 10 points are distributed logarithmically from $x_{\rm min}=10^{-3}$ to $x_{\rm mid}=0.1$ and
 50 points are 
 linearly distributed from $x_{\rm mid}=0.1$ to $x_{\rm max}=0.7$.
 
 Note that in the low-$x$ region the coverage of this constraint
 is wider than that of the available nuclear data (see Fig.~\ref{figkinplot}).
 Since proton
 PDF uncertainties, as a result of including HERA structure function data,
 are more reduced at small $x$ than in the corresponding
 nuclear case, the constraint in Eq.~(\ref{eq:chi2}) introduces
 highly non-trivial information regarding the shape of the nPDFs within
 and beyond the experimental data region.
 Moreover, we have also verified that the constraint can also be applied down
 to much smaller values of $x$, such as $x_{\rm min}=10^{-5}$, by
 taking as a proton baseline one of the NNPDF3.0 sets which include LHCb charm production
 data~\cite{Gauld:2015yia,Gauld:2016kpd,Bertone:2018dse}, as
 will be demonstrated in Sect.~\ref{sec:methstudies}.

 It is important to emphasize that the boundary condition, Eq.~(\ref{eq:constraintprotonPDFs}),
 must be satisfied both for the PDF central values and for the corresponding uncertainties.
 Since proton PDFs are known to much higher precision than
 nPDFs, imposing this condition introduces a significant amount of new information
 that is ignored in most other nPDF analyses.
 In order to ensure that PDF uncertainties are also reproduced in
 Eq.~(\ref{eq:constraintprotonPDFs}), for each nNNPDF1.0 fit
  we randomly choose a replica from the 
  NNPDF3.1 proton global fit in Eq.~(\ref{eq:chi2}).
  Since we are performing a large $N_{\rm rep}$ number of fits to estimate the uncertainties
  in nNNPDF1.0, the act of randomly choosing a different proton PDF baseline each time 
  guarantees that the necessary information contained in NNPDF3.1 will propagate into the nPDFs.
Finally, we fix the hyper-parameter to $\lambda = 100$, which is 
found to be the optimal setting together with the choice of architecture to yield $A=1$ 
distributions that best describe the central values and uncertainties of NNPDF3.1.

\paragraph{Optimization procedure.}
Having defined our $\chi^2$ function in Eq.~(\ref{eq:chi2}), we now move
to present our procedure to determine the best-fit values
of the parameters associated with each Monte Carlo replica.
This procedure begins by sampling the initial values
of the fit parameters.
Concerning the preprocessing
exponents $\alpha_f$ and $\beta_f$, they are sampled from a uniform prior
in the range $\alpha_f \in [-1,1]$ and $\beta_f \in [1,10]$ for all fitted distributions.
Note that these initial ranges are contained within the ranges from Eq.~(\ref{eq:preprocessing}) 
in which the exponents are allowed to vary.
Since the neural network can accommodate changes in the PDF shapes from
the preprocessing exponents, we find the choice of the prior range from which
$\alpha_f$ and $\beta_f$ are initially sampled does not affect the resulting distributions. 
In the end, the distributions of $\alpha_f$ and $\beta_f$ do not exhibit flat behavior, as 
is shown in Fig.~(\ref{fig:preprocessing}).

Concerning the initial
sampling of the neural network parameters, we use Xavier initialization~\cite{GlorotAISTATS2010},
which samples from a normal distribution with a mean of zero and standard 
deviation that is dependent on the specific architecture of the network.
Furthermore, the initial 
values of the neuron activation are dropped and re-chosen if they are outside two standard 
deviations.
Since a sigmoid activation function is used for the first
and second layers, this truncation of the sampling
distribution ensures the neuron input to be around the origin where the derivative is largest, 
allowing for more efficient network training.

As highlighted by Fig.~\ref{fig:nNNPDF10},
the most significant difference between the fitting methodology used
in nNNPDF1.0 as compared to previous NNPDF studies is the choice
of the optimization algorithm for the $\chi^2$ minimization.
In the most recent unpolarized~\cite{Ball:2017nwa} and polarized~\cite{Nocera:2014gqa}
proton PDF analysis
based on the NNPDF methodology, an in-house Genetic Algorithm (GA)
was employed for the $\chi^2$ minimization, while for the NNFF fits
of hadron fragmentation functions~\cite{Bertone:2018ecm} the related
Covariance Matrix Adaptation -
Evolutionary Strategy (CMA-ES) algorithm was used (see also~\cite{Rojo:2018qdd}).
In both cases, the optimizers require as input only the local values
of the $\chi^2$ function for different points in the parameter space, but never use
the valuable information contained in its gradients.

In the nNNPDF1.0 analysis,
we utilize for the first time gradient descent with backpropagation, 
the most widely used training technique for neural networks (see also~\cite{Forte:2002fg}).
The main requirement
to perform gradient descent is to be able to efficiently compute the gradients of the cost
function Eq.~(\ref{eq:chi2}) with respect to the fit parameters.
Such gradients can in principle be computed analytically, by exploiting
the fact that the relation between the structure functions and the input
nPDFs at $Q_0$ can be compactly expressed in terms of a matrix multiplication
within the {\tt APFELgrid} formalism as indicated by Eq.~(\ref{eq:ev_interp}).
One drawback of such approach is that the calculation of the gradients
needs to be repeated whenever the structure of the $\chi^2$ is modified.
For instance, different analytical expressions for the gradients are required if uncertainties
are treated as uncorrelated and added in quadrature as opposed to 
the case in which systematic correlations are taken into account.

Rather than following this path, in nNNPDF1.0 we have implemented 
backpropagation neural network training using reverse-mode automatic differentiation
in {\tt TensorFlow}, a highly efficient and accurate
method to automatically compute the 
gradients of any user-defined cost function.
As a result, the use of automatic differentiation makes it significantly easier to explore optimal
settings in the model and extend the analysis to include other types of observables
in a global analysis. 

One of the drawbacks of the gradient descent approach, which is partially
avoided by using GA-types of optimizers, is the
risk of ending up trapped in local minima.
To ensure that such situations are avoided as much as possible, in nNNPDF1.0
we use the Adaptive Moment Estimation (ADAM) algorithm~\cite{DBLP:journals/corr/KingmaB14} 
to perform stochastic gradient descent (SGD).
The basic idea here is to perform the training
on randomly chosen subsets of the input experimental data,
which leads to more frequent parameter updates.
Moreover, the ADAM algorithm
significantly improves SGD by adjusting the learning rate of the parameters using 
averaged gradient information from previous iterations.
As a result, local minima are more easily bypassed in the training procedure,
which not only increases the likelihood of ending in a global minima but also
significantly reduces the training time. 

In this analysis, most of the ADAM hyper-parameters 
are set to be the default values given by the algorithm, which have been 
tested on various machine learning problems. 
This includes
the initial learning rate of the parameters, $\eta = 0.001$, the 
exponential decay rate of the averaged squared gradients from past iterations, 
$\beta_2 = 0.999$, and a smoothing parameter $\epsilon = 10^{-8}$. 
However,
we increase the exponential decay rate of the mean of previous
gradients, $\beta_1 = 0.9 \to 0.99$, 
which can be interpreted more simply as the descent momentum. 
This choice was observed to improve the performance of the minimization
overall, as it exhibited quicker exits from local minima and increased
the rate of descent. 

Given that our neural-network-based parameterization
of the nPDFs, Eq.~(\ref{eq:param}), can be shown to be highly redundant
for the current input dataset (see also Sect.~\ref{sec:methstudies}),
we run the risk of fitting
the statistical fluctuations in the data rather than the underlying physical law.
To prevent such overfitting,
we have implemented the look-back cross-validation stopping criterion
presented for the first time in NNPDF fits in Ref.~\cite{Ball:2014uwa}.
The basic idea of this algorithm is to separate the input dataset into disjoint training
and validation datasets (randomly chosen replica by replica),
minimize the training set $\chi^2$ function, $\chi^2_{\rm tr}$, and stop the training
when the validation $\chi^2$, $\chi^2_{\rm val}$, reveals a global minimum.
In this analysis, the data is partitioned 50\%/50\% to construct each
of the two sets, except for experiments with 5 points
or less which are always included in the training set. 

The final fits are chosen to satisfy simultaneously,
\bea
\label{eq:chi2cuts}
&\chi^2_{\rm tr}/{N_{\rm tr}} < 5,\\ \nonumber
&\chi^2_{\rm val}/{N_{\rm val}} < 5,\\
&\chi^2_{\rm penalty}/(3N_x) < 5, \nonumber
\eea
where $N_{\rm tr}$ and $N_{\rm val}$ are the number of data points in the training
and validation sets, respectively, and $\chi^2_{\rm penalty}$ corresponds to the 
second term in Eq.~\ref{eq:chi2}. 
Upon reaching the above conditions during $\chi^2$ minimization, checkpoints 
are saved for every 100 iterations. A fit is then terminated 
when a smaller value for the validation 
$\chi^2$ is not obtained after 
$5\times 10^4$ iterations, or when the fit has proceeded $5\times 10^5$
iterations (early stopping).
The former is set to allow sufficient time to 
escape local minima, and the latter is set due to the SGD algorithm, which 
can fluctuate the parameters around the minimum indefinitely. 
In either case the fit is considered successful, and the parameters 
that minimize $\chi^2_{\rm val}$ are selected as the best-fit parameters
(look-back). 

While automatic differentiation with {\tt TensorFlow} is the baseline
approach used to construct the nNNPDF1.0 sets, during this work 
we have also developed
an alternative {\tt C++} fitting framework that uses
the {\tt ceres-solver} optimizer interfaced with analytical calculations
of the gradients of the $\chi^2$ function.
The comparison between the analytical 
and automated differentiation
strategies to compute the $\chi^2$ gradients and to carry out
the minimization of Eq.~(\ref{eq:chi2}) will be presented elsewhere~\cite{KhalekAnalytical},
together with a comparison of the performance and benchmarking
between these two approaches.

\paragraph{Performance benchmarking.}
While a detailed and systematic comparison between the performances
of the {\tt TensorFlow}-based stochastic gradient descent optimization
used in nNNPDF1.0 and that of the GA and
CMA-ES minimizers used in previous NNPDF analyses is beyond the scope
of this work, here we provide a qualitative estimate for improvement
in performance that has been achieved as a result of using
the former strategy.

In order to assess the performance of these two strategies,
we have conducted two Level-0 closure tests (see Sect.~\ref{sec:closuretests}
for more details) on the same computing machine.
For the first test, a variant of the nNNPDF1.0 fit was run without
the $A=1$ constraint and with the preprocessing exponents fixed
to randomly selected values within a specific range.
For the second, a variant of the NNPDF3.1 DIS-only fit was run
with kinematic cuts adjusted so that they match the value of $N_{\rm dat}$
used in the first fit. Moreover, the preprocessing exponents are 
fixed in the same manner. 
To further ensure similar conditions as much as possible with
the nNNPDF1.0-like fits, the fitting basis
of NNPDF3.1 has been reduced to that composed by only $\Sigma, T_8$ and $g$,
while the architecture of the networks has been modified so that the
number of fitted free parameters is the same in both cases.
While these common settings are suitable for a qualitative comparison
between the two optimizers, we want
to emphasize that the
results should not be taken too literally, as other aspects of the 
two fits are slightly different.

\begin{figure}[t]
  \begin{center}
    \includegraphics[width=0.8\textwidth]{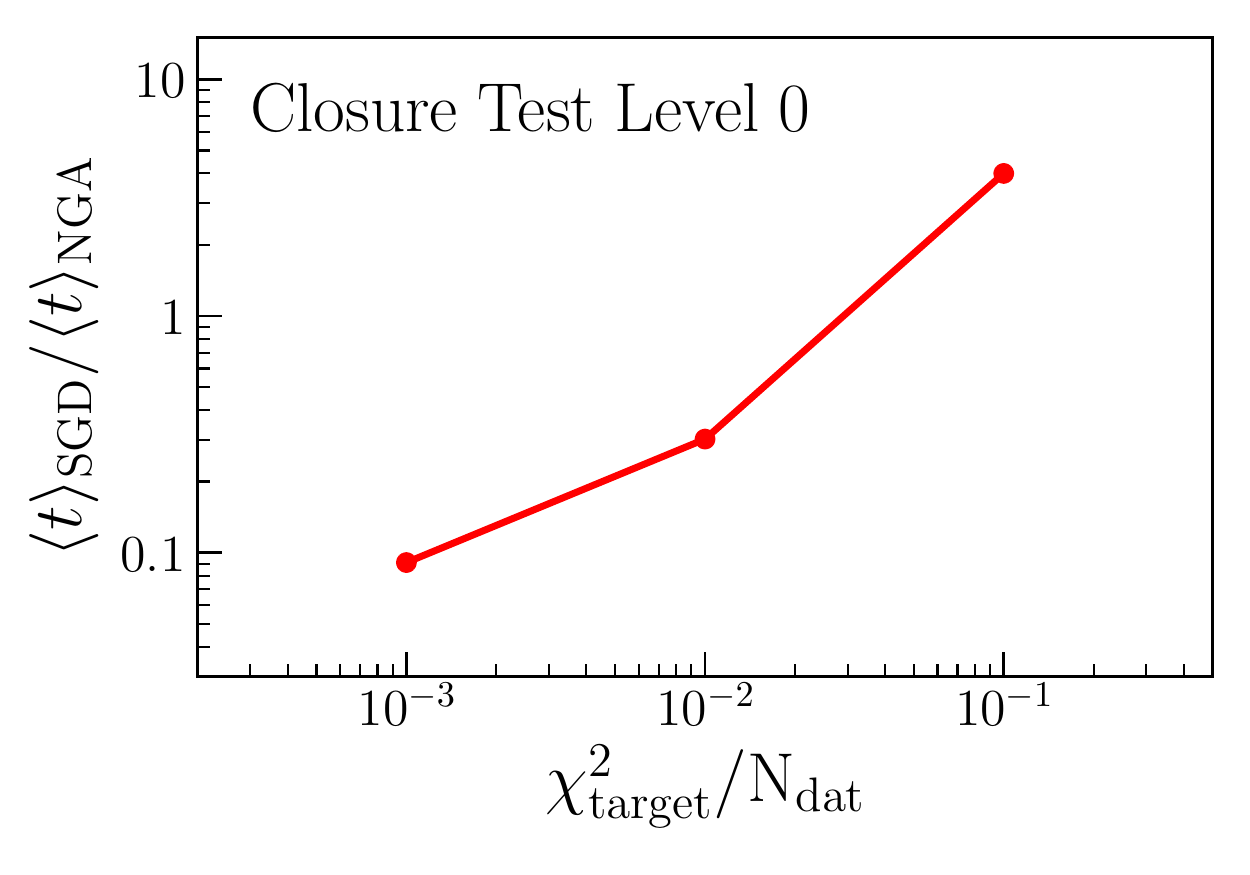}
   \end{center}
  \vspace{-0.3cm}
  \caption{\small Benchmark comparison of the performance of
    the {\tt TensorFlow}-based stochastic gradient descent optimization
    with that of the Genetic Algorithm used in most
    of the previous NNPDF fits.
    The ratio of the average SGD time over the average NGA time
    is plotted as a function of the $\chi^2_{\rm target}/N_{\rm dat}$ for
    a Level 0 closure test.
  }
  \label{fig:benchmark}
  \end{figure}

In Fig.~\ref{fig:benchmark} we show the
results of this benchmark comparison for the performance of
the {\tt TensorFlow}-based stochastic gradient descent optimization
with that of the Genetic Algorithm (labelled NGA) used in
NNPDF3.1.
Within a Level-0 closure test, we monitor the average time it takes
each optimizer to reach a given $\chi^2_{\rm target}/N_{\rm dat}$ target.
We then plot the ratio of the average SGD time with the corresponding
GA result.
For a conservative target, $\chi^2_{\rm target}/N_{\rm dat}=0.1$,
the NGA appears to perform better than SGD.
This is a well-known property of evolutionary algorithms: they manage
to rather quickly bring the parameter space to the vicinity of a minimum
of the cost function.

The real power of SGD becomes apparent once a more stringent
$\chi^2_{\rm target}/N_{\rm dat}$ target is set. As will be described in more
detail in the following section,
the figure of merit in Level 0 closure tests
can be arbitrarily reduced until asymptotically the $\chi^2 \to 0$ limit 
is reached.
We find that in this case, the average time for SGD can be significantly 
smaller than the corresponding NGA time.
For $\chi^2_{\rm target}/N_{\rm dat}=10^{-3}$, the speed improvement
is around an order of magnitude, and from the trend it is apparent
that this improvement would continue for new $\chi^2$ targets.
The benchmark comparison of Fig.~\ref{fig:benchmark} highlights how,
with the use of SGD, it becomes possible to explore the vicinity of
minima in a more efficient way than NGA, thus bringing in a considerable
speed improvement that can reach a factor of 10 or more.

\section{Closure Tests}
\label{sec:closuretests}

Since a significant amount of the fitting methodology
used to construct nNNPDF1.0 has been implemented 
for the first time in this analysis, it is important
to test its performance
and demonstrate its reliability using closure tests.
The general idea of closure tests is to carry out fits
based on pseudo-data generated with a 
known underlying theory. In this case, an
existing nPDF set is used and the fit results 
are compared with this ``true'' result
using a variety of statistical estimators.
In addition, the exercise is performed
within a clean environment
which is not affected by other possible effects
that often complicate global fits, such as limitations
of the theory calculations or the presence of
internal or external data inconsistencies.

In this section, we briefly outline
the different types closure tests that we have
performed to validate the nNNPDF1.0 fitting methodology.
The nomenclature and settings for the different levels of closure tests
follows Ref.~\cite{Ball:2014uwa} (see also~\cite{Hartland:2019bjb} for
a related discussion in the context of SMEFT analyses).

For all of the closure tests presented
in this section, the fitting methodology is identical to that used
for the actual nNNPDF1.0 analysis
and was described in the previous section.
The differences between the different closure fits are 
then related to the generation of the pseudo-data.
The underlying distributions have been chosen to be those
from the nCTEQ15 analysis, from which the pseudo-data
is constructing by
evaluating predictions for the nuclear structure
functions
using the theoretical formalism in Sect.~\ref{sec:expdata}.
Furthermore, the deuteron structure functions are constructed
from the NNPDF3.1 proton PDF set so that only the 
numerator is fitted in the $F_2^A/F_2^D$ ratios. 
The nPDFs to be determined by the fit are then
parameterized at the input scale $Q_0 = 1.3$ GeV (rather than 1 GeV) to maintain 
the consistency with the settings of nCTEQ15.
Lastly, we do not impose our boundary condition at $A=1$, since our
aim is not to achieve the smallest possible uncertainties but instead
to show that we are able to reproduce the input distributions from nCTEQ15.

\subsection{Level 0}

We start by presenting the results of the simplest type of closure test,
Level 0 (L0), and then discuss how these are modified for the more
sophisticated Level 1 (L1) and Level 2 (L2) fits.
At Level 0, the pseudo-data is generated from
the nCTEQ distributions without any additional
statistical fluctuations,
and the uncertainties are taken to be the same as the 
experimental data.
The $\chi^2$ is then defined to be the same as in the fits to real data, 
taking into account all sources of experimental uncertainties
in the $t_0$ covariance matrix Eq.~(\ref{eq:t0covmat}).
Moreover, there are no Monte Carlo replicas, and closure tests are carried
out $N_{\rm rep}$ times for different random values of the initial fit
parameters.
The variance of the $N_{\rm rep}$ fits then defines the PDF
uncertainties at this level.
By defining the closure test Level 0 in this way, there is 
guaranteed to exist at least one possible solution
for the fit parameters which result in $\chi^2=0$,
where the fitted nPDFs coincide with nCTEQ15.
Therefore a key advantage of this test is its ability to assess
the flexibility of the chosen input functional form and determine whether
the shapes of the underlying distributions are able to be reproduced.
%

\begin{figure}[t]
\begin{center}
  \includegraphics[width=0.8\textwidth]{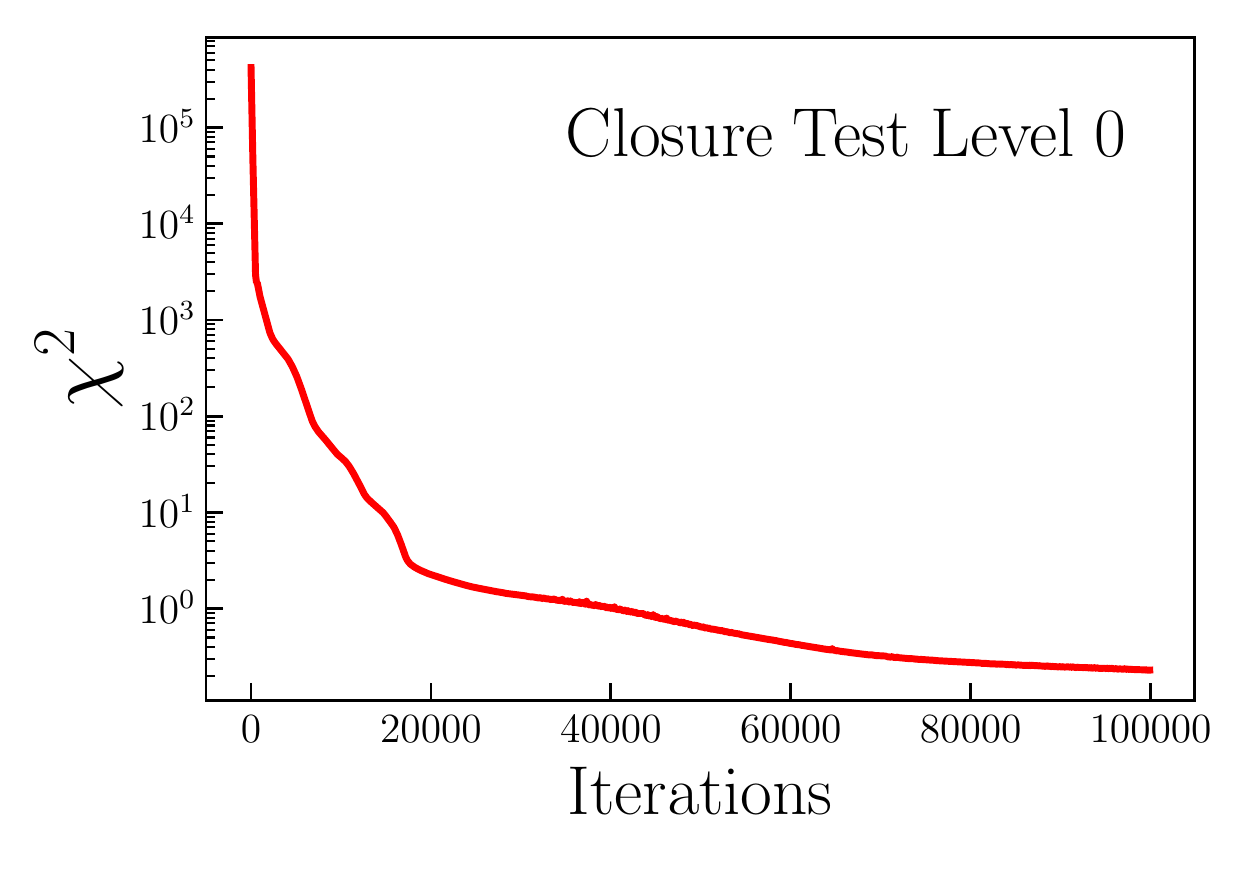}
 \end{center}
\vspace{-0.7cm}
\caption{\small The value of the total $\chi^2$ as a function of
  the number of iterations in a Level 0 minimization procedure.
  Only a single replica is shown, which represents a specific choice 
  of the initial conditions of the fit parameters.
}
\label{fig:CTL0chi2}
\end{figure}

Due to the absence of statistical fluctuations in the pseudo-data, 
overlearning is not possible.
Consequently, 
the fits are performed without cross-validation and early stopping,
and the maximum number of iterations a given fit can progress
is a free parameter 
that can be chosen to be arbitrarily large.
Although the value of the total $\chi^2$ with respect to the
number of iterations may flatten as the maximum number 
of iterations is increased and one is close to the absolute 
minimum, the $\chi^2$ should continue to vanish asymptotically 
provided the optimizer is adequately efficient. 

To demonstrate that these expectations
are indeed satisfied in our case, we display in Fig.~\ref{fig:CTL0chi2}
the value of the total $\chi^2$ as a function of
the number of iterations proceeded in the minimization process
for a specific choice of the initial conditions
of the fit parameters.
  We find the results for other initial conditions are qualitatively similar.
  For this case, the $\chi^2$ decreases monotonically with
  the number of iterations without ever saturating.
  Note also how the rate of decrease of the $\chi^2$ is high for low
  number of iterations, but becomes slower as the number of iterations
  is increased.
  This behavior is expected since it is more difficult to find directions
  in the parameter space close to the minimum that further reduce 
  the cost function.
  The final results are chosen to satisfy $\chi^2/N{\rm dat}<0.1$ after
  a maximum number of iterations of $2\times10^5$. 

\begin{figure}[t]
\begin{center}
  \includegraphics[width=0.9\textwidth]{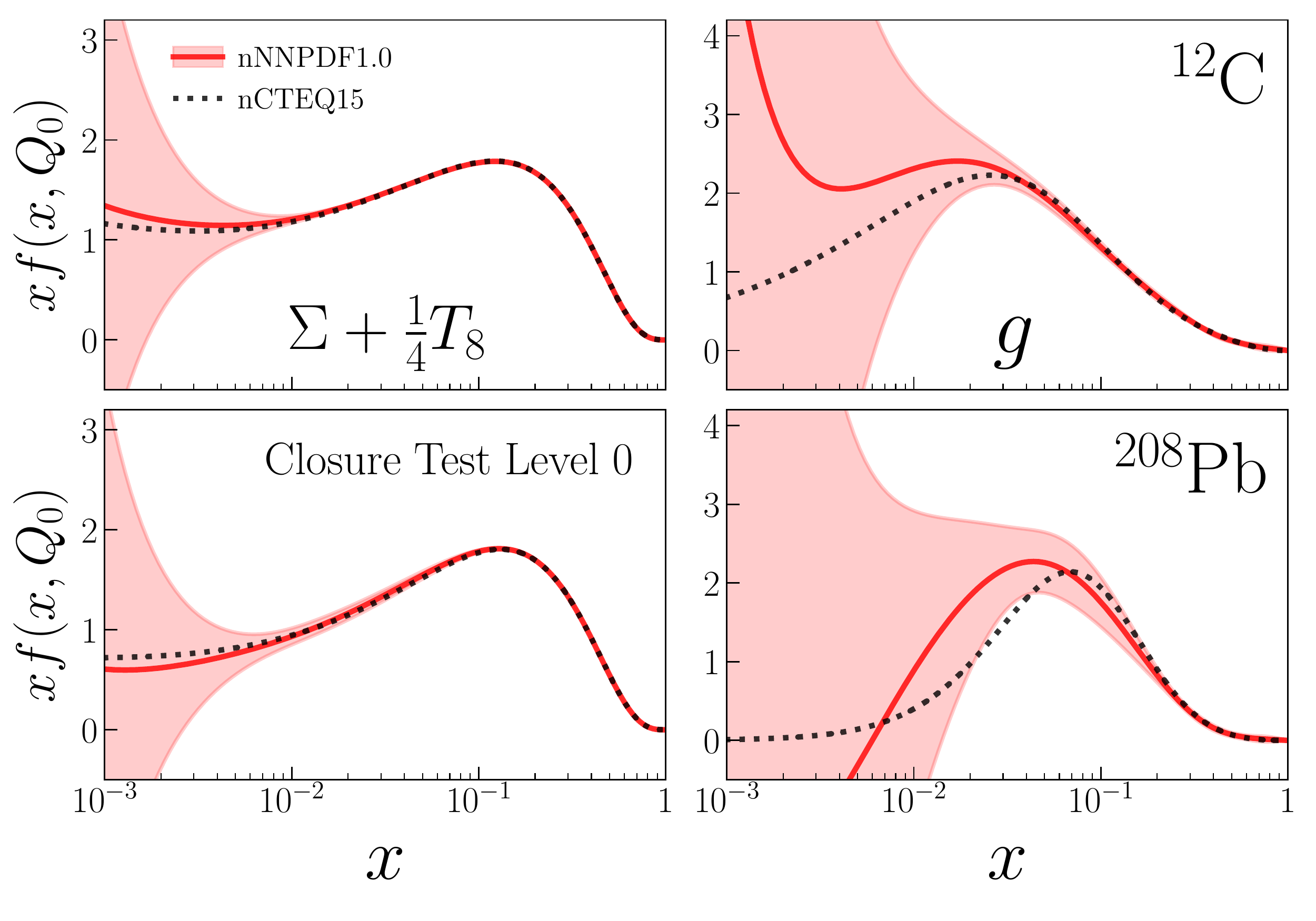}
 \end{center}
\vspace{-0.7cm}
\caption{\small Resulting nPDFs from a Level 0 closure test.
  The $\Sigma+T_8/4$ quark combination (left plots) and the gluon (right plots)
  at the initial evolution scale $Q_0=1.3$ GeV for $A=12$ (upper plots) and $A=208$
  (lower plots) nuclei.
  We compare the results of the nNNPDF1.0 L0 closure test (solid red line) and the
  corresponding 1-$\sigma$ uncertainties (shaded red band) with the central values
  of the nCTEQ15 prior (dotted black line).
}
\label{fig:PDFsCTL0}
\end{figure}

 In Fig.~\ref{fig:PDFsCTL0}, we show the resulting nPDFs from a Level 0 closure test.
  Here the $\Sigma+T_8/4$ quark combination and the gluon are plotted
  as a function of $x$
  at the initial evolution scale $Q_0=1.3$ GeV for $A=12$ and $A=208$.
  We also display the 1-$\sigma$ uncertainties
  computed over the $N_{\rm rep}$ replicas, while
  for the input nCTEQ distributions only the central values are shown.
  Since the aim of closure tests is not to reproduce the uncertainties
  of the prior distributions but rather the central values used to generate
  the pseudo-data, the nCTEQ uncertainties are not
  relevant here and therefore we do not show them in any of the results
  presented in this section.

As we can see from the comparison in Fig.~\ref{fig:PDFsCTL0}, 
there is a very good agreement between the central
values of both the quark combination $\Sigma+T_8/4$ and the gluon
with the nCTEQ15 prior in the data region.
This is especially marked for the quark distribution, given that it is directly
constrained by the structure function measurements.
Concerning the gluon nPDF, which is only constrained indirectly from data, the agreement
is somewhat better for $A=12$ than for $A=208$ for which there are very
limited experimental constraints.
This behavior can be understood by the fact that the gluon
for $A=208$ is much less constrained by the available data than for $A=12$, and thus
even in a perfect L0 closure test, with $\chi^2\to0$, one can expect small deviations
with respect to the underlying physical distributions.
Nevertheless, our results agree overall with the
nCTEQ15 central values and the L0 closure test is
considered successful. 

\subsection{Level 1}

We continue now to discuss the results of the Level 1 closure tests.
In this case, the pseudo-data is generated by adding one layer
of statistical fluctuations to the nCTEQ15 predictions.
These fluctuations are dictated by the corresponding
experimental statistical and systematic uncertainties, and are
the same that enter in the $t_0$ covariance matrix Eq.~(\ref{eq:t0covmat}).
In other words, we take the L0 pseudo-data and add random noise by 
Gaussian smearing each point about the corresponding experimental uncertainties. 
As in the L0 closure tests, the same pseudo-data set is used
to perform an $N_{\rep}$ number of fits,
each with a different initialization of
the fit parameters, and the resulting
variance defines the nPDF uncertainties.
Due to the Gaussian smearing, however, over-learning is now possible 
at Level 1 and therefore cross-validation with early 
stopping is required.
As a result, the optimal fit parameters are expected to give instead 
$\chi^2_{\rm tot}/N_{\rm dat}\simeq 1$.

\begin{figure}[t]
  \begin{center}
    \includegraphics[width=0.9\textwidth]{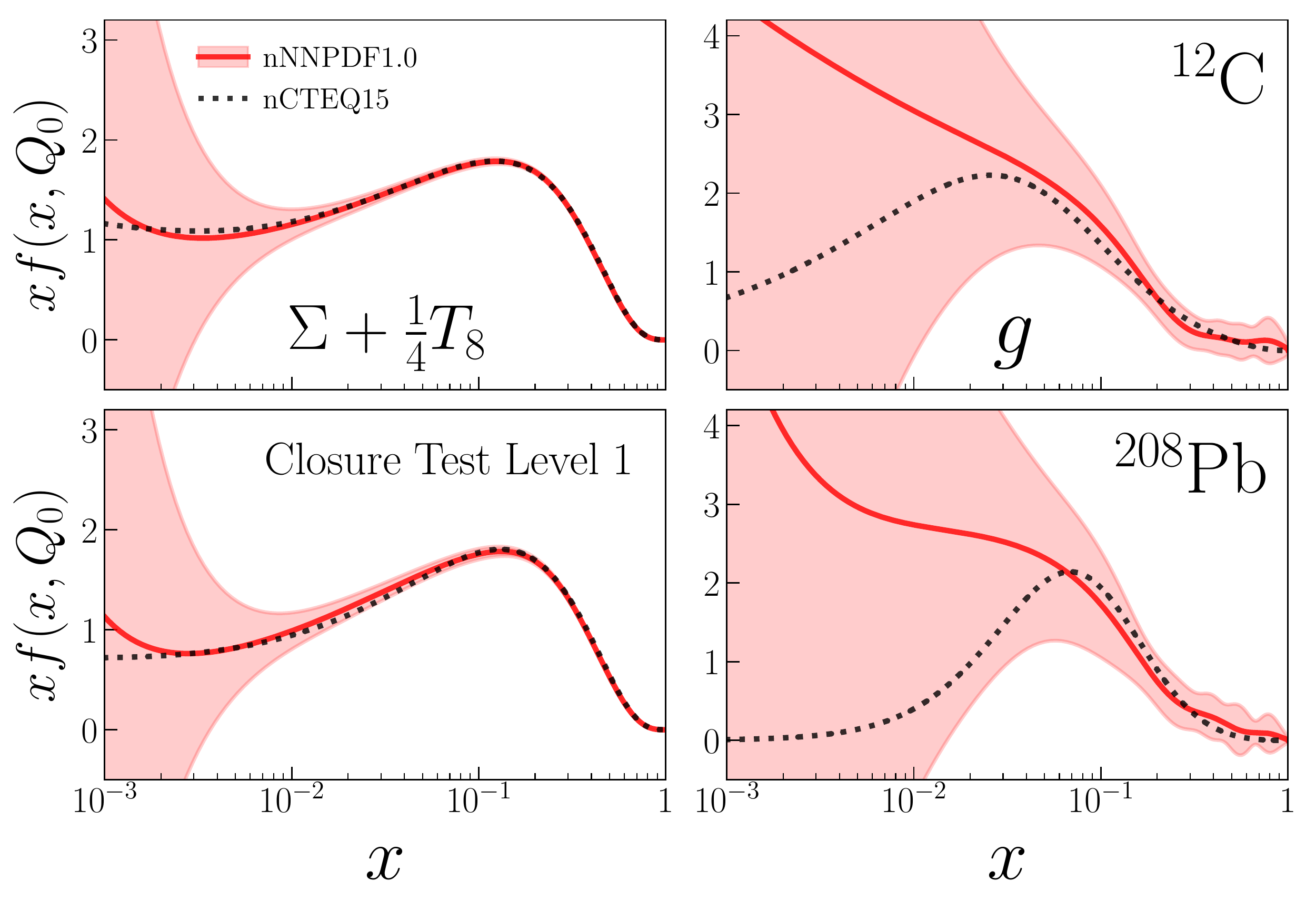}
   \end{center}
  \vspace{-0.7cm}
  \caption{\small Same as in Fig.~\ref{fig:PDFsCTL0}
    but now for the Level 1 closure tests.
  }
  \label{fig:PDFsCTL1}
  \end{figure}

In Fig.~\ref{fig:PDFsCTL1} we show a similar comparison as that of Fig.~\ref{fig:PDFsCTL0}
but now for the L1 closure tests.
While the level of agreement with the nCTEQ15 prior is similar as in the case of L0 fits,
the PDF uncertainties have now increased, especially for the gluon nPDF.
This increase at L1 reflects the range of 
possible solutions for the initial nPDFs at $Q_0$ that
lead to a similar value of $\chi^2/N_{\rm dat}\simeq 1$.
Therefore, the L1 test not only allows us to verify that the input distributions are 
  reproduced, but also that the added 
  statistical fluctuations at the level of the generated pseudo-data are 
  reflected in the resulting uncertainties.

In Table~\ref{tab:chi2table-L1CT} we list the averaged values for 
$\chi^2/N_{\rm dat}$
  in the L1 closure test compared with the corresponding
  values obtained with the prior theory.
  As expected, we find the $\chi^2$ values at L1 being
  close to those of the prior both at the level of the total dataset
  as well as that of individual experiments.
  The agreement is particularly good for datasets with a large number of points
  that carry more weight in the fit.
  In summary, the comparison in Fig.~\ref{fig:PDFsCTL1} combined with that
  in Table~\ref{tab:chi2table-L1CT} demonstrate the closure
  tests are also successful at L1.
  
\begin{table}[H]
  \centering
  \small
   \renewcommand{\arraystretch}{1.55}
   \begin{tabular}{c c c | c | c }
     \multirow{2}{*}{Experiment} &
     \multirow{2}{*}{$~~~{\rm A}_1/{\rm A}_2~~~$} &
     \multirow{2}{*}{${N}_{\rm dat}$} & \multirow{2}{*}{$\chi^2/n_{\rm dat}$ (L1CT)} &
     \multirow{2}{*}{$\chi^2/n_{\rm dat}$ (prior)} \\
     & & & & \\
 \toprule
  SLAC E-139 & $^4$He/$^2$D & 3     &  0.94       &   2.61      \\
  NMC 95, re. & $^4$He/$^2$D & 13 &    1.83       &   1.17      \\
\midrule
  NMC 95 & $^6$Li/$^2$D & 12 &    0.97        &   0.87      \\
\midrule
  SLAC E-139 & $^9$Be/$^2$D & 3 &  0.66     &   0.09    \\
  NMC 96 & $^9$Be/$^{12}$C & 14 &     1.04    &  0.99       \\
\midrule
  EMC 88, EMC 90 & $^{12}$C/$^2$D & 12 &   0.45        &    0.47     \\
  SLAC E-139 & $^{12}$C/$^2$D & 2 &   0.52        &   0.80      \\
  NMC 95, NMC 95, re.   & $^{12}$C/$^2$D & 26 &   1.79       &  1.79       \\
  FNAL E665 & $^{12}$C/$^2$D & 3 &  1.07        &   0.84      \\
  NMC 95, re. & $^{12}$C/$^6$Li & 9 &  0.71        &    0.54     \\
\midrule
  BCDMS 85 & $^{14}$N/$^2$D & 9 &   0.81     &    0.77     \\
\midrule
  SLAC E-139  & $^{27}$Al/$^2$D & 3 &   2.42        &   3.14      \\
  NMC 96 & $^{27}$Al/$^{12}$C & 14 &  1.18     &   1.26      \\
\midrule
  SLAC E-139 & $^{40}$Ca/$^2$D & 2 &   1.24       &  1.36        \\
  NMC 95, re. & $^{40}$Ca/$^2$D & 12 &   2.07       &   1.87      \\
  EMC 90 & $^{40}$Ca/$^2$D & 3 &   3.18      &  3.23       \\
  FNAL E665 & $^{40}$Ca/$^2$D & 3 &   0.23      &   0.23      \\
  NMC 95, re. & $^{40}$Ca/$^6$Li & 9 &   0.46       &   0.41      \\
  NMC 96 & $^{40}$Ca/$^{12}$C & 23 &    1.22       &  1.20       \\
\midrule
  EMC 87 & $^{56}$Fe/$^2$D & 58 &     0.60      &   0.59     \\
  SLAC E-139 & $^{56}$Fe/$^2$D & 8 &   0.66        &   0.57      \\
  NMC 96 & $^{56}$Fe/$^{12}$C & 14 &   1.35        &   1.05      \\
  BCDMS 85, BCDMS 87 & $^{56}$Fe/$^2$D & 16 &   0.82       &   0.70      \\
\midrule
  EMC 88, EMC 93 & $^{64}$Cu/$^2$D & 27 &   1.32       &   1.38      \\
\midrule
  SLAC E-139 & $^{108}$Ag/$^2$D & 2 &   0.33        &    0.28     \\
\midrule
 EMC 88 & $^{119}$Sn/$^2$D & 8 &   0.12       &   0.13     \\
 NMC 96, $Q^2$ dependence & $^{119}$Sn/$^{12}$C & 119 &  0.95     &  0.98      \\
\midrule
 FNAL E665 & $^{131}$Xe/$^2$D & 4 &    0.99       &    0.84    \\
\midrule
  SLAC E-139 & $^{197}$Au/$^2$D & 3 &  0.21      &   0.31      \\
\midrule
 FNAL E665 & $^{208}$Pb/$^2$D & 3 &    1.29      &  1.31      \\
 NMC 96 & $^{208}$Pb/$^{12}$C & 14 &   0.98      &  0.90       \\
 \midrule
 \midrule
 {\bf Total} & & {\bf 451} &   1.03        &   1.00     \\
\bottomrule
\end{tabular}
\vspace{4mm}
\caption{\small The averaged values for $\chi^2/N_{\rm dat}$
  for each experiment in a Level 1 closure test compared with
  values obtained using the nCTEQ15 distributions.
}
\label{tab:chi2table-L1CT}
\end{table}

\subsection{Level 2}

In L2 closure tests, the pseudo-data generated in the L1 case
is now used to produce a large $N_{\rm rep}$ number of Monte Carlo 
replicas.
A nuclear PDF fit is then performed for each replica, 
and look-back cross-validation is again activated 
to prevent over-fitting.
The procedure at L2 therefore matches the one applied to
real data to determine the nNNPDF1.0 set of nPDFs,
so that the statistical and systematic uncertainties
provided by the experimental measurements are 
propagated into the resulting nPDF uncertainties.
By comparing the PDF uncertainties at L2 with those at L1 and L0, one 
can disentangle the various contributions to the total nPDF
error, as we discuss in more detail below.

\begin{figure}[t]
\begin{center}
  \includegraphics[width=0.80\textwidth]{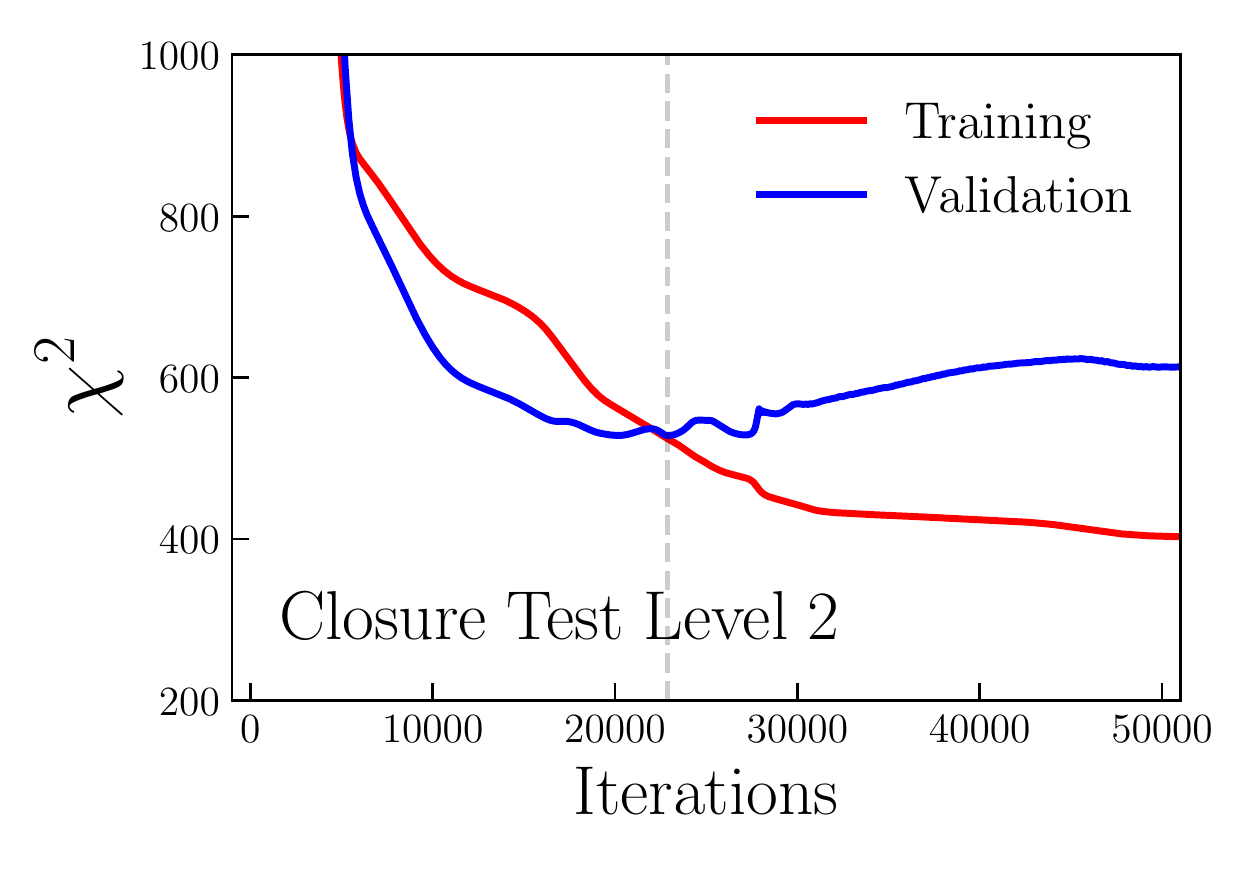}
 \end{center}
\vspace{-0.7cm}
\caption{\small Same as Fig.~\ref{fig:CTL0chi2}, but now for 
the Level 2 closure test.
  We show the separate $\chi^2$ values of the training (solid red line) and the 
  validation (solid blue line) samples, and indicate with a vertical dashed line 
  the stopping point
  for this specific replica, determined as the absolute minimum
  of $\chi^2_{\rm val}$.
}
\label{fig:CTL2chi2}
\end{figure}

Given the extra layer of statistical fluctuations introduced by the Monte Carlo
replica generation, at L2 the figure of merit
for each replica is $\chi^2_{\rm tr\,(\rm val)}/N_{\rm tr\,(val)}\simeq 2$, where
$N_{\rm tr\,(val)}$ indicates the number of data points in the training (validation) set.
In Fig.~\ref{fig:CTL2chi2}, we show
a similar plot as in Fig.~\ref{fig:CTL0chi2}
but now for a representative replica from 
the Level 2 closure test.
Here the $\chi^2$ values from the training and validation
samples are plotted separately, and the vertical 
dashed line indicates the stopping point,
defined to be the absolute minimum
of $\chi^2_{\rm val}$,
at which the optimal parameters are taken.
 Since we have $N_{\rm tr} = 239$
 and $N_{\rm val}=212$, we find
 $\chi^2_{\rm tr\,(\rm val)}/N_{\rm tr\,(val)}\simeq 2$
 as expected.
 
Fig.~\ref{fig:CTL2chi2} clearly illustrates the importance of cross-validation
stopping.
For a low number of iterations, both $\chi^2_{\rm tr}$ and $\chi^2_{\rm val}$ are
similar in size and decrease monotonically: this corresponds to the learning phase.
However, beyond a certain point the $\chi^2_{\rm tr}$ keeps decreasing
while the $\chi^2_{\rm val}$ instead begins to increase, indicating that the
statistical fluctuations rather than the underlying distributions are being fitted.
As a result of using cross-validation, we are able to avoid over-fitting and ensure that
for each replica the minimization is stopped at the optimal number of iterations.

\begin{figure}[t]
\begin{center}
  \includegraphics[width=0.9\textwidth]{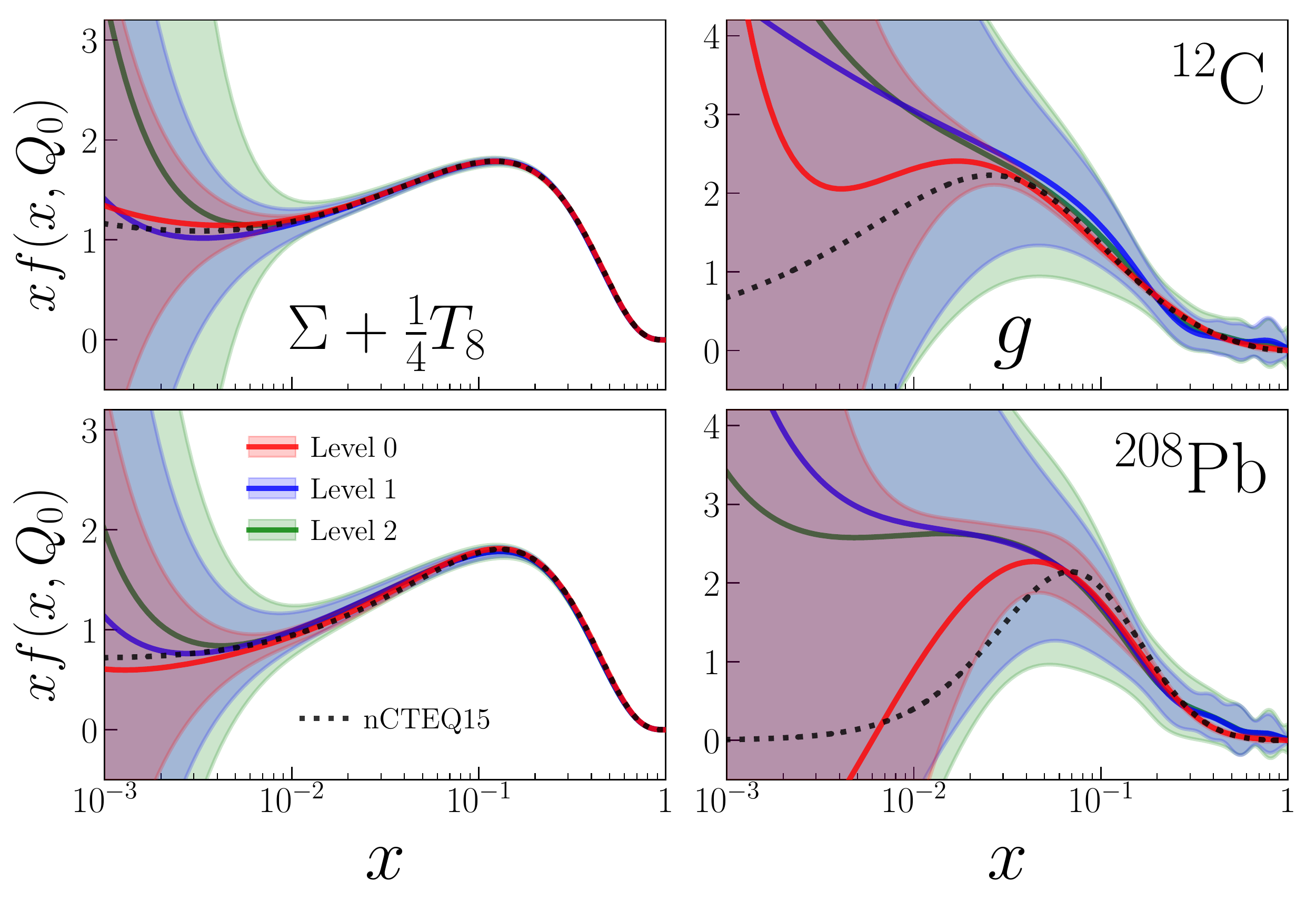}
 \end{center}
\vspace{-0.3cm}
\caption{\small Same as Fig.~\ref{fig:PDFsCTL0}, but now
  showing together the results of the L0 (red),
  L1 (blue), and L2 (green)
  closure tests.
}
\label{fig:PDFsCTL2}
\end{figure}

In Fig.~\ref{fig:PDFsCTL2},
a similar comparison as that of Fig.~\ref{fig:PDFsCTL0} is shown, where now
the nPDFs at the initial
parameterization scale $Q_0=1.3$ GeV
obtained from the L0, L1, and L2 closure tests are displayed together.
Here the nCTEQ15 prior agrees well with the central
values of all the closure tests.
Moreover, it is important to note that the nPDF uncertainties are 
smallest at Level 0 and then
increase subsequently with each level.
The comparison of the results for the different levels of 
closure tests can be interpreted following
the discussions of~\cite{Ball:2014uwa}.

First of all, 
at L0 the PDF uncertainty within the data region should go to zero as the
number of iterations is increased due to the fact that $\chi^2 \to 0$, as illustrated
in Fig.~\ref{fig:CTL0chi2}.
While the PDF uncertainties will generally decrease with the number
of iterations, this may not necessarily be true between data points
(interpolation) and outside the data region (extrapolation).
The latter is known as the extrapolation uncertainty, and is
present even for a perfect fit for which the $\chi^2$
vanishes.
In our fits, the extrapolation region can be assessed from Fig.~\ref{figkinplot},
where $x\lsim 0.01$ and $x\gsim 0.7$ are not directly constrained
from any measurement in nNNPDF1.0.

In a L1 fit, the central
values of the data have been fluctuated around the theoretical predictions from nCTEQ15.
This means that now there can exist many functional forms for the 
nPDFs at $Q_0$ that have equally good $\chi^2$ values.
The difference between the PDF uncertainties at L0 and L1 is thus 
known as the functional uncertainty.
Finally, at L2 one is adding on top of the L1 pseudo-data the Monte Carlo replica
generation reflecting
the statistical and systematic errors provided by the experimental measurements.
This implies that the difference between L1 and L2 uncertainties can be genuinely
attributed to the experimental data errors, and is therefore known
as the data uncertainty.
Comparing the resulting nPDFs for various levels of closure tests,
as in Fig.~\ref{fig:PDFsCTL2},
allows us to discriminate the relative importance of the extrapolation, function,
and data components to the total nNNPDF1.0 uncertainty band.

From the comparison of Fig.~\ref{fig:PDFsCTL2}, we see that the extrapolation uncertainty is very
small for the quarks except for $x\lsim 0.01$ where indeed experimental data stops.
The same applies for the gluon for $A=12$, while for $A=208$ the extrapolation (L0) uncertainty
becomes large already at $x\lsim 0.1$.
Interestingly, there is a big increase in uncertainties when going
from L0 to L1 for the gluon distribution: this highlights how functional uncertainties 
represent by far the dominant component
in the PDF uncertainty for most of the relevant kinematic range.
Lastly, differences between L1 and L2 are quite small, and therefore the experimental data errors
propagated by the MC replicas contributes little to the overall uncertainties.
Nevertheless, it is an important component and must be included for a robust
estimation of the nPDF uncertainties.

\section{Results}
\label{sec:results}

In this section we present the main results of our analysis, namely
the nNNPDF1.0 sets of nuclear parton distributions.
We first assess the quality of our fit by comparing 
the resulting structure function ratios with
experimental data.
This is followed by a discussion of the main features
of the nNNPDF1.0 sets, as well as a comparison with
the recent EPPS16 and nCTEQ15 nPDF analyses.
We also assess the stability of our results with respect to the 
perturbative order, which are generated using NLO
and NNLO QCD theories.
Finally, the section is concluded by presenting a few methodological 
validation tests, complementing the closure test studies discussed in
Sect.~\ref{sec:closuretests}.
Here we show that our results are stable with respect to variations
of the network architecture and quantify the impact of the $A=1$ boundary condition.

\begin{figure}[t]
\begin{center}
  \includegraphics[width=0.7\textwidth]{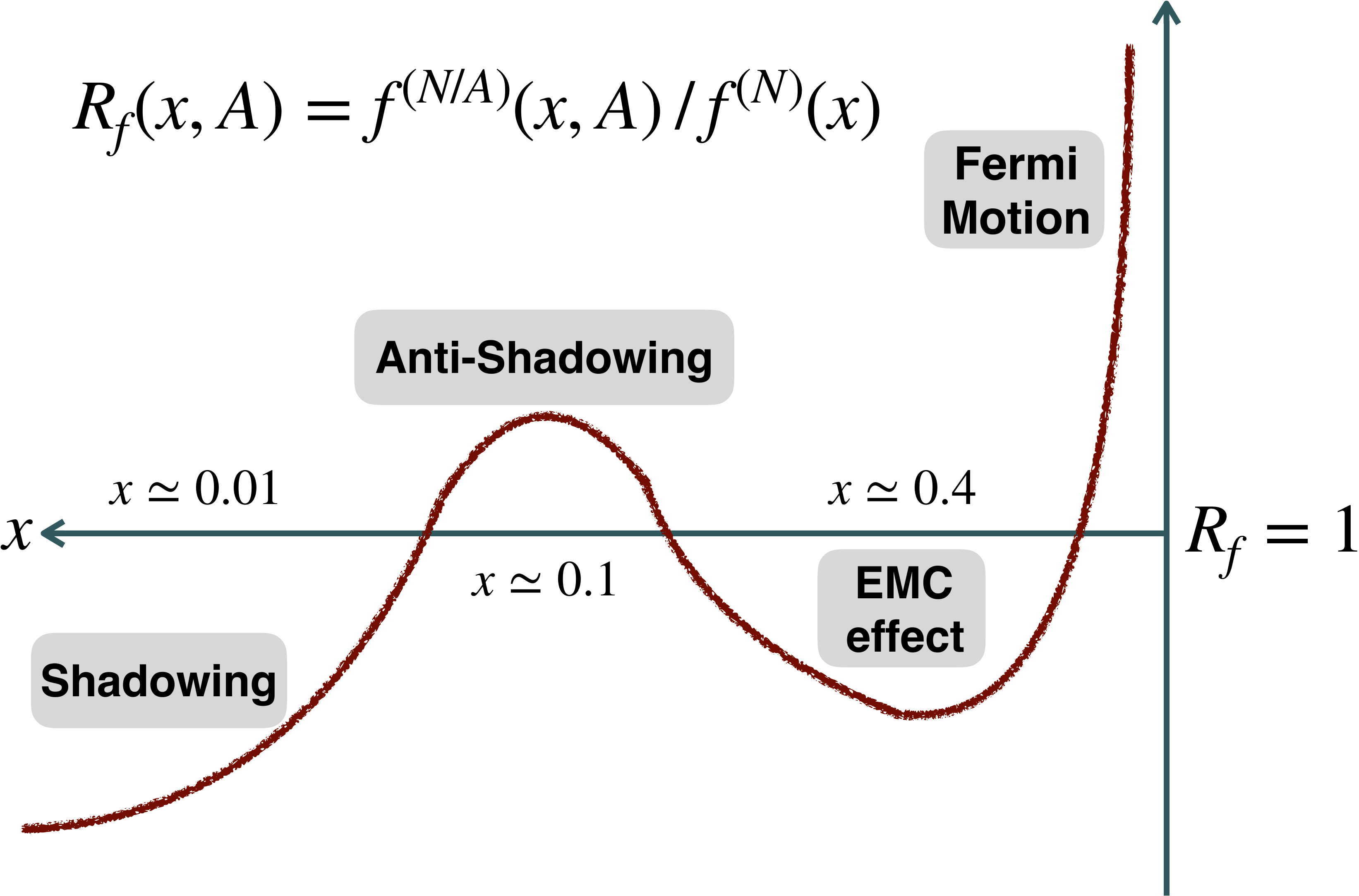}
 \end{center}
\vspace{-0.3cm}
\caption{\small Schematic representation of different types
  of nuclear modifications that
   are expected to arise in the nPDFs, $f^{(N/A)}$, when presented
  as ratios to their free-nucleon counterparts, $R_f=f^{(N/A)}\,/\,f^{(N)}$.
}
\label{fig:cartoon}
\end{figure}

Before moving forward, it is useful to illustrate qualitatively the 
expected outcome for a nuclear PDF analysis.
In Fig.~\ref{fig:cartoon} we show a
schematic representation of different types
of nuclear modifications that
are assumed to arise in the nPDFs, $f^{(N/A)}$, when presented
as ratios to their free-nucleon counterparts $f^{(N)}$,
\be
\label{eq:nPDFratios}
R_f(x,Q^2,A) \equiv \frac{f^{(N/A)}(x,Q^2,A)}{f^{(N)}(x,Q^2)} \, , \qquad
f\,=\, \Sigma+T_8/4, g \, .
\ee
The ratio $R_f$ defined here corresponds to the 
nPDF equivalent of the structure
function ratios, Eq.~(\ref{eq:Sect2Rf2}), where $R_f\simeq1$ 
signifies the absence of nuclear modifications.
Moving from small to large $x$,
a depletion known as shadowing is expected, followed
by an enhancement effect known as anti-shadowing.
For the region $x\simeq 0.4$, there is expected to be a suppression related
to the original EMC effect, while at larger $x$ there should be
a sharp enhancement as a result of Fermi motion.
In presenting the results of the nNNPDF1.0 PDF sets, the discussion
will focus primarily on whether the different nuclear effects
shown in Fig.~\ref{fig:cartoon} are
supported by experimental data
and how the various effects compare
between the quark and gluon distributions.
%

\subsection{Comparison with experimental data}

To begin, we list in Table~\ref{chi2table}
the absolute and normalized values of the $\chi^2$ for
each of the input datasets (see Table~\ref{dataset}) and for the total dataset.
The values are given for both the NLO and NNLO fits.
In total, there are $N_{\rm dat}=451$ data points that survive the kinematic
cuts and result in the overall value $\chi^2/N_{\rm dat}=0.68$, indicating
an excellent agreement between the experimental data and the 
theory predictions.
Moreover, we find that the fit quality is quite similar between 
the NLO and NNLO results.
The fact that we obtain an overall $\chi^2/N_{\rm dat}$ less than one can
be attributed to the absence of correlations between experimental
systematics,
leading to an overestimation of the total error.  
%

\begin{table}[p]
  \centering
  \small
   \renewcommand{\arraystretch}{1.55}
   \begin{tabular}{c c c | c c | c c}
     \multirow{2}{*}{Experiment} &
     \multirow{2}{*}{$~~~{\rm A}_1/{\rm A}_2~~~$} &
     \multirow{2}{*}{${N}_{\rm dat}$} & \multicolumn{2}{c|}{NLO} & \multicolumn{2}{c}{NNLO}\\
          &  &  & $\chi^2$  &   $\chi^2/N_{\rm dat}$ & $\chi^2$  &   $\chi^2/N_{\rm dat}$ \\
 \toprule
  SLAC E-139 & $^4$He/$^2$D & 3 & 1.49  &  0.50   & 1.50   & 0.50   \\ 
  NMC 95, re. & $^4$He/$^2$D & 13 & 12.81    & 1.0   & 12.79   & 0.98    \\ 
\midrule
  NMC 95 & $^6$Li/$^2$D & 12 & 10.96    &  0.91  & 10.50   &  0.88  \\  
\midrule
  SLAC E-139 & $^9$Be/$^2$D & 3 & 2.91    &  0.97  &  2.91  & 0.97   \\  
  NMC 96 & $^9$Be/$^{12}$C & 14 & 4.03    & 0.29   & 4.06   &  0.29  \\  
\midrule
  EMC 88, EMC 90 & $^{12}$C/$^2$D & 12 & 12.98    & 1.08   &  13.04  & 1.09   \\  
  SLAC E-139 & $^{12}$C/$^2$D & 2 & 0.65    &  0.33  & 0.74   & 0.37   \\  
  NMC 95, NMC 95, re.   & $^{12}$C/$^2$D & 26 & 25.12    &  0.97  & 24.81   &  0.95   \\  
  FNAL E665 & $^{12}$C/$^2$D & 3 & 3.13    &  1.04  &  3.13  &  1.04  \\ 
  NMC 95, re. & $^{12}$C/$^6$Li & 9 & 6.62    &  0.74   & 6.25   &  0.69  \\  
\midrule
  BCDMS 85 & $^{14}$N/$^2$D & 9 & 11.10    &  1.23  &  11.16  & 1.24   \\  
\midrule
  SLAC E-139  & $^{27}$Al/$^2$D & 3 & 0.52    &  0.17  & 0.65   & 0.22   \\  
  NMC 96 & $^{27}$Al/$^{12}$C & 14 & 4.34    &  0.31  &  4.31  & 0.31   \\  
\midrule
  SLAC E-139 & $^{40}$Ca/$^2$D & 2 & 2.79    & 1.40   &  2.95  &  1.48  \\ 
  NMC 95, re. & $^{40}$Ca/$^2$D & 12 & 11.75    & 0.98   &  11.86  &  0.99  \\  
  EMC 90 & $^{40}$Ca/$^2$D & 3 & 4.11    & 1.37   & 4.09   & 1.36   \\  
  FNAL E665 & $^{40}$Ca/$^2$D & 3 & 5.07    & 1.69   & 4.77   &  1.59  \\  
  NMC 95, re. & $^{40}$Ca/$^6$Li & 9 & 2.18    &  0.24  &  2.05  &  0.23  \\ 
  NMC 96 & $^{40}$Ca/$^{12}$C & 23 & 13.20    & 0.57   &  13.26  & 0.58   \\  
\midrule
  EMC 87 & $^{56}$Fe/$^2$D & 58 & 36.89    &  0.63  &  37.12  & 0.64   \\  
  SLAC E-139 & $^{56}$Fe/$^2$D & 8 & 11.01    &  1.38  & 11.20   &  1.4  \\  
  NMC 96 & $^{56}$Fe/$^{12}$C & 14 & 9.21    &   0.66 & 9.00   & 0.64   \\  
  BCDMS 85, BCDMS 87 & $^{56}$Fe/$^2$D & 16 & 9.48    & 0.6   & 9.53   & 0.6   \\  
\midrule
  EMC 88, EMC 93 & $^{64}$Cu/$^2$D & 27 & 12.56    & 0.47   & 12.63   &  0.47  \\  
\midrule
  SLAC E-139 & $^{108}$Ag/$^2$D & 2 & 1.04    &  0.52  &  1.04  & 0.52   \\  
\midrule
 EMC 88 & $^{119}$Sn/$^2$D & 8 & 17.77    &  2.22  & 17.71   & 2.21   \\  
 NMC 96, $Q^2$ dependence & $^{119}$Sn/$^{12}$C & 119 & 59.24    & 0.50   & 58.28   & 0.49   \\  
\midrule
 FNAL E665 & $^{131}$Xe/$^2$D & 4 & 1.47    & 0.37   & 1.45   & 0.36    \\  
\midrule
  SLAC E-139 & $^{197}$Au/$^2$D & 3 & 2.46    &  0.82  & 2.33   & 0.78   \\ 
\midrule
 FNAL E665 & $^{208}$Pb/$^2$D & 3 & 4.97    & 1.66   & 5.10   & 1.7   \\  
 NMC 96 & $^{208}$Pb/$^{12}$C & 14 & 5.23    &  0.37   & 5.61   & 0.4   \\  
 \midrule
 \midrule
 {\bf Total} & & {\bf 451} & {\bf 307.1}   & {\bf 0.68}   & {\bf 305.82}   & {\bf 0.68}    \\ 
\bottomrule
\end{tabular}
\vspace{4mm}
\caption{\small Same as Table~\ref{dataset}, now indicating
  the absolute and normalized values of the $\chi^2$ for
  each of the input datasets as well as for the total dataset.
  Listed are the results for both the NLO and NNLO nNNPDF1.0 sets.
}
\label{chi2table}
\end{table}

\begin{figure}[!h]
\begin{center}
  \includegraphics[width=0.99\textwidth]{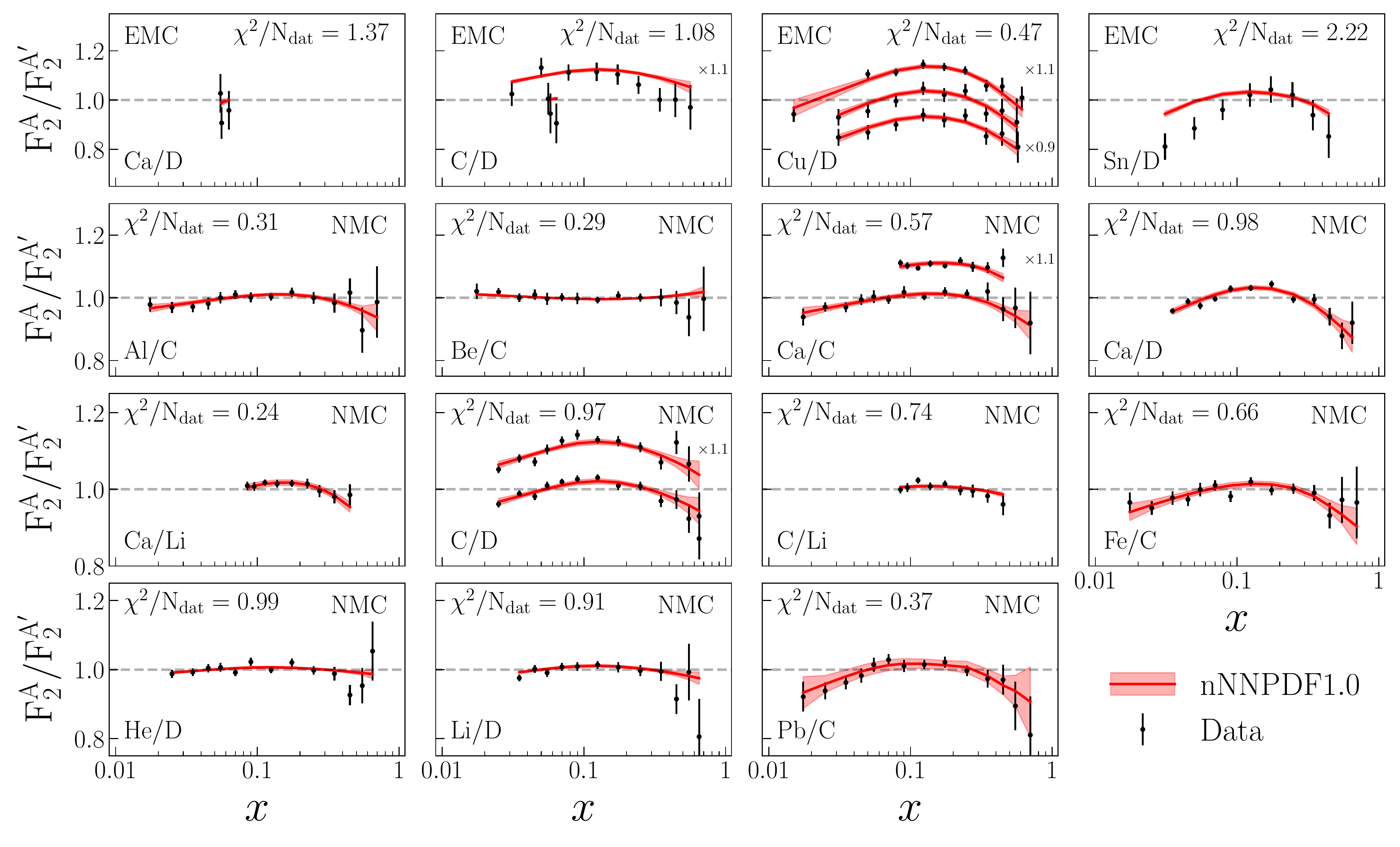}
 \end{center}
\vspace{-0.3cm}
\caption{\small Comparison between the experimental data on the structure
  function ratios $F_2^A/F_2^{A'}$ and the corresponding theoretical predictions
  from the nNNPDF1.0 NLO fit (solid red line and shaded band) for the measurements 
  provided by the EMC
  and NMC experiments.
  The central values of the experimental data points have been shifted
  by an amount determined by the 
  multiplicative systematic uncertainties and their nuisance parameters,
  and the data errors are defined by adding in quadrature the 
  uncorrelated uncertainties.
  Also indicated are the $\chi^2/N_{\rm dat}$ values for each of the datasets.
}
\label{figDvT1}
\end{figure}

\begin{figure}[!h]
\begin{center}
  \includegraphics[width=0.99\textwidth]{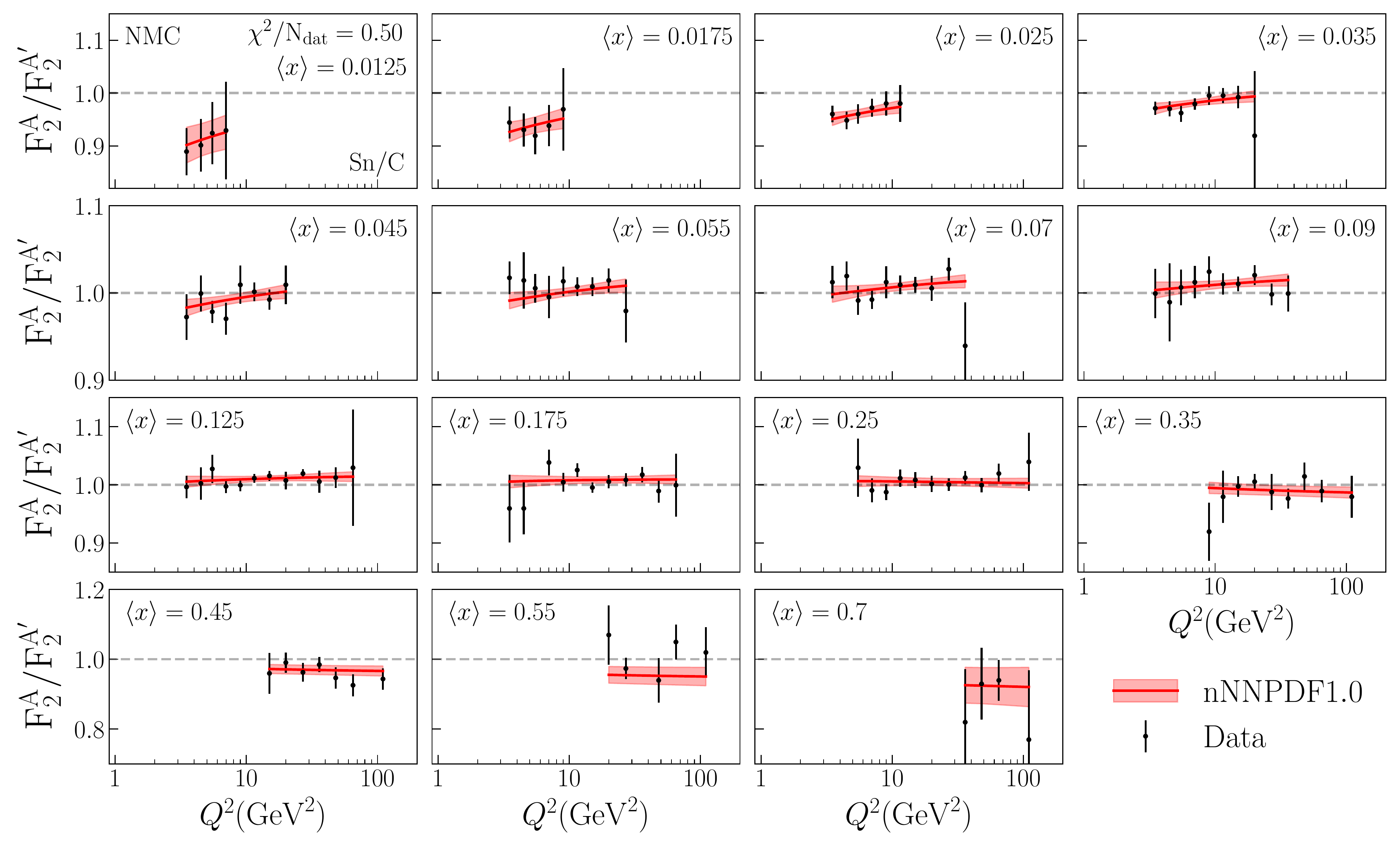}
 \end{center}
\vspace{-0.3cm}
\caption{\small Same as Fig~\ref{figDvT1} but for the 
$Q^2$-dependent structure
function ratio $F_2^{\rm Sn}/F_2^{\rm C}$ provided by the NMC experiment.}
\label{figDvT2}
\end{figure}

\begin{figure}[!h]
\begin{center}
  \includegraphics[width=0.99\textwidth]{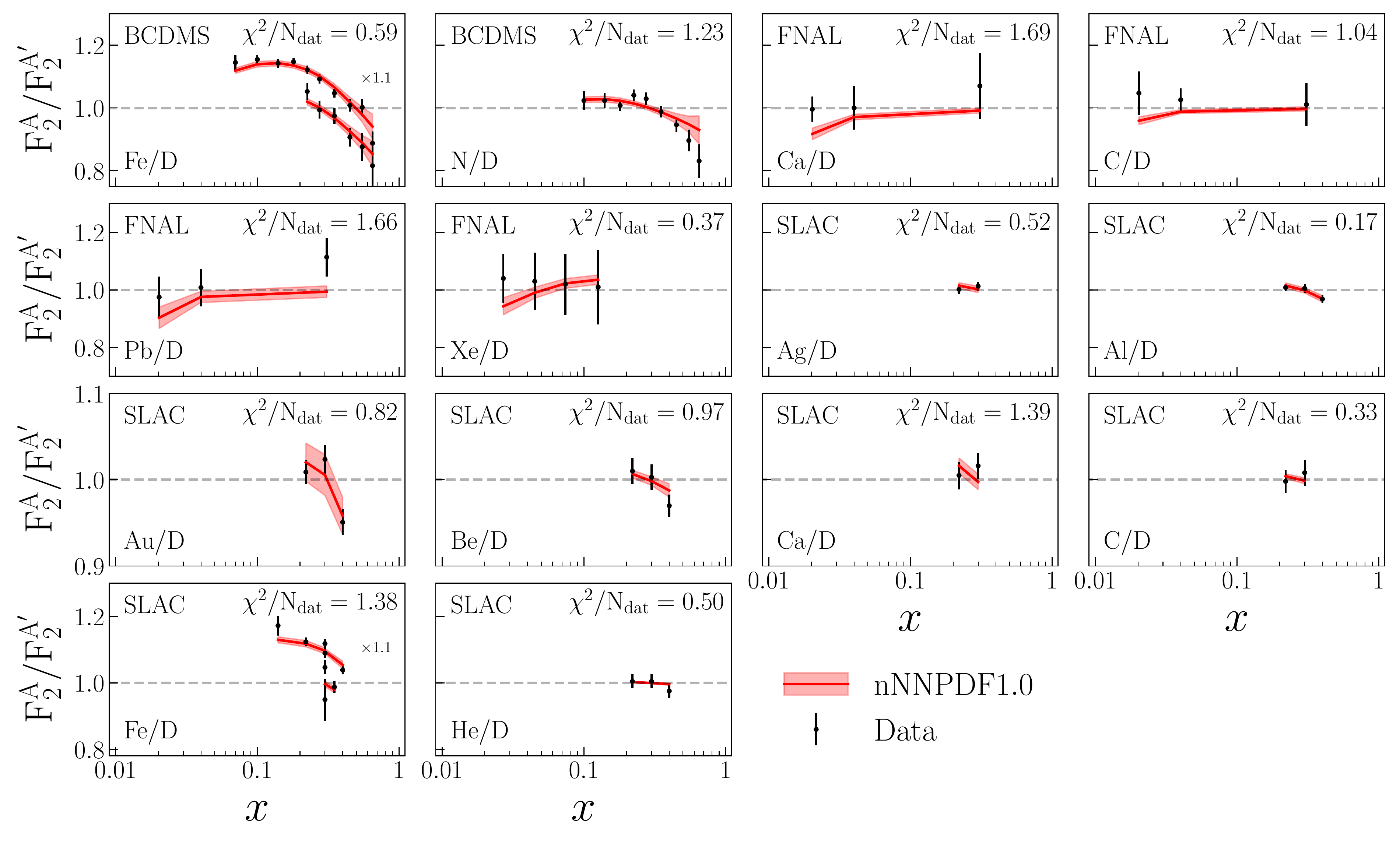}
 \end{center}
\vspace{-0.3cm}
\caption{\small Same as Fig~\ref{figDvT1} but for the data provided by the BCDMS, 
FNAL E665, and SLAC-E139 experiments.}
\label{figDvT3}
\end{figure}

At the level of individual datasets, we find in most cases a good 
agreement between the experimental measurements and
the corresponding theory calculations, with many
$\chi^2/N_{\rm dat} \lsim 1$ both at NLO and at NNLO.
The agreement is slightly worse for the ratios Ca/D and 
Pb/D from FNAL E665, as well as the Sn/D ratio from EMC, all of
which have $\chi^2/N_{\rm dat} \ge 1.5$.
The apparent disagreement of these datasets
can be more clearly understood with the visual
comparison between data and theory.
In Fig.~\ref{figDvT1} we display
the structure
function ratios $F_2^A/F_2^{A'}$ measured by the EMC and 
NMC experiments 
and the corresponding theoretical predictions
from the nNNPDF1.0 NLO fit.
Furthermore, in Figs.~\ref{figDvT2} and~\ref{figDvT3} we show
the corresponding comparisons for the
$Q^2$-dependent structure
function ratio $F_2^{\rm Sn}/F_2^{\rm C}$ provided by the NMC experiment,
and the data provided by the BCDMS, 
FNAL E665, and SLAC-E139 experiments, respectively.

In the comparisons shown in Figs.~\ref{figDvT1}--\ref{figDvT3},
the central values of the experimental data points have been shifted
by an amount determined by the 
multiplicative systematic uncertainties and their nuisance parameters, while
uncorrelated uncertainties are added in quadrature to define the total error bar.
We also indicate in each panel the value of $\chi^2/N_{\rm dat}$,
which include the quadratic penalty as a result 
of shifting the data to its corresponding value displayed in the figures.
The quoted $\chi^2$ values
therefore coincide with those of Eq.~(\ref{eq:chi2}) without
the $A=1$ penalty term.
Lastly, the theory predictions are computed at each $x$ and $Q^2$ bin
given by the data,
and its width corresponds to the 1-$\sigma$ deviation of the observable
using the nNNPDF1.0 NLO set with $N_{\rm rep} = 200$ replicas.
Note that in some panels, the theory curves (and the corresponding data points)
are shifted by an arbitrary factor to improve visibility.
  
As expected by the $\chi^2$ values listed in Table~\ref{chi2table}, 
the experimental measurements agree well with
the structure function ratios computed using the nNNPDF1.0 sets, apart
from the three observables mentioned previously.
For the FNAL data, the disagreement comes from
datasets that contain a total of 3 data points with larger
uncertainties than other experimental measurements,
and therefore do not significantly impact
the fit results. 

A similar argument can be made for the Sn/D ratio
from the EMC experiment, which has 
$\chi^2/N_{\rm dat} = 2.22$.
Here the lack of agreement between theory and data 
can be traced 
to the
low-$x$ region of the structure function ratio.
Such a deviation can also be seen in the recent nCTEQ 
and EPPS analyses, and can be attributed to a possible tension with the 
$Q^2$ dependent ratio Sn/C presented in Fig.~\ref{figDvT2}. 
While the 
comparison here is with carbon and not deuterium, the nuclei are relatively 
close in mass number and therefore the effects in the ratio are expected 
to be similar.
On the other hand, the data show a roughly $\sim15-20\%$ 
difference between EMC's Sn/D and NMC's Sn/C at $x\sim0.03$. Since 
the NMC data have smaller uncertainties than EMC, its influence on the 
fit is much stronger, driving the disagreement with EMC Sn/D at low $x$.
Overall, the agreement with NMC data is excellent, including the $Q^2$ 
dependent Sn/C data presented in Fig.~\ref{figDvT2}.

From the data versus theory comparisons, the various
nuclear effects encoded in the structure
function ratios can clearly be observed.
At small $x$ the structure functions exhibit shadowing, namely the 
depletion of $F_2(x,Q,A)$
compared to its free-nucleon counterpart (or compared
to lighter nuclei).
At larger $x$ the well known EMC effect is visible, resulting
in ratios below unity.
Between these two regimes, one finds an enhancement of the
nuclear structure functions.
However, we do not observe the Fermi motion effect, 
which gives $R_{F_2} > 1$ for large $x$ 
and increases rapidly in the $x\to1$ limit.
This is due simply to the kinematic $W^2$ cut illustrated
in Fig.~\ref{figkinplot}, which
removes much of the large-$x$ data.
Note that although the three nuclear regimes are visible at the
structure function level, such effects may not be reflected at 
the level of PDF ratios, as we will highlight in the following section.

\subsection{The nNNPDF1.0 sets of nuclear PDFs}
\label{sec:pdfcomparisons}

With the agreement between data and theory established, we present
now the results for the NLO nPDF sets.
Later,
we will assess the perturbative stability of the results by comparing to
the corresponding NNLO fit.
Unless otherwise indicated, the results presented in this section
are generated with $N_{\rm rep} = 1000$ Monte Carlo
replicas.

\begin{figure}[t]
\begin{center}
  \includegraphics[width=0.90\textwidth]{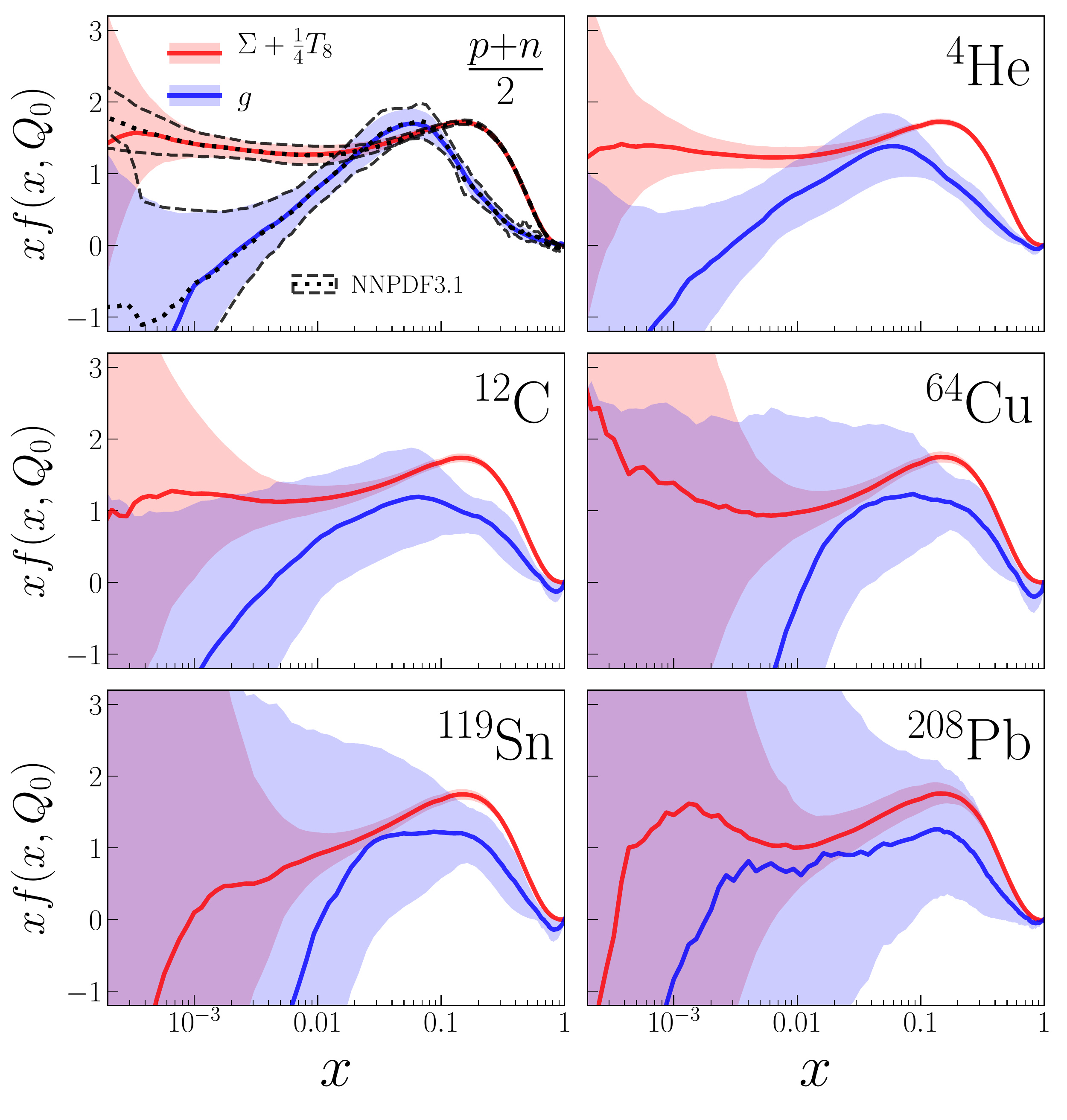}
 \end{center}
\vspace{-0.5cm}
\caption{\small The nNNPDF1.0 NLO set as a function of $x$
  at the input scale $Q_0=1$ GeV for different values of $A$.
  We show the central value for the gluon $g$ (solid blue line) 
  and the quark combination $\Sigma+T_8/4$ (solid red line)
  for $A=1$ (isoscalar nucleon), $A=4$ (He),
  $A=12$ (C), $A=64$ (C), $A=119$ (Sn), and $A=208$ Pb.
  The corresponding uncertainties (shaded bands) correspond 
  to the 90\% confidence level intervals.
  In the case of $A=1$ we also show the central value of the baseline free-nucleon
  PDF set, NNPDF3.1 (black dotted line), and its uncertainties (black dashed lines).
}
\label{figPDFsQ0}
\end{figure}

To begin, we show in Fig.~\ref{figPDFsQ0} the nNNPDF1.0 NLO set as a function of $x$
at the input scale $Q_0=1$ GeV for different values of $A$.
In this figure,
the nPDF uncertainty bands are computed as the 90\% confidence level
intervals, with the central value being taken as the midpoint of the corresponding
range.
The confidence levels presented here follow that of previous NNPDF studies~\cite{Ball:2010de}
and are computed in the following way.
For a given $x$, $Q$, and $A$, we have $N_{\rm rep}$ values of
a particular nPDF flavor $f^{(k)}(x,Q,A)$.
The replicas are then ordered such that
\be
    f_1 \le f_2 \le \ldots \le
    f_{N_{\rm rep}-1} \le f_{N_{\rm rep}} \,. 
\ee
Finally, we remove symmetrically $(100-X)\%$ of the replicas
with the highest and lowest values.
 The resulting interval defines the $X\%$ confidence level
 for the nPDF $f(x,Q,A)$ for a given value of $x$, $Q$, and $A$.
 In other words, a 90\% CL interval (corresponding to a 2-$\sigma$
 interval for a Gaussian distribution) is obtained by
 keeping the central 90\% replicas, leading to
 \be
 \label{eq:res90CL}
 \lc f_{0.05\,N_{\rm rep}}, f_{0.95\, N_{\rm rep}}  \rc \, .
 \ee

 The rationale for estimating the nPDF
 uncertainties as 90\% CL intervals, as opposed to the standard deviation,
 is that it turns out that the nNNPDF1.0 probability
 distribution is not well described by a Gaussian, in particular when ratios
 between different nuclei $A$ are taken.
 Therefore, the variance $\sigma^2$ may not be the best estimator for the
 level of fluctuations in the distribution.
 While deviations from the Gaussian approximation in the proton case are
 moderate, there are several reasons why the nPDFs may behave differently.
 First of all, there is a limited amount of experimental information, especially
 for the gluon.
 Secondly, imposing the $A=1$ boundary condition skews the $A$ dependence
 of the distribution.
 Lastly, even if the resulting nPDFs do follow a Gaussian distribution, in general 
 their ratio between different values of $A$ will not.
 Therefore, in Fig.~\ref{figPDFsQ0}, and in the remaining figures of this analysis,
 the uncertainties will be presented as the 90\% CL defined above.

We also show in Fig.~\ref{figPDFsQ0} the results
of the baseline free-nucleon PDF set, NNPDF3.1, compared
to the nuclear parton distributions evaluated at $A=1$.
As can be observed, there is an excellent match between both the central
values and the PDF uncertainties of nNNPDF1.0 and those
of NNPDF3.1 in the region of $x$ where the boundary condition is imposed,
$10^{-3} \le x \le 0.7$.
This agreement demonstrates that the quadratic penalty
in Eq.~(\ref{eq:chi2}) is sufficient to achieve
its intended goals.
In Sect.~\ref{sec:methstudies} we will discuss the importance
of implementing such a constraint, particularly
for light nuclei.

From Fig.~\ref{figPDFsQ0}, we can also see that the PDF uncertainties
increase as we move towards larger values of $A$, in particular for the
gluon nPDF.
Recall that the latter is only constrained indirectly from inclusive DIS data via
DGLAP evolution effects.
On the other hand, the quark combination $\Sigma + T_8/4$ turns
out to be reasonably well
constrained for $x\gsim 10^{-2}$, since this is the combination directly related
to the nuclear structure function $F_2(x,Q^2,A)$.
For both the gluon and the quark nuclear distributions, the PDF uncertainties
diverge in the small-$x$ extrapolation region, the beginning of which
varies with $A$.
For example, the extrapolation region for the quarks in Sn ($A=119$)
is found to be $x\lsim 5 \times 10^{-3}$, while for the gluon
PDF uncertainties become very large already for $x\lsim 5 \times 10^{-2}$.

\begin{figure}[t]
  \begin{center}
    \includegraphics[width=0.90\textwidth]{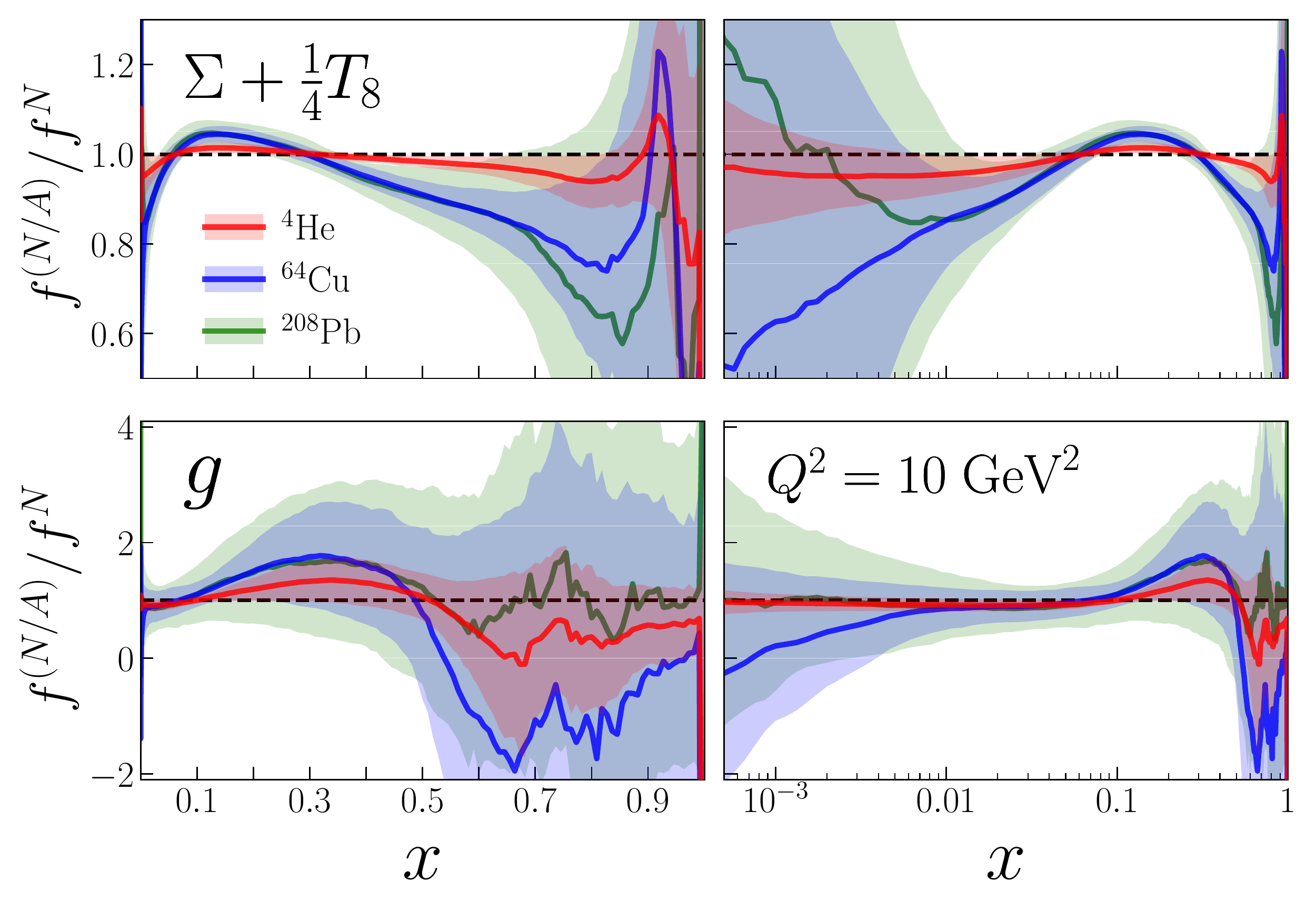}
   \end{center}
  \vspace{-0.3cm}
  \caption{\small Ratios of the nNNDPDF1.0 NLO
    distributions normalized to the $A=1$ result.
    The central values (solid lines) and uncertainties (shaded bands) for
    the quark combination $\Sigma + \frac{1}{4}T_8$ (top panels) and
    gluon (bottom pannels) are shown at $Q^2 = 10$ GeV$^2$ 
    for $^4$He (red), $^{64}$Cu (blue),
    and $^{208}$Pb (green) nuclei.
  }
  \label{figPDFsRatio}
  \end{figure}

Next, we illustrate in Fig.~\ref{figPDFsRatio} the nNNPDF1.0
PDFs normalized by the $A=1$ distributions.
Here the results for He ($A=4$), Cu ($A=64$),
and Pb ($A=208$) nuclei are shown for $Q^2 = 10$ GeV$^2$.
With this comparison, we can assess whether the different nuclear
effects introduced previously are visible at the nPDF level,
since Eq.~(\ref{eq:param}) is analagous to the structure function ratios
displayed in Figs.~\ref{figDvT1}--\ref{figDvT3}.

When evaluating ratios of nPDFs between different values of $A$,
it is important to account for the correlations between the numerator
and denominator.
These correlations stem from the fact that nPDFs at two values of $A$ are
related by the common underlying parameterization, Eq.~(\ref{eq:param}),
and therefore are not independent.
This can be achieved by computing the ratio $R_f$ for each of the $N_{\rm rep}$
Monte Carlo replicas of the fit
\be
\label{eq:nPDFratiosErr}
R_f^{(k)} =
\frac{f^{(N/A)(k)}(x,Q^2,A)}{f^{(N)(k)}(x,Q^2)} \,
\ee
and then evaluating the 90\% CL interval following the procedure that leads
to Eq.~(\ref{eq:res90CL}).
Note that a rather different result from that
of Eq.~(\ref{eq:nPDFratiosErr}) would be obtained if either the correlations
between different $A$ values were ignored (and thus the PDF uncertainties in numerator
and denominator of Eq.~(\ref{eq:nPDFratios}) are added in quadrature) or if the uncertainties
associated to the $A=1$ denominator were not considered.
Also, as discussed above, the 90\% CL interval for Eq.~(\ref{eq:nPDFratiosErr})
will in general be quite different compared to the range defined 
by the 2-$\sigma$ deviation.

From Fig.~\ref{figPDFsRatio}, we can see that for the relevant quark combination
$\Sigma+T_8/4$ in $A=64$ and $A=208$ nuclei, it possible to identify the same three
types of nuclear effects that were present at the structure function level.
In particular, the anti-shadowing and EMC effects are most evident, where 
the deviation from unity is outside the 90\% CL range. 
Moreover, shadowing behavior appears briefly in the region $x\simeq0.01$, 
particularly for copper nuclei,
before the uncertainties grow quickly in the extrapolation region.
On the other hand, the nuclear effects appear to be negligible
for all $x$ in helium nuclei within the present uncertainties. 

The situation is much worse for the nuclear gluons,
where the ratio $R_f=f^{(N/A)}/f^N$ is consistent with one
within the uncertainties for all values of $x$.
This indicates that
using only neutral-current DIS nuclear structure functions, there is limited
information that one can extract
about the nuclear modifications of the gluon PDF.
Here we find no evidence for gluon shadowing,
and the ratio $R_f$ is consistent with one for $x\lsim 0.1$.
The only glimpse of a non-trivial nuclear modification
of the gluon nPDF is found for Cu ($A=64$), where between
$x\simeq 0.1$ and $x\simeq 0.3$ there appears to be an enhancement
reminiscent of the anti-shadowing effect.

The comparisons of Fig.~\ref{figPDFsRatio}
demonstrate that, without additional experimental input, we are rather 
far from being able to probe in detail the nuclear modifications
of the quark and gluon PDFs, particularly for the latter case.
We will highlight in Sect.~\ref{sec:eic} how the present situation would
be dramatically improved with an Electron Ion Collider,
allowing us to pin down nuclear PDFs
in a wider kinematic range and with much better precision.

\paragraph{The scale dependence of the nuclear modifications.}

\begin{figure}[t]
  \begin{center}
    \includegraphics[width=0.90\textwidth]{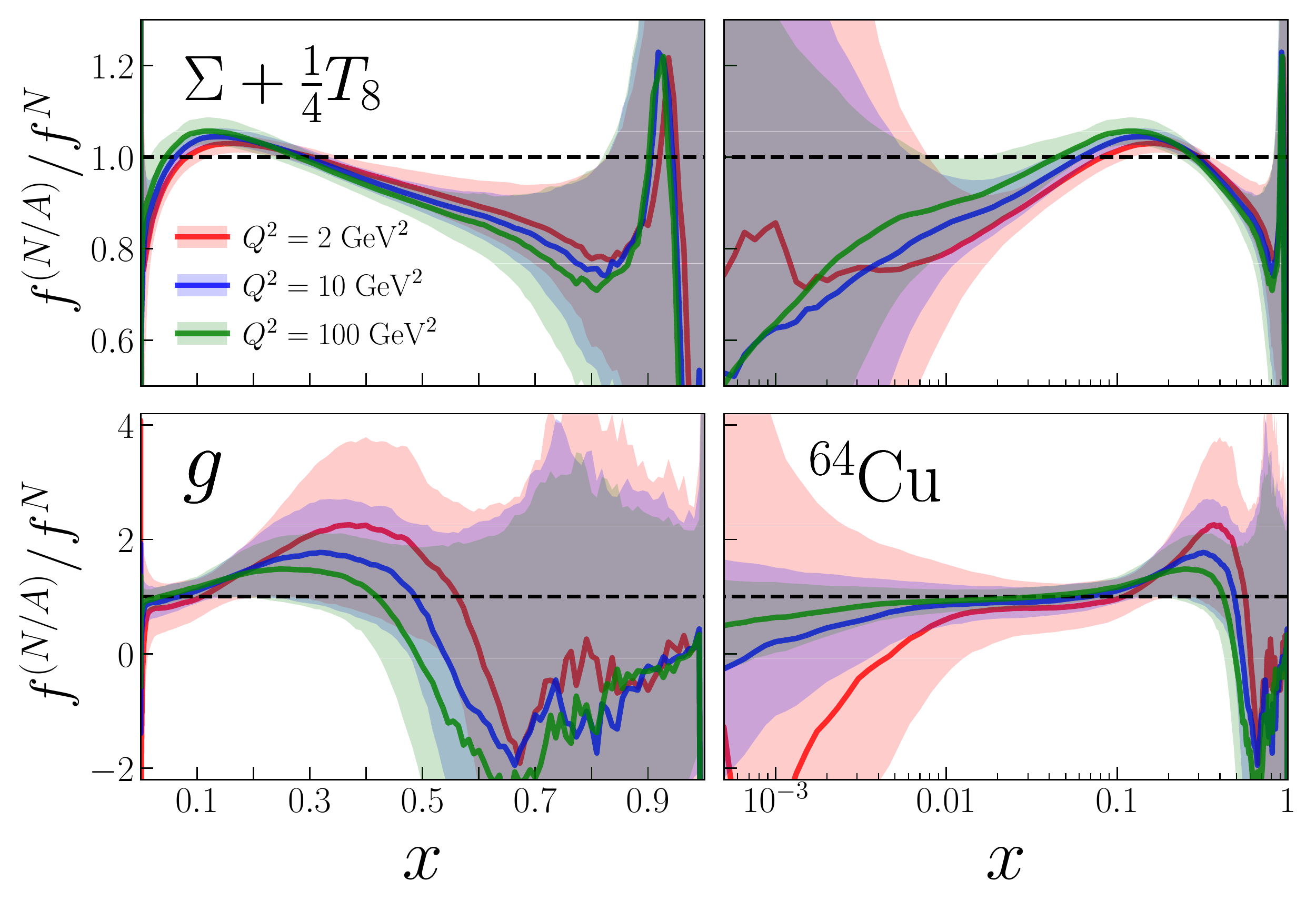}
   \end{center}
  \vspace{-0.3cm}
  \caption{\small Same as Fig.~\ref{figPDFsRatio}, but now
    for the dependence of the nuclear modifications of $^{64}$Cu
    on the momentum transfer $Q^2$.
    The ratios are given for $Q^2=1$ GeV$^2$ (red), 
    10 GeV$^2$ (blue), and 100 GeV$^2$ (green).
  }
  \label{figPDFsQ2}
  \end{figure}

In Fig.~\ref{figPDFsQ2}, we show a similar
comparison as that of Fig.~\ref{figPDFsRatio}, but now
for the $Q^2$ dependence of the nuclear modifications in $^{64}$Cu.
More specifically,
we compare the results of nNNPDF1.0,
normalized as before
to the $A=1$ distribution, for $Q^2=2$ GeV$^2$, 10 GeV$^2$, and 100 GeV$^2$.
We can observe in this case how nPDF uncertainties are reduced
when the value of $Q^2$ is increased.
This effect is particularly dramatic for the gluon
in the small-$x$ region, but is also visible for the quark
distributions.
This feature is a direct consequence of the structure of DGLAP
evolution, where at small $x$ and large $Q^2$ the results
tend to become insensitive of the specific boundary condition
at low scales as a result of double asymptotic scaling~\cite{das}.

It is important to point out that, by the same token, the sensitivity to nuclear
modifications is also reduced when going from low to high $Q^2$ in
the small-$x$ region.
Indeed, we can see from Fig.~\ref{figPDFsQ2} that the ratios $R_f$ move closer to one at
small $x$ as $Q$ is increased.
However, this is not the case for medium and large $x$, where DGLAP
evolution effects are milder.
Therefore, nuclear effects in this region can be accessible
using probes both at low and high momentum transfers.
The comparisons in  Fig.~\ref{figPDFsQ2} highlight that
the best sensitivity for nuclear modifications present
in the small-$x$ region arises from low-scale observables,
while for medium and large-$x$ modifications there is also
good sensitivity at high scales.

\begin{figure}[t]
\begin{center}
  \includegraphics[width=0.90\textwidth]{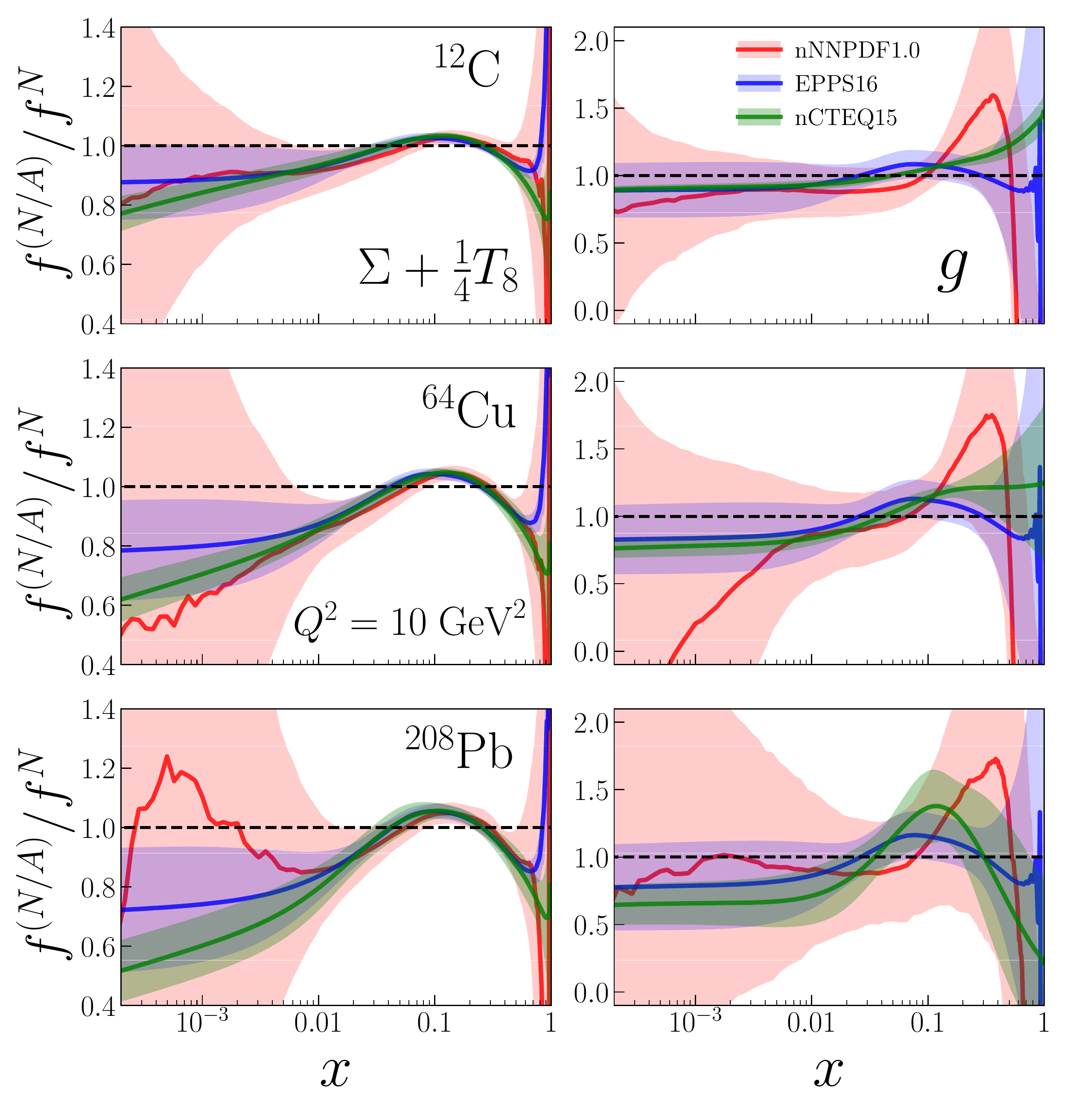}
 \end{center}
\vspace{-0.3cm}
\caption{\small Comparison between the nNNPDF1.0, EPPS16 and nCTEQ15 fits at NLO
  for $Q^2 = 10$ GeV$^2$.
  The quark combination $\Sigma + \frac{1}{4}T_8$ (left panels) and gluon (right panels)
  are normalized to the central value of each group's proton PDF baseline, and
  are shown for $^{12}$C (top panels), $^{64}$Cu (middle panels), and 
  $^{208}$Pb (bottom panels) nuclei.
  The uncertainties (shaded bands) correspond to the 90\% CL ranges computed with
  the corresponding prescription for each fit.
}
\label{figPDFsComp}
\end{figure}

\paragraph{Comparison with EPPS16 and nCTEQ15.}
We now turn to compare the nNNPDF1.0 nuclear PDFs with other recent
analyses.
Here we restrict our comparison to the EPPS16 and nCTEQ15 fits, given that they
are the only recent nPDF sets available in {\tt LHAPDF}.
In Fig.~\ref{figPDFsComp}, we display the nNNPDF1.0 NLO distributions
together with EPPS16 and nCTEQ15 at $Q^2 = 10$ GeV$^2$ for three different nuclei: 
$^{12}$C, $^{64}$Cu, and $^{208}$Pb.
The three nPDF sets have all been normalized to the central value
of their respective proton PDF baseline to facilitate the comparison.
For the nNNPDF1.0 results, the uncertainties are computed 
as before but without including the correlations with the $A=1$ distribution.
Lastly, the PDF uncertainties for EPPS16 and nCTEQ15 correspond to the
90\% CL ranges computed using the standard Hessian prescription.
%

From this comparison, there are a number of interesting similarities and
differences between the three nPDF fits.
First of all, the
three nuclear regimes sketched in Fig.~\ref{fig:cartoon}, namely shadowing, anti-shadowing,
and the EMC effect, are visible between the three sets for the quark combination $\Sigma+T_8/4$.
Interestingly, in the data region the PDF uncertainties for this quark
combination are similar between the different analyses.
Much larger differences are found in the small-$x$ and large-$x$ extrapolation
regions, particularly for nCTEQ15, where the uncertainties are smaller.
Note that the different approaches for uncertainty estimation 
have noticeable physical consequences.
For instance,
it would appear that there is rather strong
evidence for quark shadowing down to $x\simeq 10^{-4}$ for the nCTEQ15 result, 
while for nNNPDF1.0,
the nuclear modifications are consistent with zero within uncertainties for
$x\lsim 10^{-2}$.

Concerning the nuclear modifications of the gluon PDF,
here we can percieve large differences at the level of PDF errors,
with nCTEQ15 exhibiting the smallest uncertainties and nNNPDF1.0 the largest.
While nCTEQ15 indicates some evidence of small-$x$ gluon shadowing, this evidence
is absent from both nNNPDF1.0 and EPPS16.
Moreover, the three sets find some preference for a mild enhancement of the gluon at
large $x$, but the PDF uncertainties prevent making any definite
statement.
Overall, the various analyses agree well within the large
uncertainties for $x\gsim 0.3$.

While it is beyond the scope of this paper to pin down the origin
of the differences between the three nPDF analyses, one known 
reason is the choice of nPDF parameterization
together with the method of imposing the $A\to 1$ boundary condition.
Recall that in nNNPDF1.0 we adopt a model-independent parameterization
based on neural networks, Eq.~(\ref{eq:param}), with the boundary
condition imposed at the optimization level in Eq.~(\ref{eq:chi2}).
In the EPPS16 analysis, the bound nucleus PDFs are instead defined 
relative to a free nucleon baseline
(CT14) as
\be
f_i^{(N/A)}(x,Q^2,A) = R_i^A(x,Q^2)\,f_i^{(N)}(x,Q^2) \, ,
\ee
where the nuclear modification factors are parameterized at the input evolution scale
$R_i^A(x,Q_0^2)$ with piece-wise polynomials that hard-wire some of the
theoretical expectations shown in Fig.~\ref{fig:cartoon}.
In this approach, the information contained in PDF uncertainties of the free nucleon baseline is not
exploited to constrain the nPDFs.

In the nCTEQ15 analysis, the nuclear PDFs are parameterized
by a polynomial functional form given by
\be
f_i^{p/A}(x,Q^2,A) = c_0\,x^{c_1}\,(1-x)^{c_2}\,e^{c_3\,x}(1+e^{c_4}x)^{c_5} \, ,
\ee
where the coefficients $c_k(A)$ encode all the $A$ dependence.
During the fit, these coefficients are constrained in a way that
for $A=1$ they reproduce the central value of the
the CTEQ6.1-like fit of Ref.~\cite{Owens:2007kp}.
Note here that in the nCTEQ15 fit the baseline proton set
does not include the experimental measurements that have become
available in the last decade, in particular
 the information provided by the high-precision
LHC data and the HERA combined structure functions.
Moreover, as in the case of EPPS16, the information about the PDF uncertainties
in the free-nucleon case is not exploited to constrain the nPDF errors.

While these methodological choices are likely to explain the bulk of the differences
between the three analyses, a more detailed assessment could only be obtained
following a careful benchmarking exercise along the lines of those
performed for the proton PDFs~\cite{Butterworth:2015oua,Ball:2012wy,Botje:2011sn,Alekhin:2011sk}.

\begin{figure}[t]
  \begin{center}
    \includegraphics[width=0.90\textwidth]{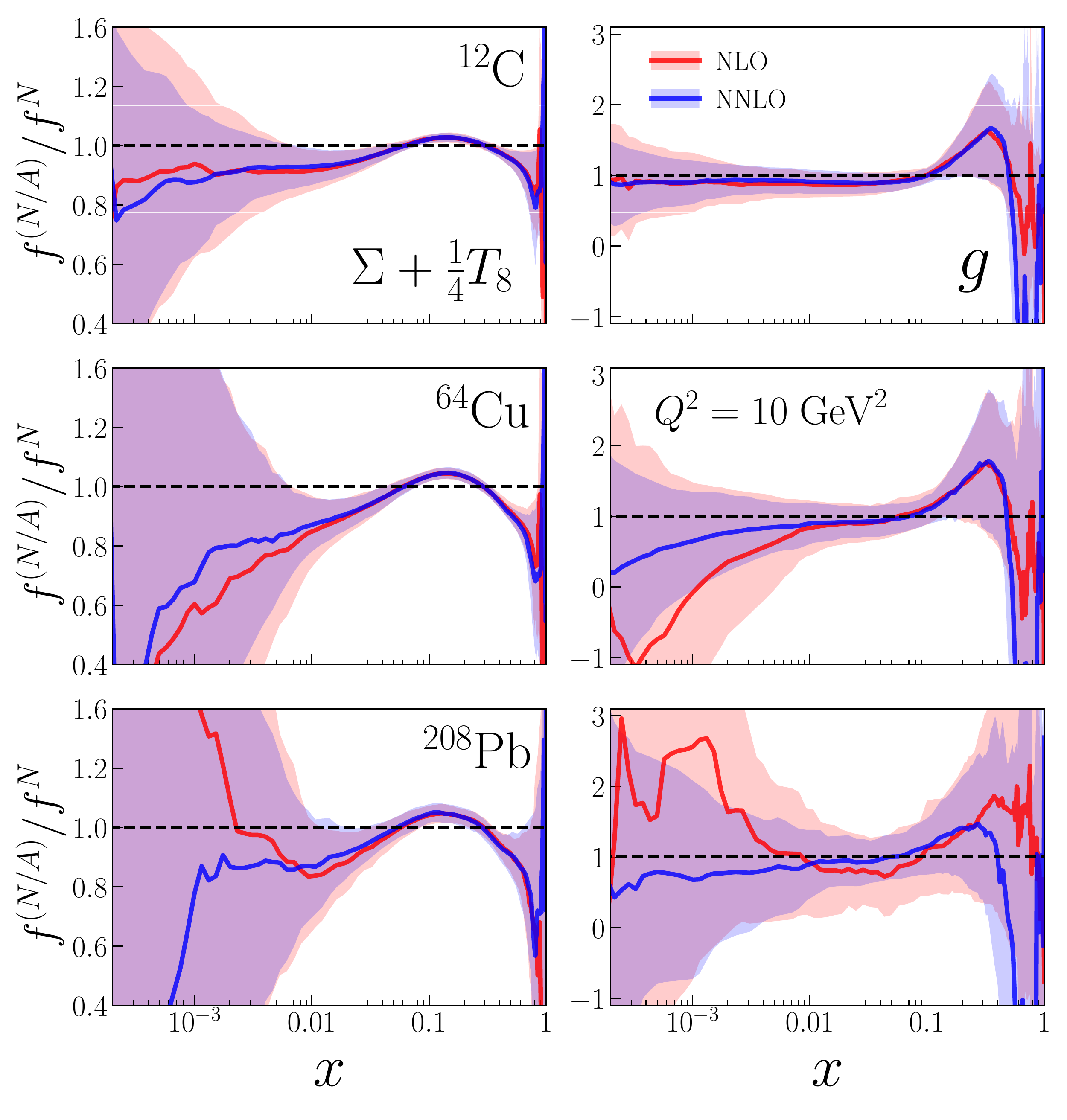}
   \end{center}
  \vspace{-0.6cm}
  \caption{\small
    Same as Fig.~\ref{figPDFsRatio}, but now comparing the results
    of the nNNPDF1.0 fits between NLO and NNLO.
  }
  \label{figPDFsCompPTO}
  \end{figure}

\paragraph{Perturbative stability.}
To conclude the discussion of the main properties of the nNNPDF1.0
fits, in Fig.~\ref{figPDFsCompPTO} we compare 
the NLO and NNLO nuclear ratios $R_f$
for the same three nuclei as in Fig.~\ref{figPDFsComp}.
The ratios are constructed using the $A=1$ distributions
from their respective perturbative order PDF set using $N_{\rm rep}=200$
replicas.
In terms of central values, we can see that the NLO and
NNLO fit results are consistent within the 90\% CL uncertainty band.
The regions where the differences between the two perturbative orders are the largest
turn out to be the small- and large-$x$ extrapolation regions, in particular as $A$ is increased.

Another difference between the NLO and NNLO nNNPDF1.0 fits concerns the size of the
PDF uncertainty band.
We find that for the gluon nPDF, the NNLO fit leads to a slight decrease
in uncertainties, perhaps due to the improved overall fit consistency when higher-order
theoretical calculations are used.
This effect is more distinct for the gluon distribution of $A=64$ and $A=208$ nuclei,
while it is mostly absent for $A=12$.
The apparent reduction of uncertainties, together with marginally better $\chi^2$
values (see Table~\ref{chi2table}), suggests that the NNLO fit is only slightly 
preferred over the NLO one.
That said, the difference is unlikely
to have significant phenomenological implications given the current level of 
uncertainties.
%

\subsection{Methodological studies}
\label{sec:methstudies}

We conclude the discussion of the
nNNPDF1.0 results by presenting some further studies that demonstrate
the robustness of our fitting methodology, complementing those based
on the closure tests discussed in Sect.~\ref{sec:closuretests}.
In particular, in the following we discuss the stability of our results with respect to variations
of the neural network architecture and the role
of the $A=1$ boundary condition in constraining the nPDF
uncertainties.
For all results presented in this section, we use $N_{\rm rep} = 200$ Monte
Carlo replicas.

\paragraph{Stability with respect to the network architecture.}
As explained in Sect.~\ref{sec:nPDFparam},
the nNNPDF1.0 fits are based on a single neural network
with the 3-25-3 architecture represented in Fig.~\ref{fig:architecture}.
This architecture is characterized by $N_{\rm par}=178$ free parameters, without
counting the preprocessing exponents.
We have verified that this choice of network architecture is redundant given our input
dataset, namely that the nNNPDF1.0 results are stable if neurons are either
added or removed from the hidden layer of the network.
To illustrate this redundancy, here we compare fit results
using the standard 3-25-3 architecture with that
using twice as many neurons
in the hidden layer, 3-50-3.
The latter configuration is characterized by $N_{\rm par}=353$ free parameters,
which is enlarged by a factor two compared to the baseline fits.

\begin{figure}[t]
  \begin{center}
    \includegraphics[width=0.90\textwidth]{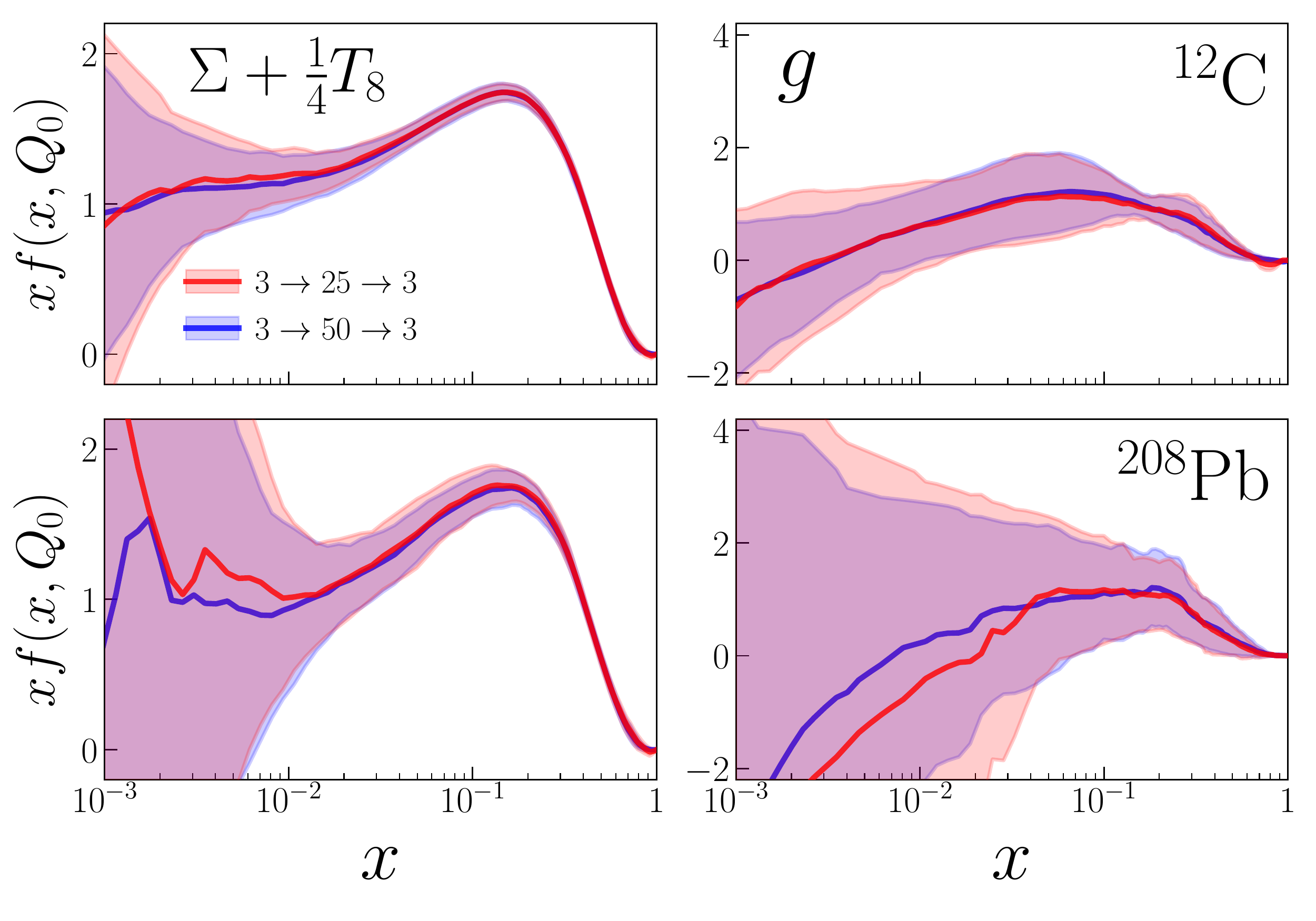}
   \end{center}
  \vspace{-0.6cm}
  \caption{\small Dependence of the nNNPDF1.0 results
    at the input scale $Q_0=1$ GeV with respect
    to the choice of neural network architecture.
    We compare the baseline results 
    obtained with a 3-25-3 architecture (solid red line and shaded band),
    with the corresponding ones using a 3-50-3 architecture (solid blue line
    and shaded band) for $^{12}$C (top panels) and $^{208}$Pb (bottom panels)
    nuclei.
  }
  \label{fig:architectureVariation}
  \end{figure}

In Fig.~\ref{fig:architectureVariation}
the nNNPDF1.0 results at the input scale $Q_0=1$ GeV for $^{12}$C
and $^{208}$Pb nuclei are shown with the two different architectures,
3-25-3 (baseline) and 3-50-3.
We find that differences are very small and consistent with statistical fluctuations.
Given that now there are twice as many free parameters as in the baseline settings,
this stability demonstrates that our results are driven by the input experimental
data and not by methodological choices such as the specific network
architecture.
Furthermore, 
we have also verified that the outcome of the fits is similarly unchanged if a network
architecture with a comparable number of parameters but two hidden layers is used. 
  
\paragraph{The role of the $A=1$ boundary condition.}
Imposing the $A=1$ boundary condition Eq.~(\ref{eq:constraintprotonPDFs}) leads to
important constraints on both the central values and the uncertainties
of nNNPDF1.0 fit, particularly for low values of the atomic mass number $A$.
Here we want to quantify this impact by comparing the baseline nNNPDF1.0 results
with those of the corresponding fit where this boundary condition
has not been imposed.
This can be achieved by performing the fits with the
hyper-parameter $\lambda=0$ in Eq.~(\ref{eq:chi2}).
Note that in this case the behavior of the fitted at nPDFs for $A=1$ is unconstrained,
since only experimental data with $A \ge 2$ is included in the fit.

\begin{figure}[t]
  \begin{center}
    \includegraphics[width=0.93\textwidth]{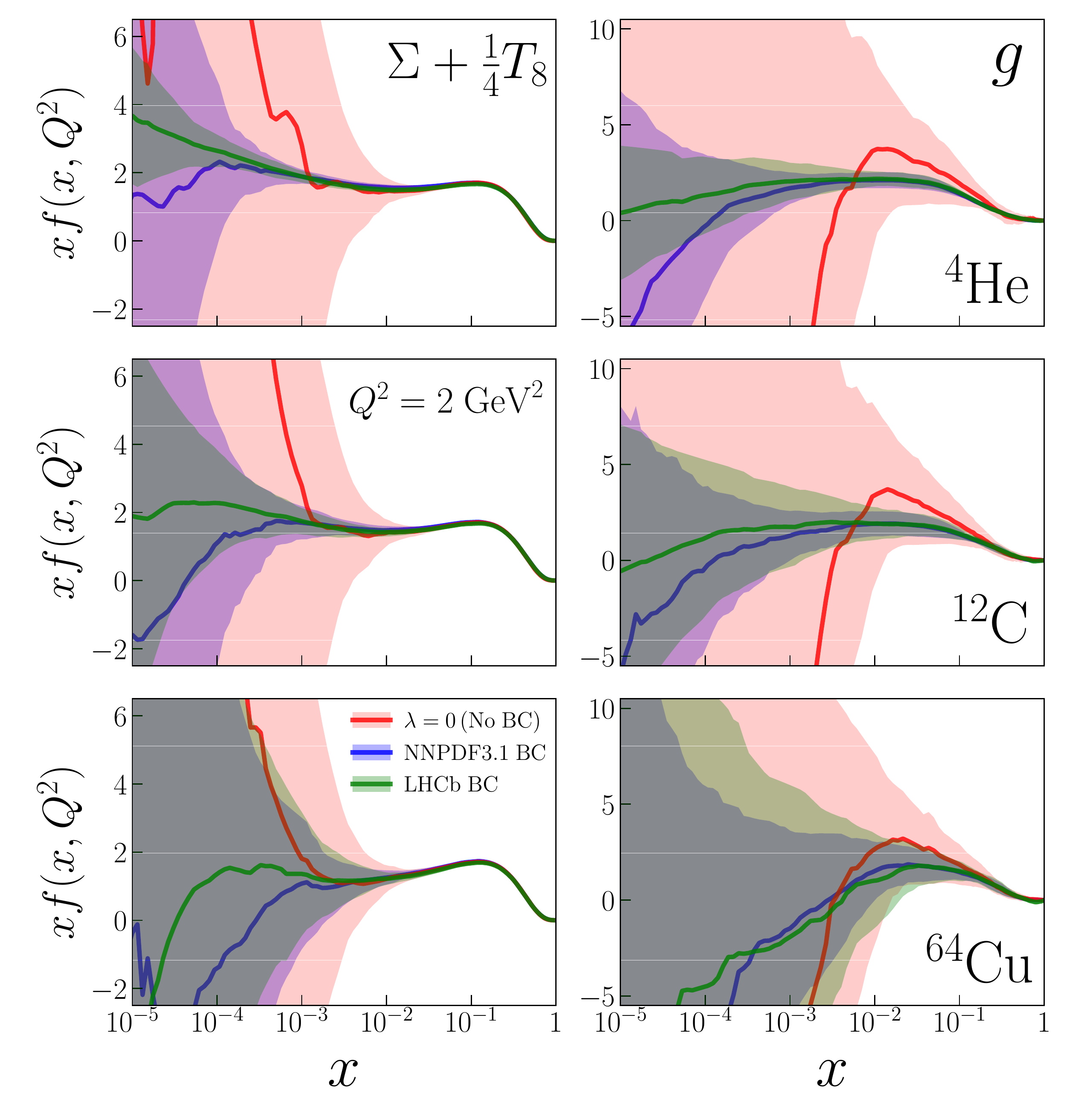}
   \end{center}
  \vspace{-0.7cm}
  \caption{\small Comparison of the nNNPDF1.0 fits for different choices of
    the $A=1$ boundary condition (BC).
    The baseline result, which imposes NNPDF3.1 as boundary condition
    between $x=10^{-3}$ and $x=0.7$ (blue), 
    are shown together with two fit variants, one 
     produced using the NNPDF3.0+LHCb set
     as boundary condition down to $x=10^{-5}$ (green), 
     and another without the boundary condition by setting $\lambda=0$ in Eq.~(\ref{eq:chi2})
     (red).
    The central values (solid lines) and uncertainties (shaded bands) are given for the
    quark combination $\Sigma + \frac{1}{4}T_8$ (left panels) and gluon (right panels) at
     $Q^2=2$ GeV$^2$ for $A=4$ (top panels), $A=12$ (middle panels) and $A=64$ (bottom panels).
  }
  \label{fig:nPDFcomp_BCA_LHCb}
  \end{figure}

In Fig.~\ref{fig:nPDFcomp_BCA_LHCb},
we show a comparison between the nNNPDF1.0 baseline, which imposes NNPDF3.1 as
the $A=1$ boundary condition
between $x=10^{-3}$ and $x=0.7$, in addition to a resulting
fit where this boundary condition is not implemented.
Moreover, we display the gluon and the $\Sigma+T_8/4$ quark
combination at $Q^2=2$ GeV$^2$ for $A=4, 12,$ and 64.
This comparison demonstrates a significant impact on nNNPDF1.0
resulting from the $A=1$ constraint, especially for helium and carbon 
nuclei where the PDF 
uncertainties are increased dramatically if no boundary condition is used.
The impact is more distinct for the gluon, where even for relatively
heavy nuclei such as $^{64}$Cu the boundary condition leads
to a reduction of the nPDF uncertainties by up to a factor two.
We can thus conclude that imposing consistently the $A=1$ limit 
using a state-of-the-art proton fit is an essential ingredient
of any nPDF analysis.

While the baseline nNNPDF1.0 fits only constrain the $A=1$ distribution between 
$x=10^{-3}$ and $x=0.7$, one in principle could extend
the coverage of the boundary condition
down to smaller values of $x$ provided a reliable proton PDF baseline is used.
Indeed, it is possible to demonstrate that we can impose
the constraint down to much smaller values of $x$,
e.g. $x=10^{-5}$.
For this purpose, we perform a fit using instead for the boundary condition
the NNPDF3.0+LHCb NLO sets constructed in Ref.~\cite{Gauld:2015yia,Gauld:2016kpd}.
More specifically we use the set based on the $N_5$, $N_7$, and $N_{13}$ normalized
distributions of $D$ meson production in the forward region
at 5, 7, and 13 TeV.
The reason is that these sets exhibit reduced quark and gluon PDF uncertainties
down to $x\simeq 10^{-6}$, and therefore are suitable to constrain 
the small-$x$ nPDF uncertainties.

The comparison between the baseline nNNPDF1.0 fit and
its LHCb variant is shown in 
Fig.~\ref{fig:nPDFcomp_BCA_LHCb}.
We now find a further reduction of the nPDF uncertainties at small-$x$, again
more notably for light nuclei.
In this case, the reduction of uncertainties is more distinct for the quarks,
which benefit from the very accurate determination of the proton's quark
sea at small-$x$ in NNPDF3.0+LHCb.
Note that, in turn, the improved nPDF errors at small-$x$ might lead
to increased sensitivity to effects such as shadowing
and evidence for non-linear evolution corrections.

\section{Nuclear PDFs at the Electron-Ion Collider}
\label{sec:eic}

As illustrated by Fig.~\ref{figkinplot}, the kinematic reach in $x$ is rather limited 
 for the available lepton-nucleus deep-inelastic 
 scattering data.
As a consequence, nPDF analyses based on these measurements will 
exhibit large uncertainties
for $x\lsim 0.01$, as was shown in Sect.~\ref{sec:results}.
However, the coverage at small $x$ and 
large $Q^2$ can be improved
with measurements in proton-ion scattering, in particular
from the p+Pb collisions provided by the LHC.
There, small-$x$ gluon shadowing can be studied
with $D$ meson production~\cite{Kusina:2017gkz} and direct
photon production~\cite{Campbell:2018wfu,Peitzmann:2018kie} in the forward region.
Furthermore, quark-flavor separation at $Q\simeq M_W$ can be disentangled using the
rapidity distributions in $W$ and $Z$ production~\cite{Khachatryan:2015pzs}.
On the other hand, access to these extended kinematic regions 
is desirable also with lepton-nucleus scattering, since 
leptons represent significantly cleaner probes in scattering processes,
as was extensively demonstrated by the HERA collider~\cite{Klein:2008di}.

Such a machine would be realized by an
Electron-Ion Collider (EIC)~\cite{Accardi:2012qut,Boer:2011fh},
currently under active discussion in the U.S.
The EIC would collide electrons with protons and nuclei using a range
of different beam energy configurations and nucleon species,
pinning down nuclear modifications of the free-nucleon
PDFs down to $x\simeq 5\times 10^{-4}$.
Such a machine would therefore significantly improve our understanding
of the strong interaction in the nuclear medium,
in a similar way as it would with the spin structure of the nucleons~\cite{Ball:2013tyh,Ball:2018odr,Aschenauer:2017jsk}.
Another option in discussion is the Large Hadron electron
Collider (LHeC)~\cite{AbelleiraFernandez:2012ni},
which would operate
concurrently with the High-Luminosity LHC and would further
extend the low-$x$ coverage of lepton-nucleus reactions 
down to $x\simeq 10^{-6}$.
Both options for future high-energy lepton-nucleus colliders have demonstrated
their potential to constrain the nuclear 
PDFs~\cite{Paukkunen:2017phq,Helenius:2015xda,Aschenauer:2017oxs}.

In this section, we quantify the constraints that future EIC
measurements of
inclusive nuclear structure functions 
would impose on the nNNPDF1.0 nuclear PDFs.
To achieve this, we generate EIC pseudo-data following
Ref.~\cite{Aschenauer:2017oxs}, where it was subsequently
interpreted in the framework of the EPPS16 
analysis of nPDFs.
The projections used here are constructed instead
with the central value of the nNNPDF1.0 NLO set
for different scenarios 
of the lepton and nucleon
beam energies, which are then added to the 
input data of this analysis listed in Table~\ref{dataset} .

The simulated EIC pseudo-data from Ref.~\cite{Aschenauer:2017oxs} is available
for both carbon ($A=12$ and $Z=6$)
and gold ($A=197$ and $Z=79$) nuclei.
We assume that the latter would be corrected for non-isoscalarity effects,
and therefore treat gold as an isoscalar nucleus with $A=197$
(see also Sect.~\ref{sec:expdata}).
The nuclear structure functions for 
carbon and gold nuclei are then
normalized by the deuteron structure functions
following Eq.~(\ref{eq:Sect2Rf2}).

\begin{figure}[t]
\begin{center}
  \includegraphics[width=0.75\textwidth]{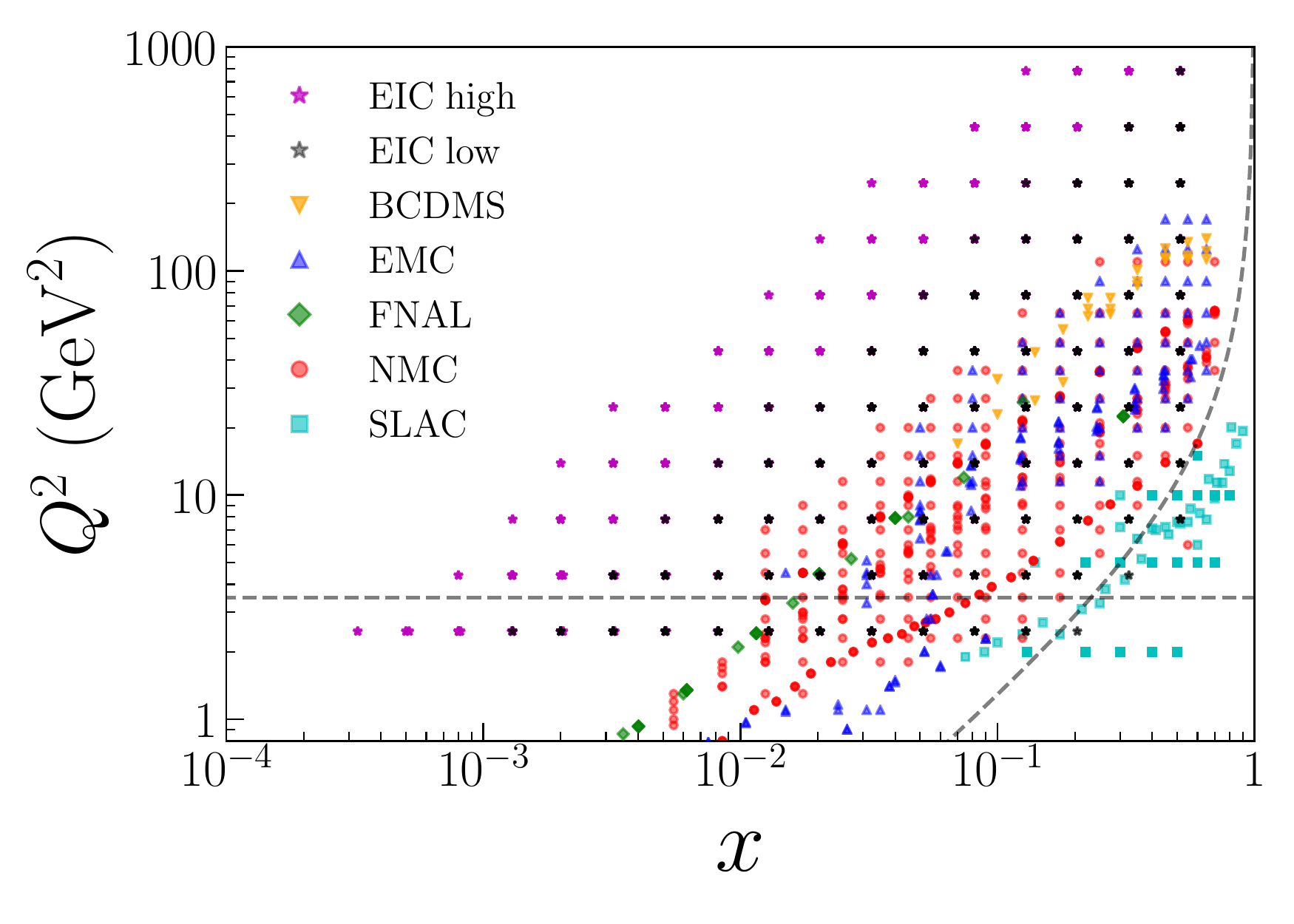}
 \end{center}
\vspace{-0.3cm}
\caption{\small Same as Fig.~\ref{figkinplot}, but now
  including also the kinematic coverage of the EIC
  pseudo-data used in this study and listed in
  Table~\ref{tab:eicpseudodata}. We indicate the
  coverage of the ``high-energy'' and ``low-energy'' EIC scenarios,
  corresponding to electron energies $E_e=20$ GeV and 5 GeV, respectively.
  \label{figkinplotEIC}
}
\end{figure}

The different scenarios
for the lepton and nucleon beam energies of the EIC pseudo-data that are considered here are
listed in Table~\ref{tab:eicpseudodata}.
    As in Ref.~\cite{Aschenauer:2017oxs}, we denote the ``low energy''
EIC scenario the one that consists of measurements taken with electron beams with energy $E_e = 5$ GeV, 
whereas the ``high energy'' EIC scenario corresponds
to measurements taken with $E_e = 20$ GeV electrons.
 We also indicate in Table~\ref{tab:eicpseudodata} the atomic mass number $A$,
 the nucleus (per nucleon) $E_A/A$ energy, the maximum and
 minimum values of $Q^2$ and $x$ of the pseudo-data, respectively, 
 and the number of pseudo-data points $N_{\rm dat}$.
 Here we restrict ourselves again to the inclusive structure functions and do
 not include EIC pseudo-data on charm structure functions.

\begin{table}[t]
   \centering
   \renewcommand{\arraystretch}{1.75}
   \begin{tabular}{l|c|cc|cc|c}
     Scenario  & $A$   & $E_e$ &   $E_A/A$  &  $Q_{\rm max}^2$  &  $x_{\rm min}$ &$ N_{\rm dat}$ \\
 \toprule
     {\tt eRHIC\_5x50C} & 12  &   5 GeV &    50 GeV  &      440 GeV$^2$ &  0.003
     & 50 \\
     {\tt eRHIC\_5x75C} & 12  &   5 GeV &    75 GeV  &      440 GeV$^2$ &  0.002
     & 57 \\
        {\tt eRHIC\_5x100C} & 12  &   5 GeV &    100 GeV  &      780 GeV$^2$ &  0.001
        & 64 \\
    {\tt eRHIC\_5x50Au} & 197  &   5 GeV &    50 GeV  &      440 GeV$^2$ &  0.003
    & 50 \\
     {\tt eRHIC\_5x75Au} & 197  &   5 GeV &    75 GeV  &      440 GeV$^2$ &  0.002
     & 57 \\
       {\tt eRHIC\_5x100Au} & 197  &   5 GeV &    100 GeV  &    780  GeV$^2$ &  0.001
       & 64 \\
       \midrule
         {\tt eRHIC\_20x50C} & 12  &   20 GeV &    50 GeV  &    780 GeV$^2$ &  0.0008
     & 75 \\
     {\tt eRHIC\_20x75C} & 12  &   20 GeV &    75 GeV  &    780 GeV$^2$ &  0.0005
     & 79 \\
        {\tt eRHIC\_20x100C} & 12  &   20 GeV &    100 GeV  &   780 GeV$^2$ &  0.0003
        & 82 \\
             {\tt eRHIC\_20x50Au} & 197  &   20 GeV &    50 GeV  &    780 GeV$^2$ &  0.0008
     & 75 \\
     {\tt eRHIC\_20x75Au} & 197  &   20 GeV &    75 GeV  &   780 GeV$^2$ &  0.0005
     & 79 \\
        {\tt eRHIC\_20x100Au} & 197  &   20 GeV &    100 GeV  &    780 GeV$^2$ &  0.0003
        & 82 \\
     \bottomrule
   \end{tabular}
   \vspace{0.3cm}
   \caption{\label{tab:eicpseudodata}
  The different scenarios for the EIC pseudo-data considered here.
  For each scenario, we indicate the atomic mass number $A$, the electron energy
  $E_e$, the nucleus (per nucleon) energy $E_A/A$, the maximum value of $Q^2$ and
  the minimum value of $x$ reached, and the number of pseudo-data points $N_{\rm dat}$.
  The upper part of the table corresponds to the ``low energy'' scenario (with $E_e=5$ GeV)
  while the lower one to the ``high energy'' scenario (with $E_e=20$ GeV).
   }
 \end{table}

In Fig.~\ref{figkinplotEIC}, we display the 
kinematic coverage of the EIC pseudo-data
compared to the existing lepton-nucleus 
scattering measurements.
Here we can see that 
the EIC would significantly extend the sensitivity to nPDFs both
in the small-$x$ and in the large-$Q^2$ regions.
This is
particularly marked
for the higher energy scenario with $E_e=20$ GeV and $E_A/A=100$ GeV, where
the kinematic coverage at small $x$ in the perturbative region $Q \gsim 1$ GeV
would be increased by a factor 20.

In this exercise, we assume that the ``true'' central value
of the EIC pseudo-data is the central value
of the nNNPDF1.0 NLO fit, which is then fluctuated according
to the corresponding experimental uncertainties.
Given that we also use NLO QCD theory for the fits including
EIC data,
by construction the resulting fits
are expected to have $\chi^2_{\rm EIC}/N_{\rm dat}\simeq 1$.
Concerning the projected experimental uncertainties for the 
EIC pseudo-data, we assume the same total relative
uncertainty as in Ref.~\cite{Paukkunen:2017phq}, which is taken to
be uncorrelated among different bins.
Moreover, each of the scenarios listed in Table~\ref{tab:eicpseudodata}
are assumed to have a $\delta_{\mathcal{L}}=1.98\%$ normalization uncertainty,
not included in the total uncertainty mentioned above.
This normalization error is taken to be fully correlated among each scenario
but uncorrelated among the different scenarios.
Note that the different nucleus energies in Table~\ref{tab:eicpseudodata} are 
statistically independent, so they can be added simultaneously to the same nPDF fit
without any double counting.

\begin{figure}[t]
\begin{center}
  \includegraphics[width=0.99\textwidth]{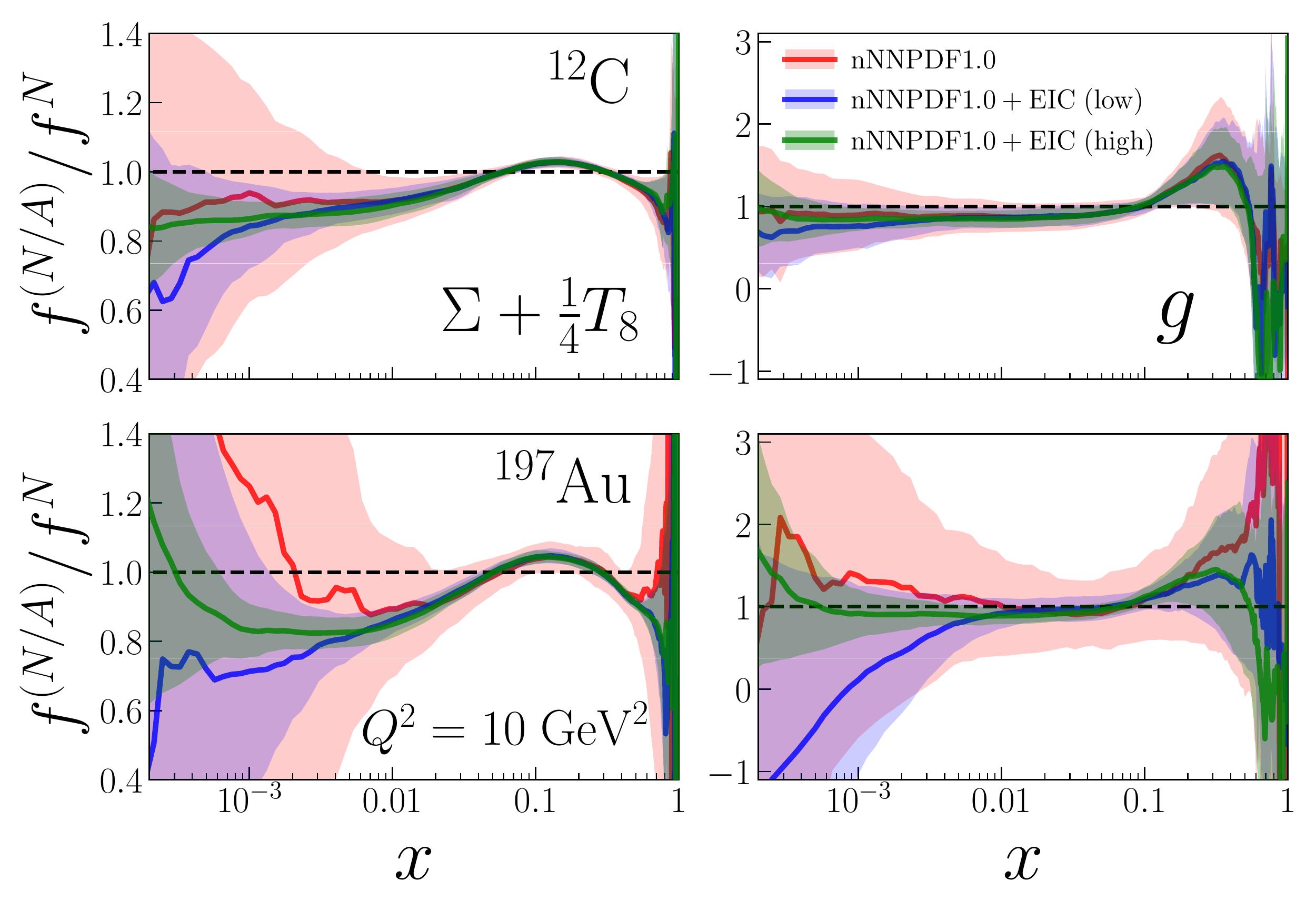}
 \end{center}
\vspace{-0.7cm}
\caption{\small Comparison between the nNNPDF1.0 NLO fit 
 (solid red line and shaded band) and
  the fits where ``low energy" (solid blue line and shaded band) 
  and ``high energy" (solid green line and shaded band) 
  EIC pseudo-data have been added.
  The the quark combination
  $\Sigma + \frac{1}{4}T_8$ (left panels) and gluon (right panels)
  ratios with respect to the corresponding $A=1$ distribution 
  are shown at $Q^2=10$ GeV$^2$ for $^{12}$C (top panels) 
  and $^{197}$Au (bottom panels) nuclei. 
 }
\label{fig:PDFs_EICcomparison}
\end{figure}

The results of the fits are given in Fig.~\ref{fig:PDFs_EICcomparison}, where a
comparison between the nNNPDF1.0 NLO fit and
the corresponding fit including the different EIC pseudo-dataset
scenarios is shown.
We show results at $Q^2=10$ GeV$^2$ for $^{12}$C and $^{197}$Au,
the two nuclei for which pseudo-data is available.
Since by construction the central values of the nNNPDF1.0
and nNNPDF1.0+EIC fits will coincide, 
the most pertinent information is the relative 
reduction of the nPDF uncertainties.
In all cases, we have verified that $\chi^2_{\rm EIC}/N_{\rm dat} \simeq 1$ is
obtained.

From the comparisons in Fig.~\ref{fig:PDFs_EICcomparison}, one finds that the EIC
measurements would lead to a significant reduction in the PDF uncertainties both
for the gluon and for the $\Sigma+T_8/4$ quark combination.
The effect is especially visible for gold ($A=197$), given that the constraint from
the proton boundary condition is much smaller there than for a lighter nuclei
such as carbon ($A=12$).
Here, the improvement can be up to an order of magnitude for $x\simeq 10^{-3}$ 
as compared to the current situation.
Therefore it is clear that such improvement will allow the EIC to carefully study
important dynamics such as the quark and gluon shadowing in addition to the
possible onset of saturation effects down to $x\simeq 5\times10^{-4}$.

From Fig.~\ref{fig:PDFs_EICcomparison} we can also observe that the ``high energy'' scenario
would constrain the nPDFs down to smaller values of $x$ better than the
``low energy'' one, again more notably for heavier nuclei such as gold.
For instance, the uncertainty for the gluon distribution in the region
$x\simeq 5 \times 10^{-4}$ would be around three times larger in the lower energy
case compared to the higher scenario.
Given that saturation effects are expected to scale by $\sim A^{1/3}$, these results
demonstrate that the  ``high energy'' scenario would provide a rather sharper
probe of small-$x$ QCD dynamics that the lower energy option.

\section{Summary and outlook}
\label{sec:summary}

In this work, we have presented a first determination of the
parton distribution functions of nucleons bound within nuclei
using the NNPDF methodology.
Using as experimental input all available measurements on
neutral-current deep-inelastic nuclear structure
functions, we have determined the nuclear gluon $g$, the quark singlet $\Sigma$,
and the quark octet $T_8$ for a range of atomic mass numbers
from $A=2$ to $A=208$.
We find an excellent overall agreement with the fitted experimental data, with stable results
with respect to the order of the
perturbative QCD calculations.
While the quark distributions are reasonably well constrained for $x\gsim 10^{-2}$,
the nuclear gluon PDFs are affected by large uncertainties, in particular
for heavy nuclei.

From the methodological point of view, the main
improvement with respect to previous NNPDF fits has been
the implementation of {\tt TensorFlow} to perform 
stochastic gradient descent
with reverse-mode automatic 
differentiation.
The application of SGD for the $\chi^2$ minimization
has lead to a marked performance improvement as
compared to the evolutionary-type
algorithms used so far in NNPDF.
Two other related developments in this study have been the use of a single
neutral network to parameterize the nPDFs rather than multiple networks, 
and the fitting of the preprocessing
exponents rather than their determination from an iterative procedure.

As opposed to other nPDF analyses, the
nNNPDF1.0 set is determined with the 
boundary condition imposed at the minimization level so that
the baseline proton PDFs (NNPDF3.1) are reproduced 
both in terms of their central values and, more importantly, their uncertainties.
Moreover, we have applied this constraint in a fully consistent way,
since the proton PDF baseline has been determined using
the same fitting methodology and theoretical settings.
We have shown that this $A=1$ constraint results in
a significant reduction of the nPDF
uncertainties, especially for low-$A$ nuclei,
and therefore represents a vital ingredient
for any nPDF analysis. 

By using nNNPDF1.0 as a baseline, we have also quantified the impact of future
e+A measurements from an Electron-Ion Collider by exploiting the projections
generated in Ref.~\cite{Aschenauer:2017oxs}.
We have demonstrated that the EIC measurements of 
inclusive nuclear structure functions would constrain the quark
and gluon nuclear PDFs down to $x\simeq 5\times 10^{-4}$, opening
a new window to study the nuclear modification of the proton substructure
in the small-$x$ region.
With future EIC measurments, it will therefore be possible to
construct a reliable nPDF set based on collinear
factorization that can identify and isolate the onset of novel
QCD regimes such as non-linear evolution effects or
small-$x$ resummation.

The main limitations of the present work are the lack
of a reliable separation between the quark flavors,
which is not possible from neutral-current DIS measurements alone, as well
as the large uncertainties that affect the nuclear gluon PDFs.
This implies that the possible phenomenological applications of nNNPDF1.0
are restricted to processes that do not require a complete quark flavor
separation, such as the analysis of EIC structure functions in Sect.~\ref{sec:eic}, or
$D$ meson production in p+Pb collisions~\cite{Kusina:2017gkz}.
To bypass these limitations, we plan to extend the present nPDF 
analysis to a global dataset including neutrino-induced
charged-current deep-inelastic structure functions
as well as inclusive jets and dijets, photons, electroweak
boson production, and heavy quark production from proton-ion collisions
from RHIC and the LHC.

\vspace{0.5cm}
\hrule
\vspace{0.5cm}

\noindent
The results of this work, namely the nNNPDF1.0 NLO and NNLO
sets for different values of $A$, are available via 
the {\tt LHAPDF} library~\cite{Buckley:2014ana}, and have also been
linked to the NNPDF website:
\begin{center}
\url{http://nnpdf.mi.infn.it/for-users/nuclear-pdf-sets/}
\end{center}  
These {\tt LHAPDF} grid files contains $N_{\rm rep}=250$ replicas each,
which are fully correlated between different values of $A$ as discussed
in Sect.~\ref{sec:results}.

Moreover, due to the lack of
a complete quark flavour separation, additional
assumptions might be required when the nNNPDF1.0 sets are used, in particular
for phenomenological applications in heavy-ion collisions.
To comply with the {\tt LHAPDF} format, we have assumed that
$u=d$ and that $\bar{u}=\bar{d}=\bar{s}=s$, namely a symmetric quark sea.
With this convention, the
only meaningfully constrained quark
combinations can be reconstructed using the flavour basis PDFs by means of
$\Sigma=2\,u+4\,\bar{u}$ and $T_8 = 2(u-\bar{u})$.

\subsection*{Acknowledgements}

We are grateful to our colleagues of the NNPDF collaboration
for illuminating discussions.
We thank Nathan Hartland and Carlota Andr\'es Casas for collaboration
in the initial stages of this work.
We are grateful to Alexander Kusina and Hannu Paukkunen for
information about the nCTEQ15 and EPPS16 nuclear sets.
We thank Hannu Paukkunen and Pia Zurita for sharing with us their
projections for the EIC pseudo-data.
J.~R. is partially supported by the European Research Council Starting
Grant ``PDF4BSM''.
R.~A..K., J.~E., and J.~R. are supported by the Netherlands Organization for Scientific
Research (NWO).
%

\bibliography{nNNPDF10}

\end{document}